\documentclass[twocolumn,longbib]{aastex7} 

\submitjournal{The Planetary Science Journal}

\usepackage{amssymb}
\usepackage{enumitem}

\usepackage{amsmath}
\usepackage{booktabs} 
\usepackage{gensymb} 
\usepackage{pifont}
\usepackage{comment}

\begin{document}

\title{
Thermal and rotational effects of giant impacts during terrestrial planet accretion}

\shorttitle{Thermal \& rotational effects of giant impacts}


\author[0009-0009-5324-4184]{Adriana N. Postema}
\affiliation{University of California, Davis, USA} 
\email[]{anpostema@ucdavis.edu}
\author[0000-0001-5365-9616]{Simon J. Lock}
\affiliation{University of Bristol, UK}
\email[]{s.lock@bristol.ac.uk}
\author[0000-0001-9606-1593]{Sarah T. Stewart}
\affiliation{Arizona State University, USA}
\email[]{sstewa56@asu.edu}

\shortauthors{Postema, Lock, Stewart}
\correspondingauthor{A.N. Postema, anpostema@ucdavis.edu}

\begin{abstract}
Terrestrial planets likely experienced one or more giant impacts during their formation that inflicted large thermal, chemical, and rotational perturbations. The early states of terrestrial planets are expected to be dominated by the thermal and rotational outcomes of giant impacts, but critical parameters that control internal processes, such as the pressures and temperatures of core formation, are not fully understood. Here we present the results from a representative suite of collisions between Moon- to super-Earth-mass bodies using the SWIFT hydrocode and updated ANEOS equations of state, allowing more robust temperature calculations. Using these results and the HERCULES planetary structure code, we calculated the contributions from thermal energy, gravitational potential energy, and post-impact rotation on the pressure-temperature conditions of the core-mantle boundary (CMB). We derived scaling laws for the efficiency of impact heating, mantle-core heat partitioning, and CMB pressures and temperatures. We find that post-impact CMB pressures are generally lower than previously assumed, due to both thermal and rotational effects. Full mantle melting is common and a substantial fraction of mantle material is heated above the Fe-MgO solvus closure temperature for impacts with modified specific energies $Q_S>10^6$ J/kg, implying that a miscible layer could form close to the CMB for many giant impacts. The comparatively low internal pressures and large regions of metal-silicate miscibility after giant impacts have significant effects on the processes of core formation, and our work indicates that metal-silicate equilibration would occur near the CMB during later post-impact cooling, consistent with Earth's geochemistry.

\end{abstract}

\section{Introduction}

Giant impacts are an important stochastic process during the formation of rocky planets \citep{wetherill_occurrence_1985,Quintana16,CHAMBERS2004241,agnor_character_1999}. 
Almost all large planetary bodies are expected to experience one or more collisions with another protoplanet, e.g., a body of moon- to Mars-mass or greater.
These energetic events can melt or vaporize substantial fractions of the mantle \citep{Tonks93,stevenson1990j,Nakajima2015,Lock17,caracas2023}. Because impacts are generally oblique, these collisions also imparted a substantial amount of angular momentum, which reduced the pressure gradients in the post-impact bodies \citep{LockStewart2019}. With increased angular momentum, a body becomes extended in the equatorial plane due to centrifugal forces and the internal pressure decreases.
This effect is amplified by the decrease in average density of the final bodies due to heating and vaporization. As a result, the giant impact phase of planet formation is characterized by stochastic episodes of decreased internal pressures even when a collision leads to an increase in mass \citep{LockStewart2019}.

As each impact creates a significant chemical and physical deviation from incremental growth, it may be possible to use them to constrain major events in a planet's history, such as the Moon-forming impact for Earth \citep[e.g.,][]{gabriel_role_2023,lock_geochemical_2020,Piet2017}. Chemical equilibration between metals and silicates may occur during and after a giant impact, depending on the event. 
Elemental partitioning between the metal and silicate phases is highly dependent on the local pressure, temperature, and fugacity conditions, as well as the composition of the equilibrating system \citep{righter_prediction_1997,fischer_sensitivities_2017,chidester_lithophile_2022}. 
Various tracers of accretion, particularly the moderately siderophile elements such as V, Cr, Mn, Co, Ni, Cu, W \citep{RUBIE201131,nimmo_tungsten_2010,deguen_turbulent_2014,fischer_effects_2018,rubie_tungsten_2025}, record the cumulative effects of core segregation in a body. The concentrations of moderately siderophile elements in Earth's present-day mantle can thus be used to construct a history of equilibration and offer important constraints on the conditions of core formation \citep{gaetani_partitioning_1997,Wade2005,Li1996}. However, relating these data to the dynamics of Earth's accretion requires understanding how giant impacts affect the thermal and pressure structures of growing planets.

Because more realistic descriptions of post-impact thermal and dynamic structures have not been available, previous core formation models have assumed that the pressure profiles of the growing bodies were solely and monotonically dependent on the total mass of each body. The effects of stochastic accretion and the timing of giant impacts have been partially included in such models by coupling $N$-body simulation results with chemical equilibrium calculations \citep[e.g.,][]{RUBIE201131,Rubie15,fischer_sensitivities_2017,brennan_timing_2022}. 
Using a simplified analytic model for CMB pressure at the liquidus temperature, \citet{Rubie15} inferred that metal-silicate equilibration occurred at a fraction (0.58-0.73) of the estimated CMB pressures at the time of each event during Earth's accretion. 
This result was consistent with previous works using simpler models \cite[e.g.][]{Li1996,Wade2005}. The equilibration pressures were interpreted to imply that the chemical signatures of siderophile element partitioning in the mantle are dominated by equilibration of a downgoing impactor core with liquid mantle material at the bottom of a partial mantle magma ocean \citep{Rubie15}.
This model has been the dominant framework for understanding core formation, but its consistency with the outcomes of impact simulations, particularly for larger collisions, is not clear.

At present we lack this detailed understanding for a broad parameter space of giant impacts to allow us to more realistically relate geochemical observables to Earth's accretion history.
The thermal outcomes, energy budgets, and heating efficiencies of giant impacts vary significantly with the impact conditions \citep{Carter2020,LockStewart2019,Stewartinprep}. For example, energy deposition is different for hit-and-run events \citep[where \added{a ``runner''} projectile escapes from the target after a grazing contact,][]{agnor_accretion_2004,asphaug_hit-and-run_2006}, compared to graze-and-merge events (where the bodies separate after an initial grazing contact, followed by an accretionary merger). The various dynamical outcomes of giant impacts, ranging from catastrophic disruption to perfect merging, are described in \citet{Leinhardt12}. 
Previous systematic studies of giant impacts have largely explored the mechanical outcomes and rarely focused on the post-impact thermal states of planets \citep[e.g.,][]{Cambioni2019,Leinhardt12,Stewart12}. 
There is fundamental work on impact outcome scaling laws as functions of impact angle (\(\theta_i\)), target-to-impactor mass ratio (\(M_T/M_i\)), and scaled impact velocity ($v_{i}/v_{esc}$) based on impact simulation results. Such studies were focused on understanding the redistribution of mass and momentum and did not investigate the resulting impact heating \citep{benz_catastrophic_1999,Leinhardt12,Movshovitz16,Marcus09,Stewart09,Stewart12}.

A key barrier to extending such works to develop realistic descriptions of post-impact thermal structures has been the unreliable temperatures in equation of state (EOS) models. EOS models describe the relationships between the thermodynamic state variables of a material, where pressure as a function of specific volume (i.e., inverse density) and specific internal energy is the minimum information needed for hydrocode simulations. Most modern EOS tables \citep[e.g., using the M-ANEOS code package,][]{thompson_s_l_2019_3525030} provide pressure, specific internal energy, specific entropy, and sound speed on a density-temperature grid. High-quality EOS models are essential to accurately determine the thermal state of bodies following giant impacts \citep{Stewart2020}. Recently, \citet{Nakajima2021} conducted impact simulations to develop scaling laws to approximate the extent of energy deposition and melting as a function of impact parameters. However, this work used older EOS models that did not include a melt curve and contained inaccurate heat capacities in the high-temperature, high-pressure liquid region. They calculated melt fractions as a post-processing step and did not develop a scaling law for post-impact pressure that included the effects of post-impact spin \citep{LockStewart2019}. As a result, we still lack a robust understanding of post-impact thermal states that can be used in core formation studies. 


In this work, we present the results of a suite of smoothed particle hydrodynamics (SPH) giant impact simulations to quantify the thermal outcomes and likely conditions of core formation during the formation of terrestrial planets. With an increasing database of laboratory measurements of shock Hugoniots (loci of states attained by shock waves from different initial states) and off-Hugoniot states \citep[e.g,][]{Kraus2012,Kraus2015,Davies2020a}, EOS models have been updated to span the wide range of pressures and temperatures encountered during a giant impact. For example, the updated M-ANEOS code can include one high-pressure phase transition, the melt curve, and an adjustment for high-pressure heat capacity \citep{ANEOS_fo,ANEOS_fe85,ANEOS_pyrolite}. In addition, \citet{wissing_new_2020} developed an analytic EOS based on a variable polytrope model with a variable Gr\"uneisen parameter, and \citet{jing_new_2011} developed a hard-sphere EOS model for silicate liquids. It is now possible to produce reliable estimates for the thermal states of giant impacts over a large parameter space. 

We varied the target-to-impactor mass ratio, target mass, impact angle, and scaled impact velocity to represent impacts that span small mutual collisions of Moon-sized bodies to super-Earth cases with impacts onto a 1.3M\(_\oplus\) target that produce bodies up to two Earth masses. The cosmochemical abundance of elements dictates that most inner solar system bodies will initially form with a similar iron mass fraction as Earth, so we held the core fraction constant at 30\% \added{\citep{Unterborn_Panero_2019}}. \added{While Earth-like core mass fractions are observed in many terrestrial exoplanets, there is a range observed in exoplanets that should be investigated in future work \citep{Lichtenberg_Miguel_2025}}. Impacts producing planets much greater than 2M\(_\oplus\) were not investigated due to the uncertainty in EOS models for planetary materials at the high temperatures and pressures achieved in large super-Earth interiors.

The bodies formed in giant impacts can evolve rapidly in the days to years following the impact. A post-impact body of a collision onto a target greater than Mars-mass is substantially vaporized, extended, and has some amount of mass in orbit. Further, as we will demonstrate in this work, the most likely impact parameters for accretionary impacts involving $>\sim$0.1M$_\oplus$ deposit sufficient energy and angular momentum to create a synestia, a body that exceeds the corotation limit (CoRoL) of a planet \citep{Lock17}. The CoRoL of a body is the combination of thermal energy and angular momentum of a body beyond which the rotational velocity at the equator intersects the Keplerian orbital velocity. Beyond the CoRoL, there is no hydrostatic solution for the internal structure of the system with a constant rotation rate, and the body does not satisfy the general concept of a planet. We adopt the definitions from \citet{Lock17} for such post-impact structures. A synestia has a corotating central mass that is continuous (in angular velocity and density) with an outer disk-like region, which has a sub-Keplerian rotational profile that is partially supported by a pressure gradient with semi-major axis. A co-CoRoL system is a planet-disk system, where the central corotating body is near the corotation limit and there is a shearing boundary (in angular velocity) between the planet and disk. For generality, we refer to the largest remnant after a giant impact as a ``body'' rather than a planet. 

As a synestia cools, a portion of the disk-like region will condense and ``fall'' into the corotating region. The redistribution of mass and thermal contraction during post-impact evolution decreases the moment of inertia and the body's rotation rate must increase to conserve angular momentum \citep{LockStewart2019,Lock2020}. The details of the process are not well studied. For this work, we will examine the thermal profiles of the bodies in the days after a giant impact and compare them to thermal profiles approximating the conditions for solidification of the mantle. For simplicity of nomenclature, we will use the term mantle to refer to the silicate particles that are corotating after the event; similarly the core refers to the corotating iron alloy particles.

We use our simulations to determine how impact energy is deposited into the largest remnant and how that energy is partitioned between the participating gravitational potential energy, kinetic energy, and internal energy in the core and mantle. 
To quantify the effect that this heating has on metal-silicate equilibration conditions in accretionary collisions, we present scaling laws for the change in temperature and pressure at the CMB from the pre-impact target planet to the resulting post-impact body. We discuss the implications of our results for the miscibility of iron and silicate materials immediately after impacts, as well as for the pressure-temperature conditions at the CMB in the days following a giant impact. 
Our results are widely applicable for estimating the thermal perturbations to rocky planets during the giant planet stage of accretion.

\section{Methods}
\label{sec:methods}
\subsection{Impact simulations with the SWIFT SPH code}
Giant impact simulations primarily utilize Lagrangian SPH methods, which can efficiently model rotating bodies and large changes in the spatial domain. The SPH method models compressible fluid flows by using a system of equal-mass particles obeying the hydrodynamic equations of motion including self-gravity. The code simultaneously solves the conservation equations for mass, momentum, and energy with the use of an EOS model that relates material pressure, density, and internal energy (see \S \ref{sec:methods:EOS}). Each particle has a dynamic smoothing length that becomes small when nearest-neighbor particles are close by and grows larger when particles are dispersed, translating to a decrease in material density. The system of particles represents a continuum fluid via the use of the kernel function dynamically applied to each particle at each time step to calculate state variables    
\citep{SPH}. 

The SPH implementation used in this study is SWIFT \citep[SPH With Inter-dependent Fine grain Tasking, v0.9.0][]{swift}, which has been modified to enable simulation of planetary giant impacts \citep{kegerreis_seagen_2019}. Like many SPH codes, in order to increase computational efficiency, SWIFT includes a fixed spatial domain bounding box and a customizable cap on smoothing lengths which limits time spent searching for nearest neighbors in extremely diffuse areas. However, this cap on smoothing length also imposes a density floor on particles within the simulation, as density cannot be resolved smaller than mass of a single SPH particle within a neighbor-less sphere with a radius roughly equal to the maximum smoothing length. This minimum density can result in low-density regions of the problem domain being incorrectly modeled if the maximum smoothing length is set too small \citep{hull_effect_2024}. In this study, the maximum smoothing length was varied with the mass resolution of the simulation such that the lowest attainable density was always less than 0.5 kg/m$^3$. The SWIFT bounding box improves computation speed by removing particles from the simulation as they leave the domain. Setting this box size too small results in significant losses of energy and information when bound clumps of particles are lost beyond the boundary. As a result, we set the box size to be 9000 \(R_\oplus\) so that no bound groups of particles would be reasonably able to exit before the end of the simulation.

In this study, we neglected material strength. \citet{emsenhuber_sph_2018,emsenhuber_sph_2024} found that material strength has little effect in the regime of giant impacts between protoplanets with mass \(\geq0.1\)M\(_\oplus\) due to the warm internal temperatures of pre-impact bodies during accretion and the substantial heating and melting that occurs during these impact events. Based on these previous works, we also do not expect strength to be a major factor in the outcome of impacts between smaller (\(<0.1\)M\(_\oplus\)) bodies with warm thermal profiles, as expected during planet formation (see discussion in \S \ref{subsec:strength}). 
Material strength will still affect the deformation of solid bodies as they approach prior to impact, which in turn perturbs the impact geometries. The details of the effects of material strength are left for future work.

The version of SWIFT used in this study follows the Monaghan form of density-energy SPH artificial viscosity to broaden shock fronts in order to avoid numerical instabilities \citep{monaghan_shock_1983,monaghan_smoothed_1992,monaghan_sph_1997}. This implementation also uses the Balsara shear switch to properly ``turn off" artificial viscosity in shearing flows \citep{balsara_von_1995}. We use the tabular equations of state feature in SWIFT with ANEOS equation of state models.


\subsection{ANEOS Equations of State}
\label{sec:methods:EOS}

A key challenge for simulating giant impact events is the wide range of pressures and temperatures experienced by the colliding bodies. The EOS models used in such simulations must capture the changes in the energy surfaces with melting and vaporization, a task made harder by the fact that geologic materials are not well represented by commonly used analytic EOS formulations \citep{emsenhuber_sph_2018}.

In this study, we use EOS constructed using the ANEOS code package \citep{Thompson1972,Melosh07,Stewart2020}. The ANEOS package is built on a collection of (semi)analytic approximations for thermodynamic variables in the equation of state that are smoothed over various regions in temperature-density space \citep{Thompson1972}. While the core ANEOS model uses analytic expressions that hold well for metals, those formulations are not able to evaluate realistic geologic materials and several modifications have been made to the ANEOS code in order to match experimental data \citep{Melosh07}. The original ANEOS code could model solid-liquid-vapor phases \citep{Thompson1972}, but did not include a complete treatment of high pressure solid phase transformations. As dense high-pressure crystal structures are important features in rocks and minerals, many ANEOS models neglected the melt transition in favor of including the high-pressure transformation \citep[e.g.,][]{Melosh07}. However, more recent developments \citep{thompson_s_l_2019_3525030} allow for inclusion of both a melt curve and the high-pressure solid phase but the code has not been extended to include a triple point solution and the high-pressure transformation extends artificially into the liquid field. We use the formulation of ANEOS developed by \citet{Stewart2020} which includes a treatment of the melting phase transition for geologic materials along with the ability to modify the high-temperature heat capacity in order to correct for deviations from the Dulong-Petit limit. The Dulong-Petit limit is the theoretical expectation for an ideal material that the heat capacity, \(C_v\rightarrow 3nR\) as temperature \(T\rightarrow\infty\) where $n$ is the number of moles and $R$ is the gas constant. We emphasize that the heat capacity modification is essential to attaining more realistic temperatures in numerical simulations. Prior to this development, EOS model shock temperatures were routinely much greater than those found experimentally. Inferences about the thermodynamic outcomes of giant impacts are much more robust from simulations using the newer EOS models.

We used EOS models for an iron alloy, to represent the cores of colliding bodies, and pyrolite, to represent the mantles. These EOS take advantages of the developments in the ANEOS code package described above and recent laboratory data on these materials \citep{ANEOS_fe85,ANEOS_pyrolite}. Pyrolite is a model chemical composition for the Earth's bulk mantle \citep{MCDONOUGH1995223} and the pyrolite EOS model was developed as a proxy for a terrestrial magma ocean \citep{ANEOS_pyrolite}. Pyrolite is denser than pure forsterite (Mg$_2$SiO$_4$, which has been used in many previous studies of giant impacts) because it includes iron and more accurately represents a mixture of iron-bearing phases in the mantle (e.g., olivine, bronzite, or bridgmanite). ANEOS is limited to modeling single-component systems and the pyrolite melt curve was chosen to fall between the solidus and liquidus of the mantle.  
Using an Fe-85,Si-15wt\%
alloy \citep{ANEOS_fe85} accounts for the fact that planetary cores incorporate light elements including silicon during accretion and are so lighter than pure iron \citep[e.g.,][]{li_experimental_2007,liu_hydrogen_2024,wanke_new_1997}. Neither the pyrolite or iron alloy ANEOS models include high-pressure solid phases. The melt curves of the two EOS are shown in Figure~\ref{fig:rubie_ICs} (black solid and dashed lines).

Tabulated version of the ANEOS models were used by the codes in this work. The tables are documented in the GitHub repository for each material: ANEOS pyrolite \citep[version SLVTv0.2G1,][]{ANEOS_pyrolite}, and ANEOS Fe-85,Si-15wt\% alloy \citep[version SLVTv0.2G1,][]{ANEOS_fe85}. SWIFT calculates thermodynamic values from EOS tables by using bilinear interpolation with log values for density, specific internal energy, and specific internal entropy; and bilinear interpolation with linear values for pressure and sound speed. Default ANEOS tables include a solid tension region at low temperatures. 
This version of SWIFT does not implement tension, so all negative or zero pressure values included in EOS tables that are input into SWIFT's table validation code are forced to be positive, non-zero, and monotonically increasing. Negative and zero pressure values replaced by very small positive values on the order of C machine precision \citep{swift}. Removing the tension regions in ANEOS tables prior to using SWIFT's table validation routine results in larger entropy and energy errors, as the validation routine will attempt to double-correct some parts of the tension region, or attempt to fix regions of the no-tension table that would not need to be adjusted for SWIFT if the tension region was left intact.
Our tabulated ANEOS tables are composed of 860 density points between 0 and 29800 kg/m$^3$ with 744 temperature points between 1 and 8.83$\times10^6$ K for the pyrolite mantle, and 986 density points between 0 and 99300 kg/m$^3$ with 1028 temperature points between 1 and 9.55$\times10^6$ K for the Fe-85,Si-15wt\% iron alloy core. The EOS tables are included in the Supplementary GitHub Repository.

In comparison to analytic EOS models, tabulated versions do produce some amount of interpolation error, especially around phase boundaries. As a result, the ANEOS tables used here are very finely gridded around melt curves and other phase boundaries. Total energy errors associated with the simulations included in this study were \added{usually} on the order of tenths of a percent or less. \added{A small number of simulations obtained energy errors on the order of a percent due to either hit-and-run particles leaving the simulation box, or simulations in which the baryon softening SPH setting was left slightly too low.} Overall, energy errors were influenced by anomalous gravitational interactions of particle pairs resulting from limitations of the SPH method instead of EOS table interpolation. \added{Data on energy conservation and all other extracted simulation quantities is included in a data table file in the companion GitHub repository.}

\subsection{SPH Planet Initialization}
\label{sec:methods:SPH_init}
\begin{figure}
    \centering
    \includegraphics[width=0.45\textwidth]{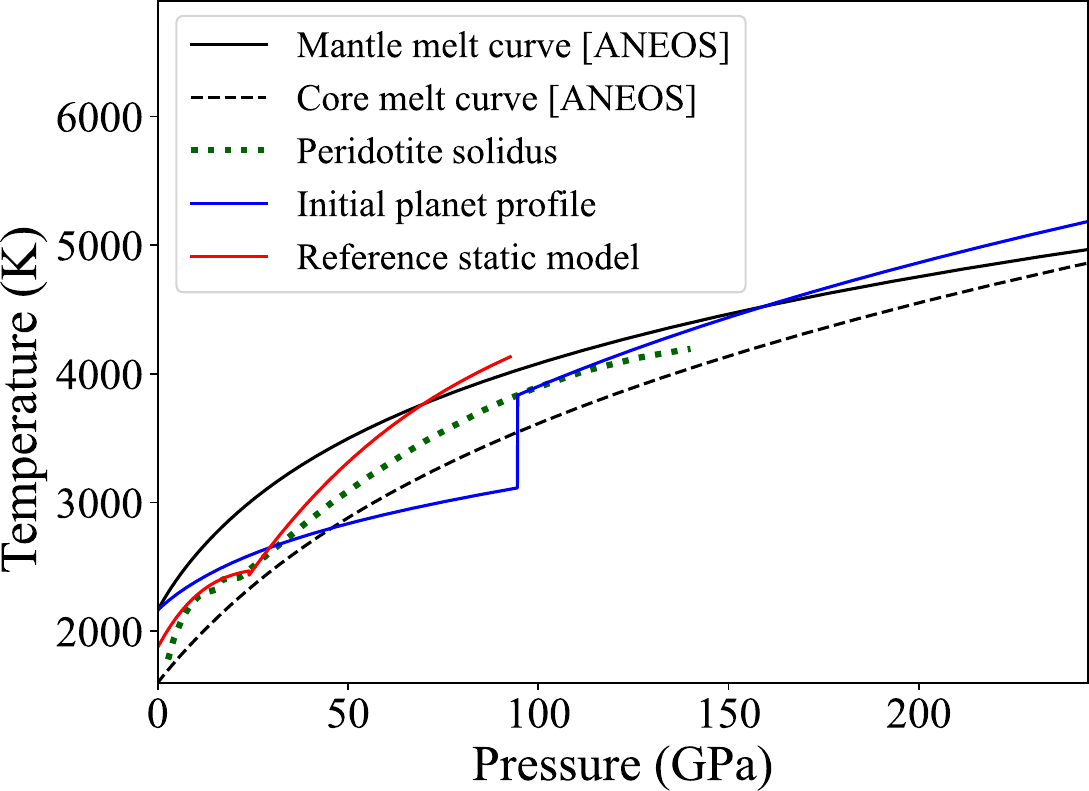}
    \caption{Example initial thermal profile for a 0.71\(M_\oplus\) planet (blue line) compared to the peridotite solidus \citep{Fiquet2010}, ANEOS pyrolite melt curve (black line), and ANEOS iron alloy melt curve (dashed line). Core-mantle equilibration calculations typically use the static model thermal profile (Equation \ref{eqn:rubie}, red line for same body as blue line). Since the reference static model only concerns itself with the CMB temperature, the core is modeled as just a single temperature reservoir that meets the mantle at the CMB, and as such is not shown here.}
    \label{fig:rubie_ICs}
\end{figure}

All of the initial SPH planet models in this study were differentiated metal-silicate bodies with 30\% core material by mass. An initial estimate of the structure of the planets was found by iterating a one-dimensional hydrostatic structure calculation over various central pressures until the resulting planet converged to a desired total mass. Their mantle and core thermal profiles were isentropes defined by potential temperatures at the surface and CMB, respectively. The chosen mantle pressure-temperature profiles (blue line in Figure \ref{fig:rubie_ICs}) met the modeled pyrolite melt curve (black line in Figure \ref{fig:rubie_ICs}) at the surface pressure (defined as 1 bar) of our planet models (at $\sim$2200 K), as mantle thermal profiles likely fall just below the solidus during accretion \citep{Solomatov1993, Andrault2016}. Core pressure-temperature (P-T) profiles were set so as to meet the peridotite solidus of \citet{Fiquet2010} (dotted green line in Figure \ref{fig:rubie_ICs}) at the CMB, or the mantle isentrope if the CMB was at less than \(\sim 33\) GPa in order to prevent the mantle from being hotter than the core in smaller bodies. Core temperature profiles that meet the mantle solidus at the CMB are likely long-lived states during accretion \citep{Andrault2016}. An example initial thermal profile that meets the peridotite solidus at the CMB is shown in Figure~\ref{fig:rubie_ICs} (blue line). 

The motivation of this initial thermal profile is based on the expected changes in heat transfer through the planet. After dissipation of any initial thermal stratification \citep{Lock17}, convective heat transfer through an adiabatic magma ocean is orders of magnitude faster than through a solid mantle. When both the mantle and core are fluid, the thermal boundary layer is small. The subsequent freezing of the base of the magma ocean drastically reduces the efficiency of heat flow out of the core such that the outer core boundary temperature remains relatively close to the mantle solidus after the mantle solidifies \citep{Andrault2016}. 


Our model planets were initialized from these P-T profiles at a resolution of \(N_{SPH}\sim2\times10^5\) particles using SEAGen \citep{kegerreis_seagen_2019} with spherically concentric particle placement. The final mass and radius of the initialized SPH planet does not quite match that initially specified due to accommodations made for particle placement by SEAGen and these values vary slightly with particle resolution. Each body was then relaxed by running an SPH simulation of the isolated body for $1.20\times10^5$~s, after which the initial concentric shells placed by SEAGen were smoothed into a continuum, oscillations in thermodynamic quantities had converged, and residual velocities became small (typically \(<15\)m/s). During the initialization simulation, the entropy profiles of the core and mantle were fixed to their desired values using the SWIFT ``--enable-planetary-fixed-entropy'' compilation option with additional modifications to remove viscous heating and add velocity damping (see Supplementary Materials GitHub repository). 

At material interfaces, the pressure, density, and temperature profiles of the resulting planet developed discontinuities arising from assumptions about the differentiability of the density field inherent in standard SPH \citep[e.g.,][]{hosono_comparison_2016}. Some theoretical methods to address these discontinuities exist \citep[e.g.,][]{ruiz-bonilla_dealing_2022,sandnes_remix_2025}, but robust verification of thermal outcomes in implementations of these methods are not available. After initialization, the exact radius of the planet often shifts slightly due to the variation of the internal pressure field. 
Because the radii vary with the number and initial placement of particles, we use the radii of the analytic profiles when determining the initial placement of colliding planets in order to be independent of resolution. When initializing impact simulations, the planets are given an initial velocity and placed a distance apart such that the collision occurs at approximately one hour of simulation time at the specified velocity and impact angle. These conditions are calculated using the WoMa Python package \citep{ruiz-bonilla_woma_2020}, which does not account for the effects of tidal deformation as the bodies approach each other. While the physical effects due to pre-impact deformation warrant further investigation, the heating contributions are negligible compared to the thermal effects of the impact itself.

\subsection{Giant impact parameter space}
The choice of impact parameter space explored in our study (Table~\ref{tab:parameters}) was motivated by collisions that occur in a wide range of Solar system formation scenarios, from dynamically cold (low orbital eccentricities and inclinations) to excited cases like the Grand Tack model \citep{Walsh11,Carter15}. Sample impact probability distributions were sourced from $N$-body simulations performed by \citet{Carter15} and \citet{Carter2022}, and expanded to higher values of target-impactor total mass to allow for applications to exoplanet systems that contain super-Earths. Collision angles \(\theta_i\) ranged from zero to near-90 degrees, and scaled impact velocities \(V_i/v_{esc}\) ranged from 1.0 to 2.0, where \(V_i\) is absolute impact velocity (speed of the projectile in the frame of the target when the projectile and target first make contact) and 
\begin{equation}
  v_{esc}=\sqrt{2G(M_T+M_i)/(R_T+R_i)}  
\end{equation}
is the mutual escape velocity. \(G\) the gravitational constant, \(M\) and \(R\) are planet mass and radius with subscripts \(T\) and \(i\) indicating the target and impactor bodies, respectively. This parameter space covers the majority of accretionary collisions \citep[collisions that result in non-erosive mergers or graze-and-merge events,][]{Stewart12}. Hit-and-run outcomes are also sampled by this parameter space, though not as extensively. Hit-and-run impacts can lead to significant thermal processing in the target bodies, but the much larger velocity distribution that they cover is beyond the scope of this study, as is the degree of thermal processing on the runner. The target planets for these impact simulations ranged from lunar-sized (\(0.01M_{\oplus}\)) to proto-super-Earths (\(1.3M_{\oplus}\)), with target-impactor mass ratios of 1:2, 1:3, and 1:6. All of our initial planets were non-rotating as including pre-impact spin \added{was beyond our computational resources as it would increase the size of the parameter space by an order of magnitude}. A complete list of the impact scenarios included in this study is presented in Appendix Table \ref{tbl:ICs}.

All of our impact simulations were run for at least 48 hours of in-simulation time, after which the bound angular momentum of most of the resulting bodies had stabilized. Some simulations resulted in graze-and-merge impacts, where an initial glancing blow is followed by a slower, sub-escape-velocity accretionary collision, which required longer run times to capture the completion of the collision. An important part of our SWIFT simulation parameters was requiring the maximum baryon softening length (a correction preventing point mass particles from becoming non-physically close to each other and producing an unrealistically large gravitational interaction) to be roughly equal to, or slightly larger than, the inter-particle spacing distance (\(d=2R_p/(N_{SPH}^{1/3})\), where $N_{SPH}$ is the total number of particles in the SPH simulation) of the resulting planets, as values smaller than this produce spurious heating \citep{borgani_hot_2006}.

\begin{table}[]
    \centering
    \begin{tabular}{l|l|l}
         Parameter & Symbol & Values \\ \hline
         Target-impactor & $M_T/M_i$ & 2, 3, 6 \\
          mass ratio && \\
         Target size & $M_T$ & 0.01, 0.1, 0.5, 0.91, 1.3 $M_\oplus$ \\
         Impact angle & $\theta_i$ & 0, 20, 30, 45, 60, 75 [degrees] \\
         Scaled impact  & $V_i/V_{esc}$ & 1, 1.15, 1.5, 2.0 \\ 
         velocity && \\
         Simulation time & $T$ & $\geq$48 hours\\ \hline
         
    \end{tabular}
    \caption{Initial impact parameters that were varied as part of this study and the main values that were used for each. Some impact parameters were tested at more values than the basic ones listed here; see Table \ref{tbl:ICs} for the full list of impact conditions included in this study.}
    \label{tab:parameters}
\end{table}

\subsection{Static planet models}\label{subsec:Splanet} 
The dynamics and timing of the cooling of post-impact bodies is uncertain. Immediately after an impact, these bodies often have extended super-critical atmospheres that lack a traditional magma ocean-atmosphere interface \citep{caracas2023}. Eventually, these bodies cool to become more classical magma oceans overlain by a volatile-dominated atmosphere \citep{Lock2020}, although the details of the evolution of the thermal profiles are not yet known. Previous models did not consider core formation during this supercritical post-impact state, but instead found that the elemental abundances in the mantle and core were potentially set at the bottom of a (partial mantle) magma ocean \citep[e.g.,][]{de_vries_impact-induced_2016,siebert_terrestrial_2013,wood_accretion_2006} as it was freezing at temperatures close to the mantle liquidus \citep{Rubie15}, though other studies have explored super-liquidus models \citep[e.g.,][]{badro_core_2015}. We refer to these freezing magma ocean profiles from previous works as ``static models''
where static indicates that the planet has no rotation.

To model the conditions of metal-silicate equilibration near the mantle liquidus, we follow the prescription of \citet{Rubie15}, where the mantle thermal profile $T(P)$ for these static models is given by \citet[Equation 8 in][]{Rubie15}:
\begin{eqnarray}
    T(P) & = &
        \begin{cases}
            1874+55.43P-1.74P^2+0.0193P^3 \\
            \quad \text{     for } P < 24 \, \text{GPa, and } \\
            1249+58.28P-0.395P^2+0.00113P^3 \\
            \quad \text{     for } P \geq 24 \, \text{GPa.}
        \end{cases}
    \label{eqn:rubie}
\end{eqnarray}
The pressure conditions for the core and mantle are calculated after the form of \citet{turcotte_geodynamics_2002} and given by: 
\small \begin{eqnarray} \label{eqn:T&S12}
    P(r)=
        \begin{cases}
             \frac{2}{3}\pi G\rho_{\text{core}}^2(r_\text{CMB}^2-r^2) +\frac{2}{3}\pi G\rho_{\text{mantle}}^2(r_{\text{surf}}^2-r_{\text{CMB}}^2)\\
            \quad+\frac{4}{3}\pi G\rho_{\text{mantle}} r_{\text{CMB}}^3(\rho_{\text{core}}-\rho_{\text{mantle}})\left(\tfrac{1}{r_{\text{CMB}}}-\tfrac{1}{r_{\text{surf}}}\right)  \\
            \quad\quad\text{   for }r\leq r_\text{CMB}, \,\text{  and} \\
            \frac{4}{3}\pi G\rho_\text{mantle}r_\text{CMB}^3(\rho_\text{core}-\rho_\text{mantle})\left(\frac{1}{r}-\frac{1}{r_\text{surf}}\right)\\
            \quad +\frac{2}{3}\pi G \rho_\text{mantle}^2(r_\text{surf}^2-r^2)\\
            \quad\quad\text{ for }r>r_\text{CMB}
        \end{cases}
\end{eqnarray}
\normalsize \begin{eqnarray} \label{eqn:T&S1}
    r_{\text{CMB}}&=&\left(\frac{M_{\text{core}}}{\tfrac{4}{3}\pi \rho_{\text{core}}}\right)^{1/3} \\
    r_{\text{surf}}&=&\left(\frac{M_{\text{mantle}}+\tfrac{4}{3}\pi \rho_{\text{mantle}}r_{\text{CMB}}^3}{\tfrac{4}{3}\pi \rho_{\text{mantle}}}\right)^{1/3}
   \end{eqnarray}
where \(M_\text{mantle}\) and \(M_\text{core}\) are the mass of the mantle and core of each planet while $\rho_{mantle}$ and $\rho_{core}$ represent density estimates for the mantle and core, calculated after the method of \citet{turcotte_geodynamics_2002}: \(\rho_{core}=2.5\rho_{mantle}\), where $\rho_{mantle}$ is described by a linear relationship as a function of planet mass between the mantle densities of Earth (4500 kg/m$^3$) and Mars (3550 kg/m$^3$) after the fashion of \citet{Rubie15}. This produced the equation \(\rho_{mantle}=(1063.83\textrm{ kg/m}^3)(M_{planet}/M_{\oplus})+3436.17\textrm{ kg/m}^3\). The CMB conditions of these static model profiles are given when \(r=r_\text{CMB}\).

Figure \ref{fig:rubie_ICs} presents a comparison between this static profile (red line) and an example pre-impact thermal profile used as an initial condition in the SPH simulations in our study (blue line) described in \S \ref{sec:methods:SPH_init}. We note that this static mantle profile is not isentropic and is likely to overturn while freezing from the bottom up \citep{boukare_timing_2018,Morison2019}. In previous work \citep{Rubie15}, the static core profile was effectively modeled as an isotherm whose temperature intersected the static model mantle profile at the CMB pressure. 
We note that this profile is only concerned with the CMB temperature and thus models the core as a single-temperature reservoir (which is why a core temperature profile is not depicted in Figure \ref{fig:rubie_ICs}). The main utility of this thermal profile is to determine the $P-T$ conditions for equilibration at the CMB in a non-rotating magma ocean. 


For each impact simulation, we calculated the static model thermal profile for each post-impact planet, using the final core and mantle masses \citep[following][]{turcotte_geodynamics_2002,Rubie15}. A more detailed description of these static profiles is included in \S  \ref{sec:rotational effects}.

\subsection{Rotating planetary structures using the HERCULES code}\label{subsec:meth-HERC}
Oblique giant impacts impart a significant amount of angular momentum to the resulting planetary body.
In order to de-convolve the effects of thermal heating versus rotational forces on post-impact pressure profiles, a comparison must be made between our post-impact SPH bodies and cooler, fully condensed planets with the same mass and angular momenta. 
In this work, we report post-impact angular momenta as a fraction of the angular momentum of the Earth-Moon system, \(L_{EM}=3.5 \times 10^{34}\text{ kg m}^2/\text{s}\) \citep{Canup2004}.
To do this, we use the publicly available HERCULES code to determine the internal structure of corotating planets \citep{Lock17,Lock2018}. The HERCULES code models planets as a series of concentric, constant-density spheroids following equipotential surfaces. The code iteratively solves for the equilibrium structure of the planet, self-consistently updating the thermal, density, and pressure profiles using specified EOS models. The pressure profile of the body is calculated using a first-order integration including both gravitational and centrifugal contributions. 

In our calculations, the initial thermal state for each calculation was the reference static model profile corresponding to each impact simulation (i.e., based on the mantle and core mass of the post-impact body). The initial thermal state of the mantle and core were set from that static model profile by enforcing a mass fraction-entropy relation in the planet. As described in the previous subsection, mantles were modeled as the non-isentropic thermal state described in Equation \ref{eqn:rubie}, and in an attempt to match the conditions described by \citet{Rubie15} as closely as possible, cores were effectively modeled as a single-temperature isothermic reservoir (isentropes with constant density).
The angular momentum of each planet was iteratively increased to match the angular momentum of the corotating region of the post-impact SPH body. As the body is spun up, the internal structure expands with entropy-mass-fraction relationships held constant and the resulting material densities (and pressures) are calculated from the same ANEOS equation of state tables used in our SPH simulations (\S \ref{sec:methods:EOS}). This allows for easy comparisons to SPH simulation results.

While the mantle and core mass-fraction-entropy relationships were held fixed (i.e., each spheroidal layer maintained a constant entropy value) during each HERCULES calculation, the internal pressure profile of each planet changed due to rotational forces and the resulting redistribution of mass. After adding angular momentum to the system, the mantle P-T profile no longer exactly matched the original P-T relationship described in Equation~\ref{eqn:rubie} due to the change in material densities and therefore internal pressure.

We refer to these resulting planets and their thermal profiles as ``rotating models'' as they represent the same reference states for likely conditions of equilibration as the static models, except with added rotation. In effect, these are cooled magma ocean planets where post-impact heating has dissipated but the rotational contributions remain due to the conservation of angular momentum.
Note that the mass of the post-impact body in the disk-like regions was not included in the rotating models because, in some cases, the angular momentum of the entire bound mass exceeded the corotation limit of the cooler, equivalent-mass planet. We identified the inner edge of the disk-like region by observing where the equatorial angular velocity-radius trend of each planet begins to deviate from the constant value of the corotating inner region, which is also the radius where the kinetic energy is greatest. Since some disk mass and angular momentum will eventually be accreted to the final body, our calculations provide a lower limit on the effects of rotation on the cooled planetary profiles.

\section{Results}
\subsection{Giant impact outcomes}
\begin{figure*}
    \centering
    \includegraphics[width=1.0\textwidth]{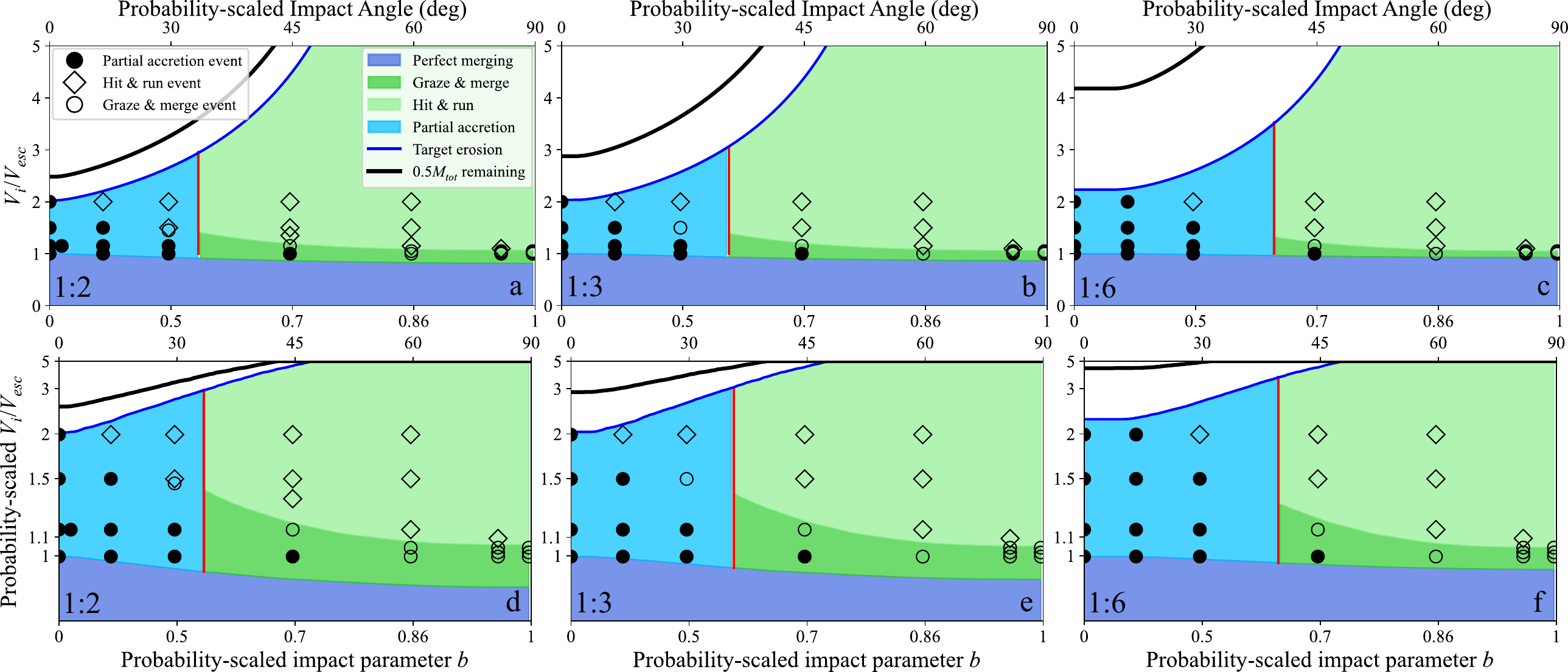}
    \caption{Analytic collision outcome maps for the three impactor-to-target mass ratios (1:2, left column; 1:3, center column; 1:6, right column) included in this study. The horizontal axis of all panels represents impact angle (top axis) or impact parameter (bottom axis) scaled by probability \citep[from][]{shoemaker1962interpretation}, while  the lower row of panels also scale the vertical axis of scaled impact velocity by probability \citep[from N-body data collected in][]{Carter2022}, and thus depict plot area relative to impact event probability \citep[following][]{Stewart12}. 
    The colored regions correspond to the outcomes predicted by \citet{Leinhardt12}, dark blue for perfect mergers, darker green for graze \& merge, light green for hit \& run, and cyan for partial accretion, with white spaces indicating erosive impacts. Symbols denote the collision outcomes observed in our SPH simulations: filled circles for accretionary events, empty circles for graze-and-merge events (a subset of accretionary events), and empty diamonds for hit-and-run events. These collision outcomes do not depend on the total mass of the target and impactor involved, only the mass ratio.
    The blue line indicates the boundary of the erosion regime while the thick black line within the erosion region denotes the catastrophic disruption threshold where the largest remnant is half the total system mass.}
    \label{fig:vscaled_all}
\end{figure*}

The main outcomes of giant impacts are usually categorized by the relative mass of the largest remnant post-impact body \citep{Leinhardt12,Cambioni2019}. The probability of a particular outcome depends on the mass ratio of the colliding bodies and can be visualized in collision outcome maps as shown in Figure~\ref{fig:vscaled_all}. Here, the impact parameter and impact velocity axes are stretched by their respective probability distributions such that the plotted area of each outcome region, denoted by the background color, is proportional to its probability \citep[as in][]{Stewart12}. The velocity probability scaling is based on giant impacts (involving Moon-mass bodies and greater) calculated in the $N$-body terrestrial planet accretion simulations from \citet{Carter15} and \citet{Carter2022}. The impact angle probability distribution is from \citet{shoemaker1962interpretation}.

Perfect merging events (darker blue region) form a final largest remnant body with mass close to the sum of the target and impactor for velocities near or below the mutual escape velocity. Partial accretion of the smaller body onto the larger body (cyan region) occurs at greater velocities for more head-on collisions.
Collisions are erosive when the largest remnant planet is less massive than the original larger body (white region). Catastrophic disruption is defined as when the largest remnant is less than 50\% of the mass of target-impactor system (thick black line denotes the 50\% contour). 

For more oblique collisions (impact parameters greater than the red line denoting the transition from non-grazing to grazing events), hit-and-run events (light green region) are the most common outcome, where the impactor grazes the target planet but its velocity is not reduced sufficiently to be captured. Such collisions can be both slightly erosive or accretionary, as mass can be exchanged during the impact and the largest remnant can have a mass greater or less than the original target. Generally, the unbound impactor remnant, referred to as the runner, decreases in mass with increasing impact velocity. In the presence of multiple planets, about half of runners return to impact the same target at a lower velocity, leading to a higher probability of merging \citep{emsenhuber_fate_2019}.

The graze-and-merge regime (darker green region) occurs when the impactor separates from the target after the initial collision but enough kinetic energy has been lost that the smaller body remains gravitationally bound to the larger body. After a short time period, a secondary merging collision occurs below the mutual escape velocity of the two bodies to form one resulting planet. In this study, we simulate graze-and-merge impacts until the final merger occurs and group the outcomes with accretionary collisions for analysis.

In the analysis of thermal outcomes, we divided our impacts into two broad groups: accretionary and hit-and-run collisions. Using the formalism of \citet{asphaug2010}, we describe the outcomes of our collisions by the accretion efficiency, \(\xi=(M_{bnd}-M_T)/M_i\), where \(M_{bnd}\), \(M_T\), and \(M_i\) are the bound mass of the largest remnant planetary body, the initial target, and the impactor, respectively. \(\xi\) approaches a value of one for perfectly accretionary impacts, is equal to zero for perfect hit-and-run collisions, and becomes negative with erosive impacts. For our dataset, we divide the outcomes into efficient accretion for \(\xi>0.8\), partial accretion for \(0.8>\xi>0.2\), and hit-and-run for \(\xi<0.2\). Note that graze-and-merge outcomes have no distinct \(\xi\) signature value, and we manually identified graze-and-merge events for purposes of outcome categorization. 

Comparing our simulation results to the analytic outcome model from \citet{Leinhardt12}, we see general agreement but some differences (Figure \ref{fig:vscaled_all}). In the \citet{Leinhardt12} model, demarcation between grazing and non-grazing impacts (vertical red line) was given by the critical impact angle from \citet{asphaug2010}, 
\begin{equation}  
    \theta_{\text{crit}}=\text{arcsin}\left(\frac{R_T}{R_T+R_i}\right),
\end{equation}
where \(R_T\) and \(R_i\) are the radii of the target and impactor respectively. Impacts with impact angles greater than \(\theta_{crit}\) are considered grazing impacts, and \(\theta_{crit}\) is a constant for each impactor-to-target mass ratio in the model. However, the impact outcomes we have observed in this study indicate that the boundary between grazing and non-grazing impacts also varies with impact velocity. We observe accretionary events without an initial graze-and-merge impact at angles slightly larger than \(\theta_{crit}\) close to escape velocity, as well as hit-and-run and graze-and-merge events for higher velocity impacts at angles smaller than \(\theta_{\text{crit}}\). 
An explanation for the discrepancies between the analytic models and our data is that the model was based on simulations that began  with tangent planets (the impact begins at time \(t=0\)) which does not allow for pre-impact deformation to occur.

Our results are in agreement with the machine learning method from \citet{Cambioni2019} and \citet{Emsenhuber2020}. They found that angles below \(\theta_{crit}\) can produce hit-and-run outcomes at \(v/v_{esc}\geq2\), though there are some cases where our data does not totally reproduce graze-and-merge collisions expected by the machine learning model of \citet{Emsenhuber2020}. The models of \citet{Cambioni2019} and \citet{Emsenhuber2020} began with the planets ``several radii apart'', which is more comparable to simulations in this study beginning an hour apart. 
The SPH code used by \citet{Cambioni2019} and \citet{Emsenhuber2020} contains a partial implementation of material strength which will lead to different patterns of deformation and fragmentation than those observed in a purely fluid SPH implementation (like SWIFT). This could potentially lead to finer differences in outcomes than focused upon here and notable differences in the state of the runner after hit-and-run impacts, which this study does not focus on. 


While our study contains initial collision states that are likely to occur in the inner solar system, it does not equally sample that probability space as shown in Figure \ref{fig:vscaled_all}, as our data set focuses on accretionary impacts.
$N$-body simulations of the early solar system and impact surveys designed to accurately represent all impact probability space both find that the majority of giant impacts that occur during planet formation result in hit-and-run collisions \citep{Carter2022,emsenhuber_sph_2024}. In contrast, our data contains roughly 50\% accretionary collisions and fewer than 40\% hit-and-run events. We did not include catastrophically disruptive collisions in our study. We should note that a large portion of hit-and-run impacts are part of ``collision chains'' that can eventually become accretionary events \citep{emsenhuber_fate_2019}.

Another collision outcome is the formation of synestias, bodies with sufficient heating and rotation to exceed the co-rotation limit and form a continuum between the corotating region and the disk-like region of the post-impact structure \citep{Lock17}. In our study, we manually determined whether each simulation produced a synestia using the criteria from \citet{Lock17}. Post-impact structures with a clear offset between the corotating regions and disk in angular-velocity-equatorial-radius space are considered planetary bodies with disks. Post-impact structures where the angular velocity-radius data was continuous between a constant corotating region and the Keplerian curve have a disk ``connected'' to the body and are referred to as co-corotation-limit (co-CoRoL) bodies, a minor form of synestia. Bodies where the angular velocity-radius data are continuous and smooth from the corotating region to a sub-Keplerian curve, are referred to as super-CoRoL synestias. Most other accretionary impacts in this study exceeded the corotation limit as a whole but resulted in a clearly demarcated corotating planet and Keplerian disk region with a discontinuity at the planet-disk interface, and so were not described as synestias. In this survey, the largest collision remnants are synestias in 63 out of 163 accretion events and 18 out of 116 hit-and-run events, while the largest remnants formed Co-CoRoL bodies in 33 out of 163 accretion events (see supplemental Table \ref{tbl:ICs}). This study did not focus on the end states of runners in hit-and-run impacts, though there is nothing that precludes second-largest remnants from forming synestias as well. Our results demonstrate that synestias or synestia-like bodies are a common outcome in the giant impact stage of planet formation.


\subsection{Thermal effects: Impact heating efficiency}

The efficiency by which impact energy is converted into heat in giant impacts is of major interest for understanding the thermal and chemical evolution of growing planets, as is the partitioning of impact heating between the core and mantle. Previous works have not provided a wide-ranging formulation for giant impact heating. \citet{Nakajima2021} developed a parameterization of heating within the target mantle, and it has been assumed that heat is partitioned between the core and mantle in proportion to the ratio of their masses \citep[e.g.,][]{jacobson2017}. The exchange between gravitational potential energy, kinetic energy, and thermal energy during giant impacts has been investigated by \citet{Carter2020}, but that work only considered a small parameter space of giant impact events.

In order to understand the exchange of energy during a giant impact, we must take into account the appropriate contribution from the large amount of negative gravitational potential energy (GPE) present in the system.
In this work, we adopt the concept of participating gravitational potential energy (PGPE) from \citet{Carter2020}, which defines the amount of potential energy that was converted into other forms of energy during the impact event. For each SPH simulation, we identify the most negative potential energy value for the whole event, which usually occurs with the closest initial approach of the cores. This value represents the maximum depth of the gravity well (GPE\textsubscript{min}). To convert the negative potential energy to a positive value in the total energy budget, the minimum GPE is subtracted from the potential energy of every particle (PGPE=GPE-GPE\textsubscript{min}). Thus, the total participating energy of the system, $E_{\text{tot}}$, is the sum of every particle's internal energy (IE), kinetic energy (KE), and participating gravitational potential energy (PGPE):
\begin{equation}
    E_{\text{tot}} = \text{IE}_{\text{tot}}+\text{KE}_{\text{tot}}+\text{PGPE}_{\text{tot}}.
\end{equation}

We use an empirical analytic measure of the specific impact energy, \(Q_S\), as a major parameter for the mechanical \citep{Leinhardt12} and thermal \citep{Lock17,LockStewart2019} outcomes from a giant impact. $Q_S$ is the specific kinetic energy of the collision in the center of mass frame modified by the interacting mass of the smaller body \citep[see][]{Leinhardt12,Lock17}:
\begin{eqnarray}
        Q_S &= & (1-\text{sin}\,\theta_i)\left(1+\frac{M_i}{M_T}\right)\left(\frac{\alpha M_i+M_T}{\alpha M_i M_T}\right)\frac{\mu^2 V_i^2}{2M_{tot}} \\
        l & = & (R_T+R_i)(1-\text{sin}\,\theta_i) \\
        \alpha & = &
            \begin{cases}
                \frac{3R_i l^2-l^3}{4 R_i^3}   & \quad \text{if } l < 2R_i \\
                1  & \quad \text{otherwise}
            \end{cases}
\end{eqnarray}
where \(\mu=(M_T M_i)/(M_T+M_i)\) is the reduced mass, 
\(\theta_i\) is the impact angle where \(\theta_i=0\) for a head-on impact, and $\alpha$ is the interacting mass fraction of the impactor. 
$Q_S$ is meant to describe the amount of energy in an impact that is transferred into the shock pressure field of the colliding bodies and is accurate over a wide range of initial impact conditions \citep{Lock17}. However, it becomes inaccurate at extremely glancing impact angles (see Appendix \ref{subsec:QS}). While this study performed impact angles extremely close to 90$\degree$ (\(\theta_i=89.7\degree\)) as endmember cases, these collisions were not included in our fitted scaling relations because the formulation for $Q_S$ does not correctly capture the deformation and true interacting mass at such highly oblique angles. The near 90$\degree$ impact outcomes are tabulated in the supplementary materials.

\begin{figure}
    \centering
    \includegraphics[width=0.5\textwidth]{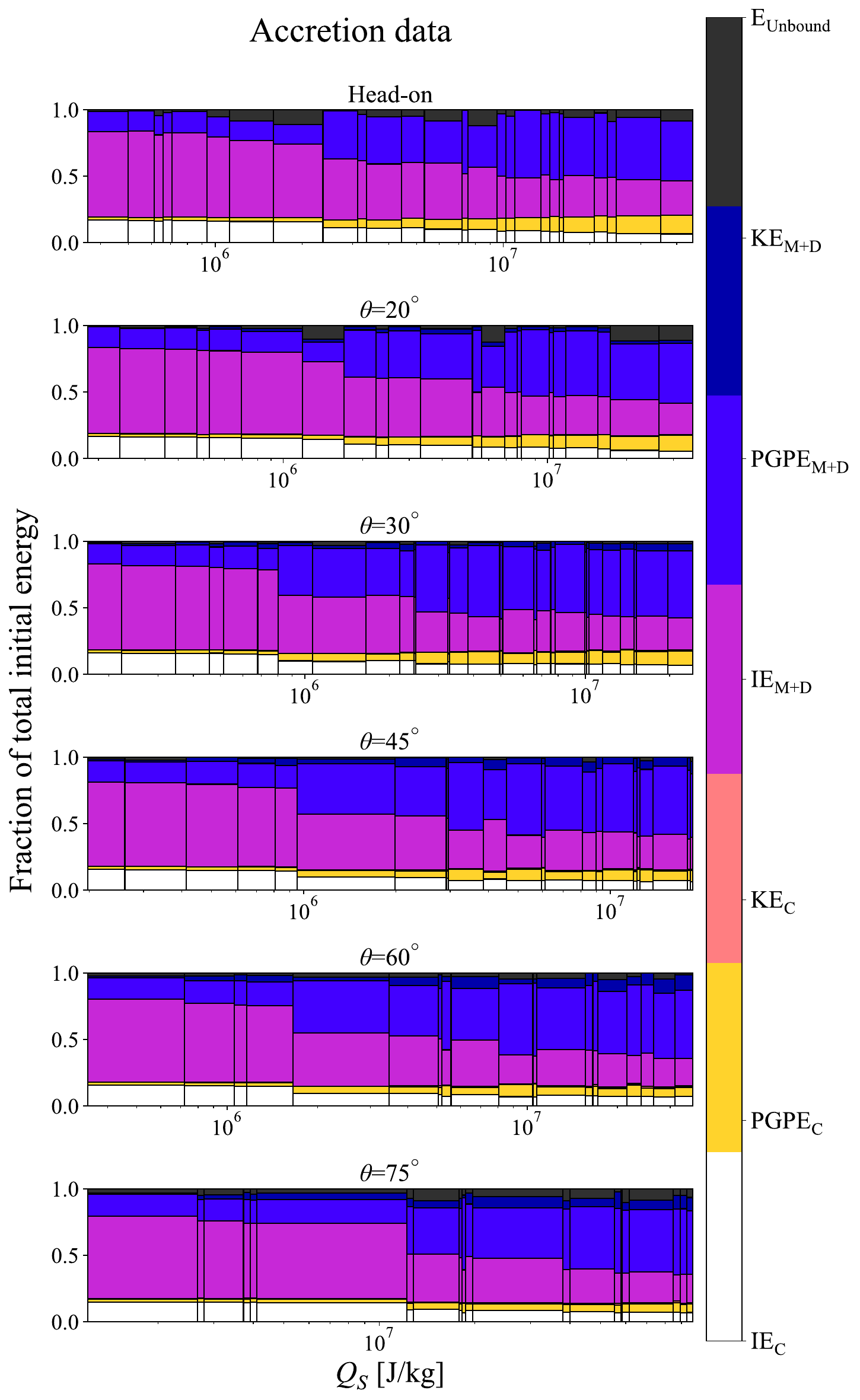}
    \caption{
    Energy budgets for the largest collision remnant after accretionary impact events, separated by impact angle. As specific impact energy $Q_S$ increases, the portion of the energy budget allotted towards gravitational potential energy increases at the expense of internal energy, while the partitioning of total energy between the core and mantle remains relatively constant. The energy budgets are separated into 3 categories for each the mantle and the core: internal energy (IE), participating gravitational potential energy (PGPE), and kinetic energy (KE), along with an amount of energy that was lost from the final largest remnant ($\text{E}_\text{Unbound}$).}
    \label{fig:IE_plus_PGPE}
\end{figure}

In accretionary collisions, most of the initial total participating energy ($E^0_{\text{tot}}$) is deposited into the largest post-impact body. The components of the energy budget are presented in Figure~\ref{fig:IE_plus_PGPE}. 
Overall, the fraction of total energy deposited in the largest remnant as IE, KE, or PGPE is approximately the same over more than two orders of magnitude of specific impact energy, as is the partitioning between mantle and core energy budgets.
As the specific impact energy increases, more energy is partitioned into the gravitational potential energy (PGPE). This increase is primarily offset by a reduction in the internal energy budget. 

This tradeoff in energy terms is primarily due to the post-impact 
bodies having significantly expanded states and lower internal pressures than cooler, slowly rotating planets of the same mass \citep{LockStewart2019}. In fact, most of the post-impact bodies are synestias, which are so expanded that they exceed the corotation limit. As oblique impacts become more energetic and the absolute amount of heating increases (absolute values of internal energy in the mantle and core still monotonically increase with increasing impact energy despite the efficiency of heating decreasing), a given mass layer of a planet will be less dense. In addition, impacts between the same bodies with the same non-zero impact parameter but higher velocity, and therefore more energy, also impart more angular momentum and so produce more rapidly rotating post-impact bodies. Both of these effects cause a given mass layer to move further from the central potential well, requiring an increase in the potential energy budget at the expense of heating. As a result, the amount of impact energy partitioned into potential energy in the post-impact body shows the opposite trend to heating, increasing with specific impact energy. In fact, the combined fraction of total initial energy that ends up in either internal or potential energy is notably more stable compared to internal energy alone for all impacts we considered. This suggests that the main features of post-impact energy budgets is a tradeoff between internal and potential energy (Figure \ref{fig:IE_plus_PGPE}) \citep{Carter2020,Stewartinprep}. As the post-impact body cools, some of this potential energy budget could be converted to kinetic energy as the body contracts and begins to rotate faster, but our comparison between the post-impact bodies and the rotating models indicates that most is converted to heating upon contraction.

\begin{figure}
    \centering
    \includegraphics[width=0.5\textwidth]{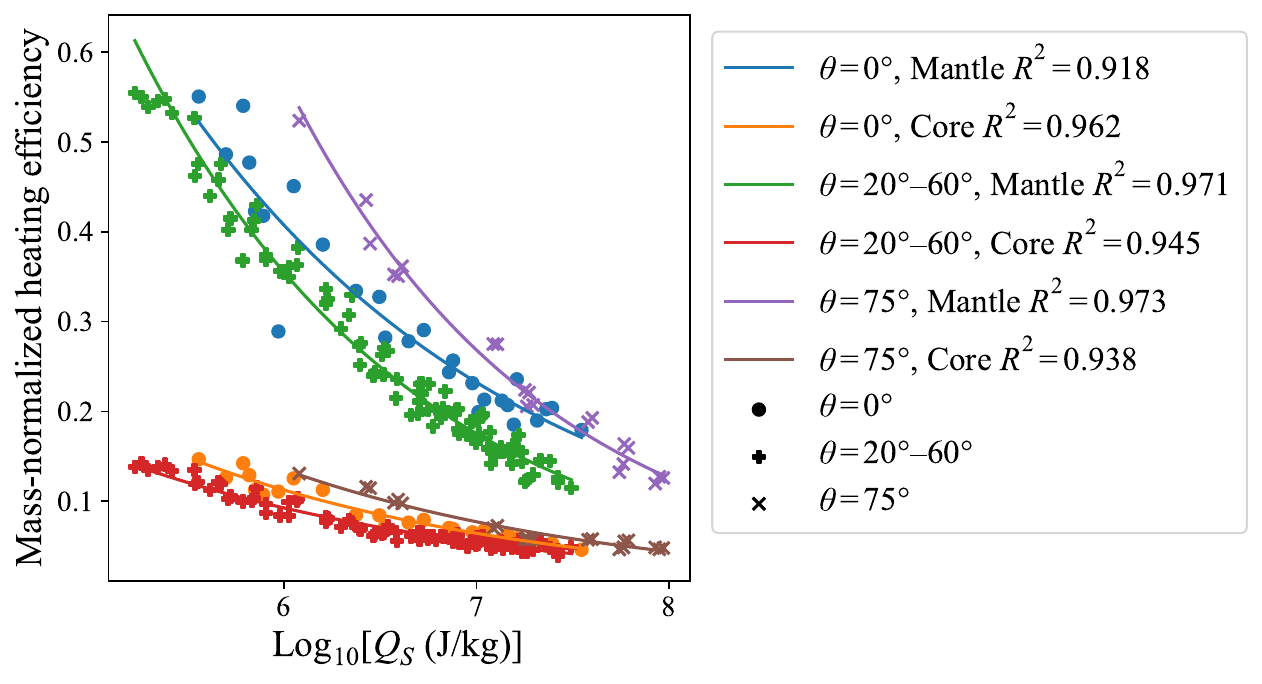}
    \caption{
    Accretionary impact heating efficiency relationships as a function of the log modified specific impact energy $Q_S$: the final mantle or core internal energy as a fraction of total initial system energy 
    normalized by the mass ratio of the final bound mass to the initial target (\((M_T/M_{\text{bnd}})(\text{IE}_{\text{core or mantle}}/E_{tot})\)). We identified three main groupings that behaved as power laws: the bulk of impact angles tested ($\theta=20\degree$ to $60\degree$), head-on impacts (which physically vary due to the lack of rotational energy), and highly oblique impacts ($\theta\geq75\degree$), for which $Q_S$ cannot totally capture the amount of interacting mass in the collision due to the importance of pre-impact tidal deformation (see \S\ref{subsec:QS}. The coefficients for the power law fits displayed here are included in Table \ref{tbl:HE_coeff}.}
    \label{fig:Qs_IE_hnr}
\end{figure}

For both of our major outcome groupings (accretionary and hit-and-run), we find that the efficiency of converting impact energy into heating of the largest remnant decreases somewhat with increased specific impact energy \(Q_S\) (Figure \ref{fig:Qs_IE_hnr}). For hit-and-run impacts we find that the relationship between the internal energy of the final body (as a fraction of total initial impact energy) and the specific impact energy is well fit by a power law (Appendix \ref{sec:subfigs} Figure \ref{fig:Qs_IE_hnr_supp}). 
For accretionary impacts, an additional normalization by the ratio of the initial target mass to the final bound mass (mass-normalized (M-N) heating efficiency):
\begin{equation}
   \text{M-N heating efficiency =}\left ( \frac{M_T}{M_{\text{bnd}}}\right ) \left ( \frac{\text{IE}_{\text{core or mantle}}}{E_{tot}} \right) \; ,
\end{equation} 
is required in order to provide a higher quality fit (squared correlation coefficient \(R^2>0.9\)). The need for this additional normalization indicates that the final amount of heating has an additional dependence on the mass of the original planet, possibly due to pre-compression in larger mass body and the larger gravitational potential well.

Head-on impacts were separated from the bulk of our fits for accretionary impacts as they show different behavior from oblique impacts and are also vanishingly unlikely in realistic accretion scenarios. Our results and discussion for head-on impacts are presented in Appendix \ref{subsec:head-on}. In addition, highly oblique impacts were also separated, as large impact angles begin to have slightly larger than expected $Q_S$ values (for \(\theta_i\approx75\degree\)), while near-tangent impacts (\(\theta_i\rightarrow90\degree\)) are omitted entirely due to extremely large $Q_S$ values (see Appendix \ref{subsec:QS}).

Scaling laws for the heating efficiency of the mantle and core during accretionary impacts for the bulk of impact angles tested had \(R^2\) values of 0.971 and 0.945, respectively. These corresponded to fit errors in log-log space (\(\sqrt{\sigma^2_{res}/n}\), where \(\sigma^2_{res}\) is the variance of $n$ residuals) of $\pm$0.032 and $\pm$0.034, or percent errors in non-log space of $\pm$7.1\% and $\pm$7.5\%, respectively. Mantle and core fits for hit-and-run impacts both produced \(R^2\) values of 0.929 and 0.922, respectively. These corresponded to errors in log-log space of $\pm$0.071 and $\pm$0.065, or percent errors in non-log space of $\pm$15.0\% and $\pm$13.9\%, respectively. The fit coefficients for our scaling laws are listed in Table \ref{tbl:HE_coeff}.

\begin{table}
\centering
\begin{tabular}{lll}
\multicolumn{3}{l}{{\bf A.} \(\text{IE}/E_{\text{tot}}=(M_{\text{bnd}}/M_T)\cdot10\string^[A(\text{Log}_{10}[Q_S])+B]\)} \\ \hline
\textbf{Accretion}, $20\degree\leq\theta_i\leq60\degree$ & $A$& $B$ \\  \hline 
Mantle & -0.30602565 & 1.38630987  \\ 
Core & -0.23114965 & 0.3537509 \\ \hline 
\textbf{Accretion}, head-on & $A$ & $B$ \\  \hline
Mantle & -0.24374435 & 1.0724414 \\ 
Core & -0.24232102 & 0.50535294  \\ \hline 
\textbf{Accretion}, $\theta_i=75\degree$ & $A$ & $B$ \\  \hline
Mantle & -0.32851918 & 1.72897198 \\ 
Core & -0.23994513 & 0.56381691  \\ \hline \\ 
\multicolumn{3}{l}{{\bf B.} \(\text{IE}/E_{\text{tot}}=10\string^[A(\text{Log}_{10}[Q_S])+B]\)} \\ \hline 
\textbf{Hit-and-run}, $20\degree\leq\theta_i\leq60\degree$ & $A$ & $B$ \\ \hline  
Mantle & -0.43046148 & 2.17150644 \\ 
Core & -0.37738919 & 1.25230422 \\ 
\textbf{Hit-and-run}, $\theta_i=75\degree$ & $A$ & $B$ \\ \hline 
Mantle & -0.43597471 & 2.38459241  \\ 
Core & -0.33194853 & 1.14111375 
\end{tabular}%

\caption{Scaling law fits for internal heating in the largest post-impact body shown in Figure \ref{fig:Qs_IE_hnr}. \textbf{A.} Coefficients for mass-normalized heating efficiency for oblique accretionary collisions as a function of $Q_S$ (in units of J/kg). The data, fit lines, and $R^2$ values for these scaling laws are shown in Figure \ref{fig:Qs_IE_hnr}: mantle and core data for $0\degree=\theta_i$ are shown in blue and orange, respectively;
mantle and core data for $20\degree\leq\theta_i\leq60\degree$ are shown in green and red; mantle and core data for $75\degree=\theta_i$ are shown in purple and brown.
\textbf{B.} Coefficients for heating efficiency for hit-and-run collisions. 
}
\label{tbl:HE_coeff}
\end{table}

Our simulations capture the states of the post-impact planets a few hours after impact. The thermal and dynamical structure, and hence energy budgets, of the bodies will evolve as they cool and the angular momentum of the bodies could be changed by tidal interaction with moons. The radius of the body will decrease and compression of the mass layers increase as the outer layers condense. The gravitational potential energy released during this process will likely buffer cooling or be converted into internal energy potentially millions of years into the planet's post-impact recovery \citep{Lock2020}. In our simulations, we find that the kinetic energy of the post-impact SPH bodies is mostly maintained through to the rotating planet models, indicating that the gravitational potential energy that returns to the planet will likely do so in the form of heating.

\begin{figure}
    \centering
    \includegraphics[width=.45\textwidth]{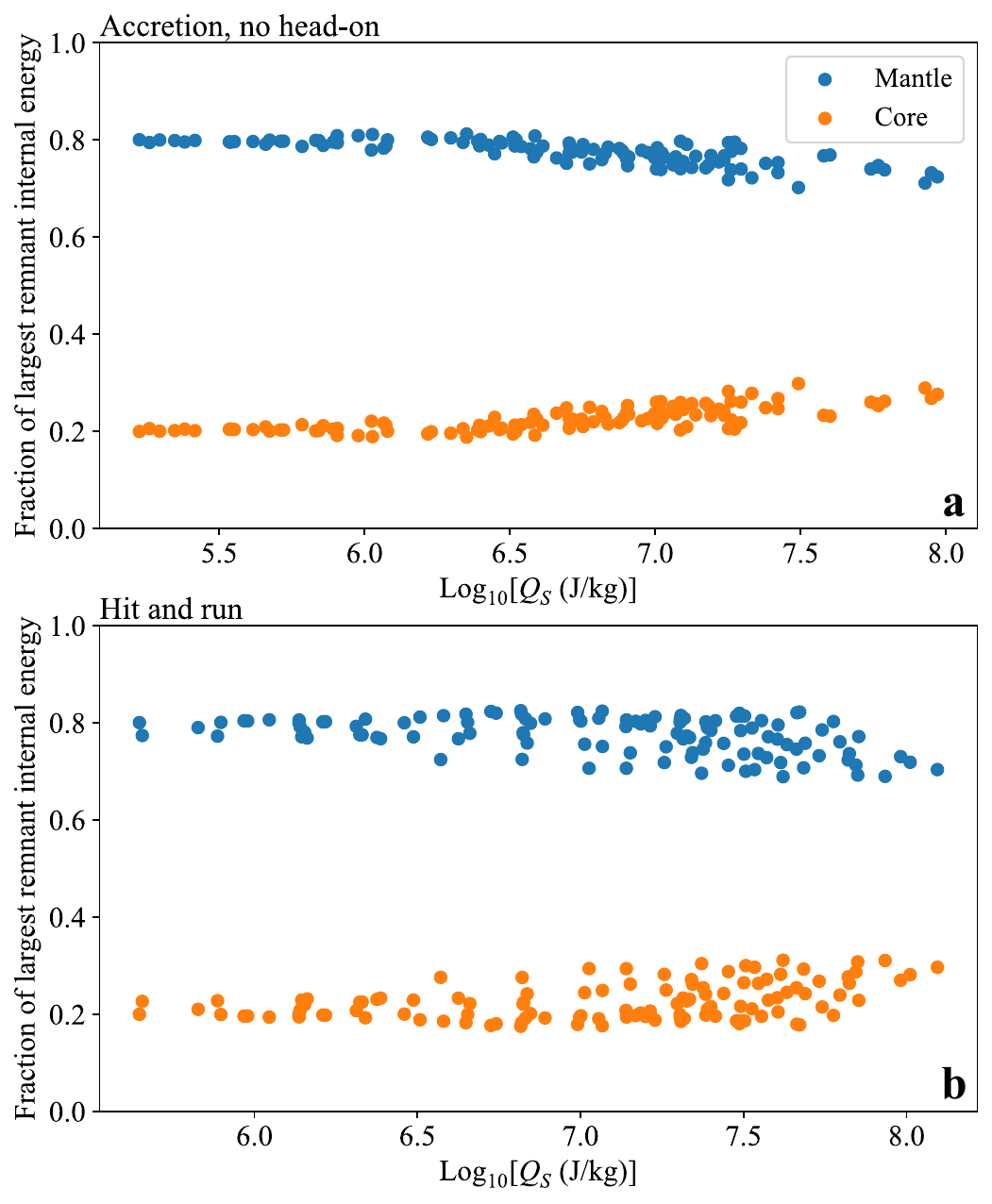}
    \caption{Internal energy partitioning for oblique accretionary collisions (excluding head-on, Panel \textbf{a}) and hit-and-run collisions (Panel \textbf{b}). In both outcome regimes, the mantles contain roughly 80\% of the planets' internal energy budgets for low- to mid-energy impact events, while approaching 70\% for higher energies, more comparable to the mantle-core mass ratios.}
    \label{fig:IE-partition}
\end{figure}

Both accretionary and hit-and-run cases show similar internal energy partitioning (Figure \ref{fig:IE-partition}). In both outcome regimes, post-impact mantles contain roughly 80\% of the planets' internal energy budgets for most low- to mid-energy impact events. This means heating is preferentially distributed into the mantle compared to the case where heat is distributed equal to mass (i.e., roughly 70\% mantle and 30\% core). At higher impact energies, the partitioning of heat approaches a ratio similar to the mantle-core mass ratio (70\%). For all cases, this partitioning is more preferential to the mantle than the pre-impact mantle/core internal energy ratios (which vary with planet mass).
When comparing accretionary and hit-and-run collisions in isolation, we see that while the shape of the trend is very similar, there is notably more spread in the hit-and-run data (Figure \ref{fig:IE-partition}B). The general trend of heating with impact energy does not hold in one isolated case: efficient accretion from head-on impacts, which display a flat trend with 80\% of heating partitioned into the mantle (see Supplementary Figure \ref{fig:IE-partition-0}). Though again, head-on impacts are highly unlikely to actually occur in planet formation so this exception is likely to have little effect on the thermal evolution of growing planets (see \S\ref{subsec:head-on}).

\subsubsection{Scaling laws for P-T change at the core-mantle boundary}
The pressures and temperatures at the CMB (or at some fixed fractional depth in the mantle) are often used as an estimate for the conditions of metal-silicate equilibration after giant impacts \citep{RUBIE201131,Rubie15}. We find large and systematic changes in both temperature and pressure at the CMB due to accretionary collisions. 

We parameterize changes in temperature separately for the mantle and core side of the CMB as the change in conditions between the initial mantle or core isentrope in the pre-impact target (\S \ref{sec:methods:SPH_init}) and the mean post-impact conditions of the 100 equatorial particles (defined as within a vertical distance of one-tenth of a planetary radius from the equatorial plane) with radii nearest the CMB radius in the SPH largest remnant. For our purposes, we define the equatorial radius of the CMB as the mean radius of the silicate SPH particles closest to the center of mass of the post-impact body's core (in order to accurately describe the center of the planet in a case where moonlets are present, which can skew the total center of mass of the system). In almost all cases, expanding this number to the innermost 500 particles on the mantle or core side produces average temperature values within the 25\textsuperscript{th} to 75\textsuperscript{th} percentile of the innermost 100 particle values (typically within $<$2-5\% of the mean for the mantle side, approaching $\sim$10\% on the core side). 

Parameterizing CMB pressure is more complex. Without any modification for density discontinuities, standard SPH formulations can produce artifacts in some physical quantities (including pressure) near material boundaries with significant density contrasts, such as the CMB \citep[e.g.,][]{ruiz-bonilla_dealing_2022,hosono_comparison_2016,hosono_giant_2016}. This leads to ``wiggles'' in the final pressure profiles of some of our simulations. We avoid this in our analysis by fitting two polynomial curves for the core and mantle pressure-equatorial radius profiles outside of the pressure artifact, then extrapolating tangent lines from the ends of those curves across the CMB, a methodology further described in \S \ref{subsec:polynomial}. We then assume the pressure profile of the planet follows those curves, switching between the extrapolations from the mantle and core at the point they intersect. We find that the intersection point is not necessarily at the boundary of where metal and silicate SPH particles coexist, and can often occur at a smaller radius than the location of any silicate particles. For the purposes of using CMB pressures as proxies for the conditions of metal-silicate equilibration, we use the equatorial mantle particles with the smallest radii to define the core radius, as the metal-silicate equilibration point would not be further inward where no silicates are present. We take the pressure at the CMB to be that of the interpolated pressure profile at the average radius of the 100 innermost equatorial mantle particles.

\begin{figure}[h]
    \centering
    \includegraphics[width=.45\textwidth]{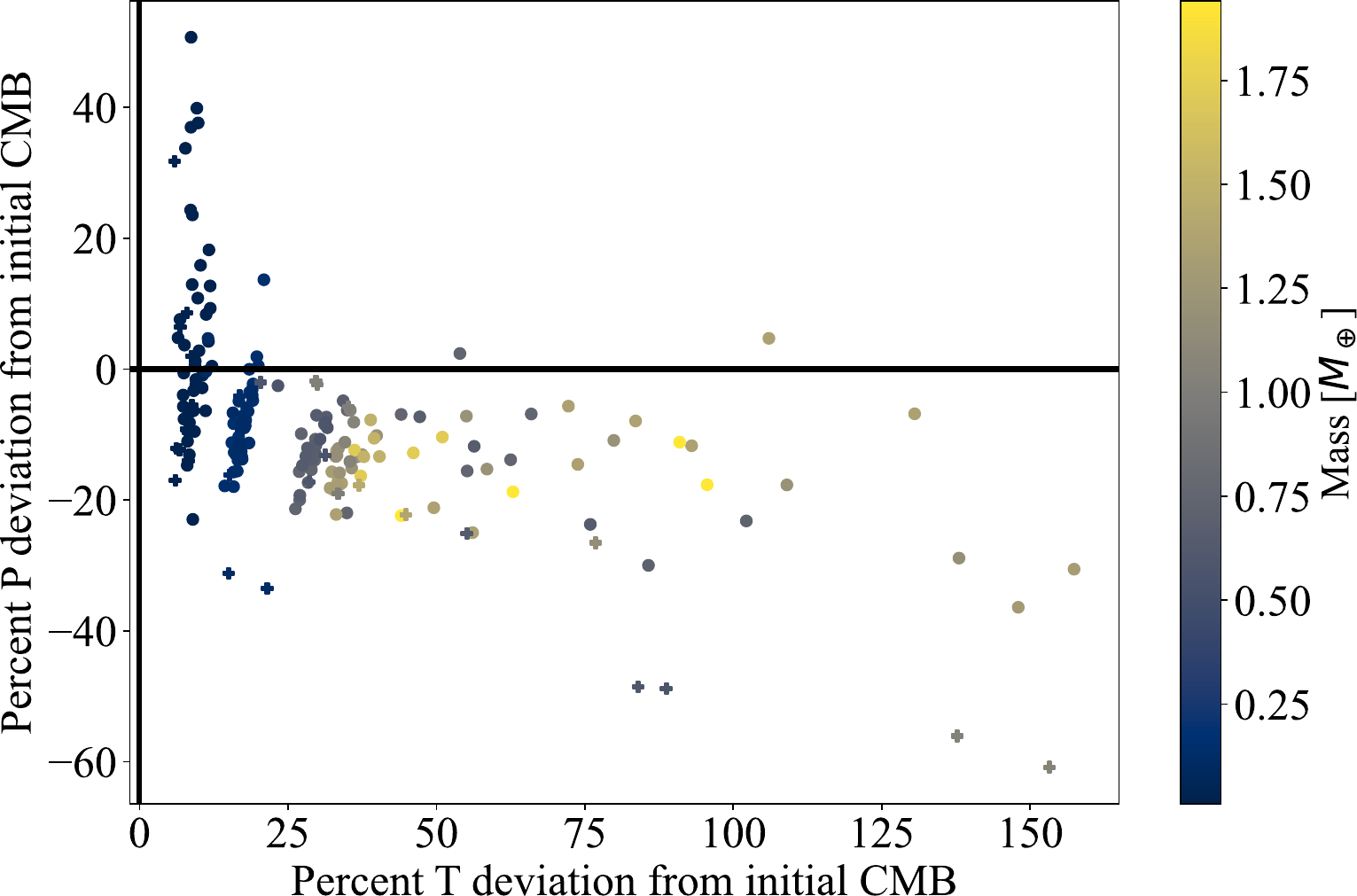}
    \caption{Difference in pressure and temperature at the core-mantle boundary between the initial target planet and the final largest remnant for accretionary collisions (i.e., those for which the post-impact body is more massive than the largest colliding body). Points are colored according to the bound mass of the resulting planet. All cases showed temperature increases at the CMB while pressures generally decreased in all but the smallest-mass cases.}
    \label{fig:dPT_CMB}
\end{figure}

We find that temperatures at the CMB increase in all impacts (\(\Delta T>0\)) and that in most cases pressures decrease, though small moon-sized collisions see pressures increasing in some cases (Figure \ref{fig:dPT_CMB}). It is important to note that Figure \ref{fig:dPT_CMB} shows only accretionary impacts, i.e. those for which the post-impact body is more massive than the largest colliding body, for which pressure would conventionally be expected to have increased. Pressures in post-giant-impact bodies are lower than before the impact due to a combination of the heated material being less dense and the effect of rapid rotation resulting from the increased angular momentum of the post-impact body, both of which counter the increased gravitational attraction from added mass \citep{LockStewart2019}. The smallest bodies are thermally cold enough that they are not heated beyond the melt curve and densities do not change significantly, thus the increased gravitational attraction from the added mass dominates over the rotational effect. 

\begin{figure*}
    \includegraphics[width=0.95\textwidth]{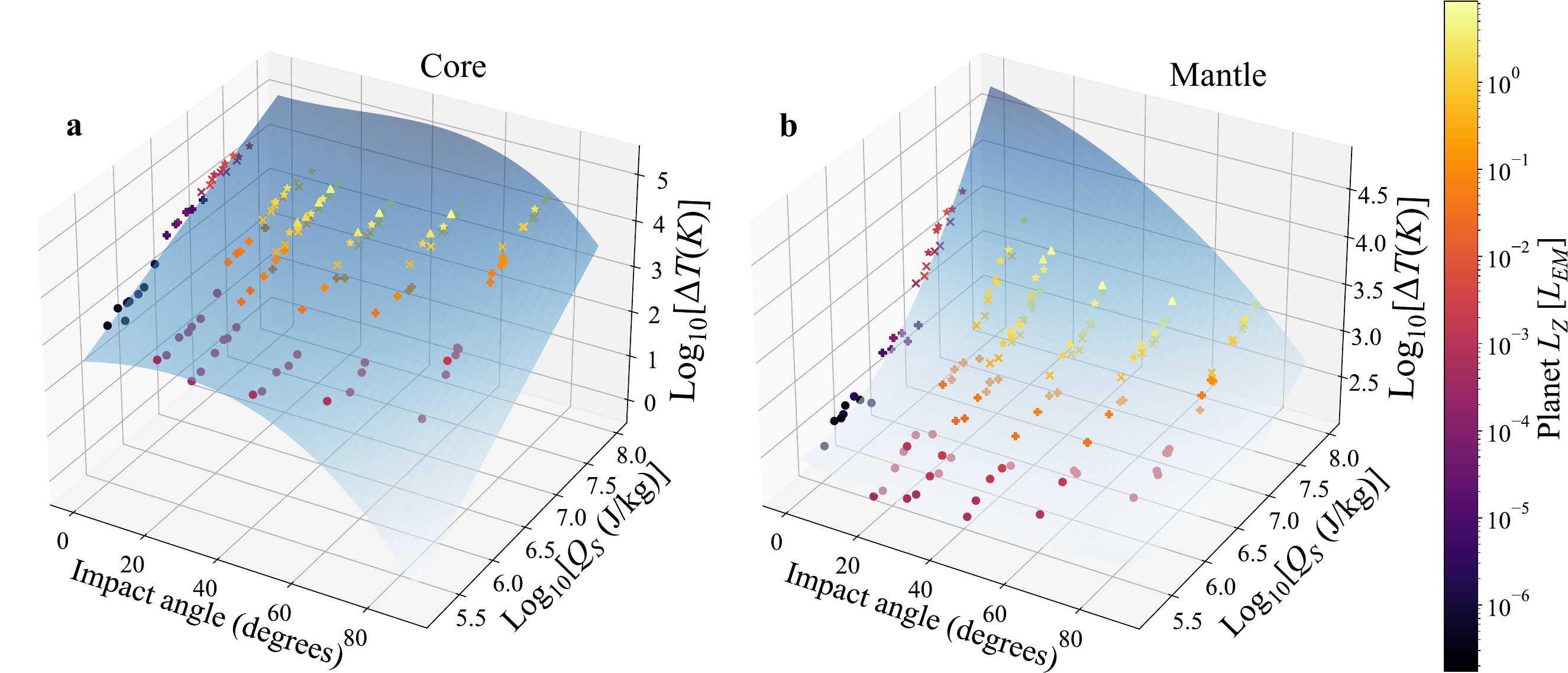}
    \caption{Core-mantle boundary core- (left) and mantle-side (right) temperature change in accretionary collisions compared to our scaling-law fits (blue surfaces) as a function of specific impact energy \(Q_S\) and impact angle \(\theta_i\), color-coded by the final angular momentum in units of $L_{\rm EM}$ (the angular momentum of the Earth-Moon system). The shape of the data point markers indicates the size of the initial target planet in the simulation: $\bullet$ -- moon mass (\(0.01 M_\oplus\)), \ding{58} -- Mars (\(0.1M_\oplus\)), $\times$ -- half-Earths (\(0.5 M_\oplus\)), $\star$ -- proto-Earths (\(0.91M_\oplus\)), and $\blacktriangle$ -- super-Earths (\(1.3 M_\oplus\)). Residuals for these fits are depicted in Supplementary Figure \ref{fig:CMB_T_resids}.}
    \label{fig:3D_dT_fit}
\end{figure*}

In order to produce scaling laws for temperature and pressure change at the CMB, we take an approach similar to that of \citet{Nakajima2021}, and parameterize the change in each thermal quantity as a polynomial function of impact angle and an energy term. However, rather than using a simple analytic expression for kinetic or potential energy as the energy term, we use specific impact energy \(Q_S\) as a single term to describe the impact energy coupled to the target in the collision, as suggested by \citet{Lock17}. We used least-squares regression to fit \(\Delta T_{CMB}\) and \(\Delta P_{CMB}\) as a function of \(Q_S\) and \(\theta_i\) over all accretionary events in our survey. For the mantle-side temperature change, \(\Delta T_{\text{CMB,Mant}}\), we fit the data with a quadratic function of $Q_S$ (in units of J/kg) and $\theta_i$ (in degrees) in log-log space, such that the resulting fit equation is of the form:
\begin{equation}
\begin{split}
    \Delta T_{\text{CMB,Mantle}} =& 10\string^ [K_{\text{M:}2}(\text{log}_{10}[Q_S])^2+\\
    & K_{\text{M:}1}\text{log}_{10}[Q_S]+K_{\text{M:}0}] \label{eqn:Tfitfn} \\
    K_{\text{M:}j}[\theta_i] =& A_{\text{M:}j}\theta_i^2+B_{\text{M:}j}\theta_i+C_{\text{M:}j}
\end{split}
\end{equation}
where \(K_j\) is a fitting term for the \(j\)th degree term of \(\text{log}_{10}[Q_S]\) as a quadratic function of impact angle \(\theta_i\) and constants \(A_j\), \(B_j\), and \(C_j\). The \(\theta_i\) dependence of this fitting form was motivated by fitting Equation \ref{eqn:Tfitfn} for the subsets of accretionary collisions occurring at each impact angle as function of \(Q_S\) alone. 
Fitting the combined form over all accretionary impacts in log-log space produced an \(R^2\) value of 0.935, with a corresponding fit error of \(\pm0.107\) in log-log space, or 21.8\% in non-log space. This error is on the scale of, or smaller than, the 2$\sigma$ spread of the temperatures of individual SPH particles near the CMB in most of our simulations. The coefficients for this fit are given in Table \ref{tbl:mant_fit} and the match to data shown in Figure \ref{fig:3D_dT_fit}.

Our approach to fitting the core-side temperature change, \(\Delta T_{\text{CMB,Core}}\), was similar but, instead of a quadratic fit in both \(\text{log}_{10}[Q_S]\) and \(\theta_i\), we use a linear form in \(\text{log}_{10}[Q_S]\) (again in units of J/kg) and a cubic in \(\theta_i\) (again in degrees):
\begin{equation}
\begin{split}
    \Delta T_{\text{CMB,Core}} =& 10\string^ [K_{\text{C:}1}\text{log}_{10}[Q_S]+K_{\text{C:}0}]\label{eqn:Tfitfncore} \\
    K_{\text{C:}j}[\theta_i] =& A_{\text{C:}j}\theta_i^3+B_{\text{C:}j}\theta_i^2+C_{\text{C:}j}\theta_i+D_{\text{C:}j}
\end{split}
\end{equation}
Fitting this combined form over all accretionary impacts in log-log space produced a comparable \(R^2\) value to the mantle fit of 0.907, corresponding to an error \(\pm0.176\) in log-log space, or 33.3\% in non-log space. Again, this error is on the scale of the spread in SPH particle temperatures, as the core particles show more variance in temperature than the mantle side. The coefficients for this fit are given in Table \ref{tbl:core_fit} and the match to the data shown in Figure \ref{fig:3D_dT_fit}.

\begin{figure}[h]    \includegraphics[width=0.45\textwidth]{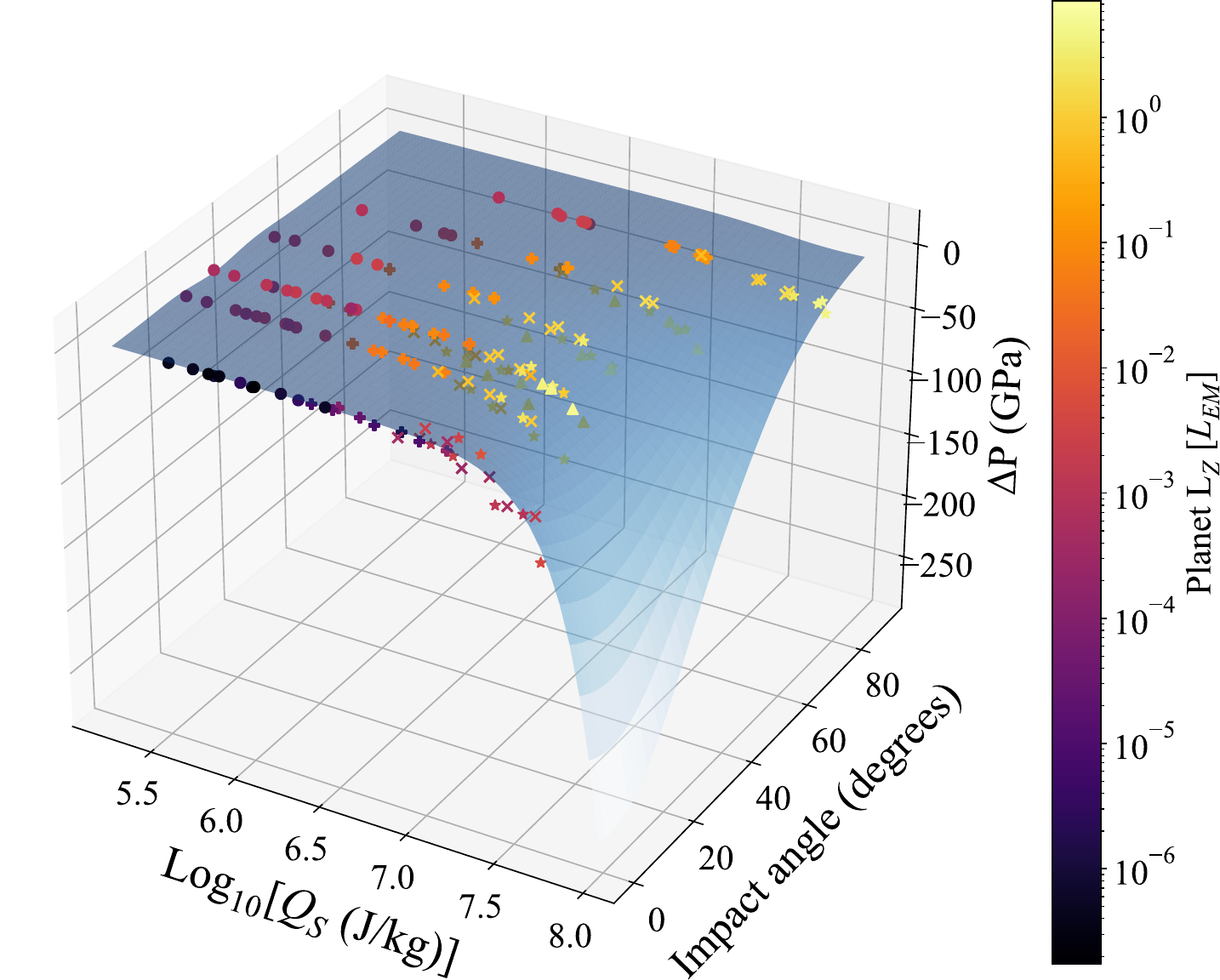}
    \caption{Core-mantle boundary pressure change data in accretionary collisions compared to our scaling-law fit (blue surface) as a function of specific impact energy \(Q_S\) and impact angle \(\theta_i\), similar to Figure \ref{fig:3D_dT_fit}. Note that the x- and y-axes are flipped compared to Figure \ref{fig:3D_dT_fit} to better illustrate data trends. The shape of the data point markers again indicates the initial target size of the collision, see Figure \ref{fig:3D_dT_fit} for a description of individual symbols. Residuals for this fit are depicted in Supplementary Figure \ref{fig:CMB_P_resids}.}
    \label{fig:3D_dP_fit}
\end{figure}

Two modifications to this approach were required for fitting \(\Delta P_{CMB}\). The foremost difference between the CMB temperature and pressure changes is that while CMB temperature changes are universally positive, pressure changes are both positive and negative. We therefore have included an offset term in pressure change ($Z$, expressed as a quadratic function of $\theta_i$ in log space), and fit the function to the data in \(\Delta P_{CMB}\)-\(\text{log}_{10}[Q_S]\) space rather than in log-log. Otherwise, the fitted function was the same as for \(\Delta T_{\text{CMB,Mant}}\) with
\begin{equation}
\begin{split}
    \Delta P_{CMB} = & Z[\theta_i]-10\string^[K_{2,\theta_i}(\text{log}_{10}[Q_S])^2+\\ 
    &  K_{1,\theta_i}\text{log}_{10}[Q_S]+K_{0,\theta_i}]\label{eqn:Pfitrn} \\
    K_{j}[\theta_i] = & A_j\theta_i^2+B_j\theta_i+C_j \\
    Z[\theta_i] = & 10\string^[Z_2\theta_i^2+Z_1\theta_i+Z_0] 
\end{split}
\end{equation}
%
The pressure change data shows noticeably more variation than the temperature data, likely due to the necessity of extrapolating a pressure profile across the CMB. As a result, the fit for pressure change over all accretionary data produced a lower \(R^2\) value of 0.870, with a corresponding fit error of \(\pm\)4.52 GPa. While this error estimate is still quite accurate for a broad range of collisions with larger CMB pressure changes, it is skewed by the magnitude of those data and the exponential trend. Looking at the quality of our scaling law for only the data with absolute values of CMB pressure change smaller than the overall error, we find a better fit error of \(\pm\)1.82 GPa. Correspondingly, for the subset of pressure change data larger than the overall error, the fit error increases slightly to \(\pm\)6.12 GPa. The coefficients for this fit are included in Table \ref{tbl:mant_fit_P} and depicted in Figure \ref{fig:3D_dP_fit}.

Our pressure and temperature fits are based on quadratic functions in $Q_S$ and should therefore be used with great caution for planets with masses smaller or larger than those considered in our study. The impacts with the most massive targets in our dataset involved targets of \(1.3 M_\oplus\) and produce final planets close to \(2 M_\oplus\) as a high endmember. On the lower end, our collision dataset approaches the limit of what SPH simulations can accurately resolve without necessitating treatments for strength \citep{emsenhuber_sph_2024}. Our scaling relations are therefore applicable to accretionary, gravity-dominated giant impacts that occur during the early solar system, where we expect no terrestrial planet to grow beyond 1 Earth mass, or for exosystems with similar mass planets as the solar system. More SPH impact data is needed to reliably expand the range of these scaling laws for higher-mass collisions, though the ability to obtain this data is limited by the reliability of existing EOS models in extremely high temperature and pressure regimes.

\begin{table*}[]
    \centering
        \begin{tabular}{llll}
        \hline
        &\(A_{\text{M:}j}\) & $B_{\text{M:}j}$ & $C_{\text{M:}j}$ \\
        \(K_{\text{M:}j=2}\) & 3.99156398e-05 & -7.27176201e-03 & 3.79648158e-01 \\
        \(K_{\text{M:}j=1}\) & -6.63134111e-04 & 9.83899569e-02 & -4.17146409e+00 \\
        \(K_{\text{M:}j=0}\) & 2.46994904e-03 & -3.23985262e-01 & 1.38863332e+01  \\ \hline
        \end{tabular}%
    \caption{Scaling law coefficients for Equation \ref{eqn:Tfitfn}: change in core-mantle boundary (CMB) temperature on the mantle side as a function of specific impact energy \(Q_S\) (in units of J/kg) and impact angle \(\theta_i\) (in units of degrees). See Figure \ref{fig:3D_dT_fit}B.}
    \label{tbl:mant_fit}
\end{table*}

\begin{table*}[]
    \centering
        \begin{tabular}{lllll}
        \hline
        & \(A_{\text{M:}j}\)& $B_{\text{M:}j}$ & $C_{\text{M:}j}$ & $D_{\text{M:}j}$ \\
        \(K_{\text{M:}j=1}\) & -3.77584405e-06 & 4.47891647e-04 & -1.06571082e-02 & 8.90451508e-01 \\
        \(K_{\text{M:}j=0}\) & 2.15390466e-05 & -3.00543848e-03 & 8.25130389e-02 & -2.10733234e+00 \\ \hline
        \end{tabular}%
    \caption{Scaling law coefficients for Equation \ref{eqn:Tfitfncore}: change in core-mantle boundary (CMB) temperature on the core side as a function of specific impact energy \(Q_S\) (in J/kg) and impact angle \(\theta_i\) (in degrees). See Figure \ref{fig:3D_dT_fit}A.}
    \label{tbl:core_fit}
\end{table*}

\begin{table*}[]
    \centering
        \begin{tabular}{llll}
        \hline
        & \(A_{j}\)& \(B_{j}\) & \(C_{\text{M:}j}\) \\
        \(K_{\text{M:}j=2}\) & -1.33096709e-03  & 8.18403983e-02 & -9.94945628e-01  \\
        \(K_{\text{M:}j=1}\) & 2.00254714e-02 & -1.24145795e+00  & 1.63634505e+01 \\
        \(K_{\text{M:}j=0}\) & -7.55677546e-02  & 4.72160801e+00 & -6.52285365e+01  \\ \hline
         & \(Z_2\) & \(Z_1\)  & \(Z_0\) \\ 
        \(Z:\) & -5.44944846e-03 & 3.00001259e-01 & -3.50452700e+00 \\ \hline
        \end{tabular}%
    \caption{Scaling law coefficients for Equation \ref{eqn:Pfitrn}: CMB pressure change as a function of specific impact energy \(Q_S\) (in J/kg) and impact angle \(\theta_i\) (in degrees). See Figure \ref{fig:3D_dP_fit}. }
    \label{tbl:mant_fit_P}
\end{table*}


\subsubsection{Melt Generation} \label{sec:melting}
The volume of mantle material melted in a giant impact is the main control on the development of a global magma ocean. Classically, a giant impact is assumed to initially produce a spatially constrained melt pond \citep[e.g.,][]{Tonks93} that then spreads globally due to isostatic adjustment \citep{Reese2006}. Metal-silicate equilibration is often treated as occurring as the impactor core goes downwards, at the base of these spherically symmetric magma oceans, such that the basal pressure of the magma ocean is the pressure of equilibration. Recent studies have used SPH simulation surveys to explore the effects of impact angle, velocity, target size, and target-impactor mass ratio on the generation of impact melting in order to produce melt volume and geometry scaling laws \citep{Nakajima2021}, though these studies did not use an EOS model with a melt phase treatment. While developing robust scaling laws for melt generation is beyond the scope of this paper, we will describe general trends observed in our study. These melt fraction trends for post-impact planets are depicted in Figures \ref{fig:mant_melt} and \ref{fig:melt_hnr}, for accretionary and hit-and-run collisions respectively.

A large majority of our simulations produced totally melted mantles in accretionary impacts, with almost all simulations that involved a Mars-sized or larger target (\(M_T\geq 0.1M_\oplus\)) at impact angles \(\theta_i\leq 45\degree\) ending with zero silicate particles in the solid region of the pyrolite ANEOS phase diagram (see Figure \ref{fig:mant_melt}). Even for Moon-sized targets, only a small handful of the lowest energy collisions at impact angles \(\theta_i\leq 30\degree\) resulted in solid mantle fractions between 10\% and 25\%. 

The non-solid portions of the mantle were primarily composed of partial melt (material with temperatures and densities between the solidus and liquidus of the pyrolite ANEOS model) and fully liquid portions (temperatures and densities above the pyrolite liquidus) in lower-mass collisions, while larger-mass collisions were primarily a mix between liquid and supercritical fluids. This corroborates the findings of \citet{caracas2023} for canonical Moon-forming impacts, with the addition that even smaller collisions producing \(\sim0.7 M_\oplus\) final mass bodies can see significant supercritical regions in post-impact mantles. Oblique and highly oblique accretionary collisions (\(\theta_i\geq60\deg\)) do produce significant solid fractions in most low- to mid-energy collisions but still can leave a supermajority ($>67$\%) of the planet in a non-solid phase. However, these impacts represent a much smaller portion of the outcome probability space for accretionary events, indicating that the majority of collisions producing planets larger than \(0.1 M_\oplus\) will result in totally melted mantles. 

Notably, non-grazing hit and run impacts (\(\theta_i\leq 30\degree\)) also do not produce solid post-impact mantles in almost all cases, while also producing significant supercritical mass fractions in higher-mass collisions (Figure \ref{fig:melt_hnr}). If the vast majority of accretionary collisions result in insignificant solid mantle mass fractions in post-impact bodies, whether metal-silicate equilibration occurs at the bottom of a melt pond or after hydrostatic adjustment to a magma ocean makes no difference
as the melt pond contains the entire mantle.

\begin{figure*}[b] 
\centering
\includegraphics[width=0.8\textwidth]{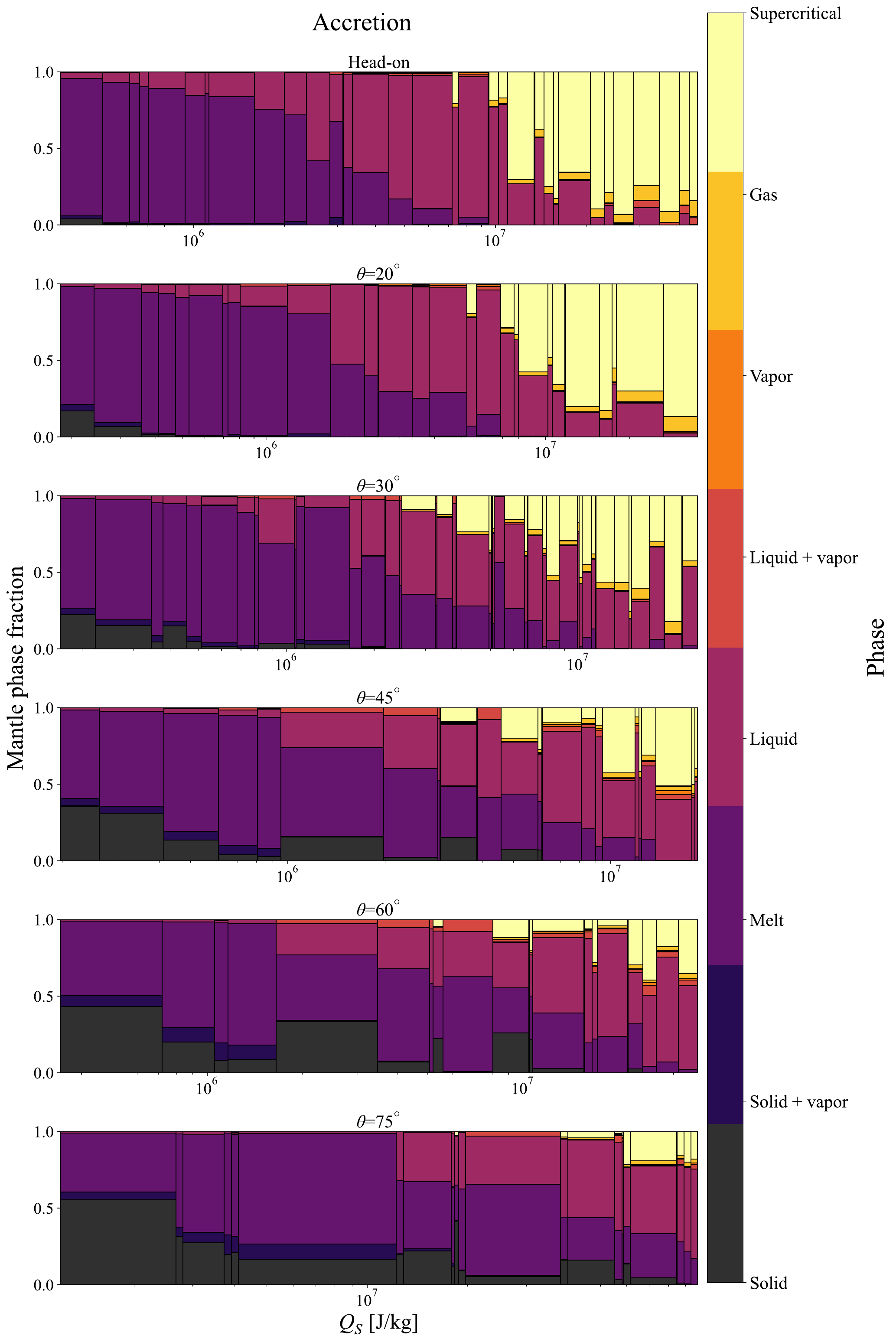}
\caption{\textit{Caption on next page.}}
\end{figure*}
\addtocounter{figure}{-1}
\begin{figure*}[t]
    \centering
    \caption{\textit{Continued from previous page.} Mantle phase mass fractions for accretionary collisions in this study at each impact angle tested using ANEOS pyrolite phase boundaries \citep{ANEOS_pyrolite}. Each bar is centered on one simulation in our dataset, where the width of each bar only indicates the density of data points, with wider bars correlating to sparsely sampled regions. At lower impact angles (which produce the bulk of accretionary outcomes), the mantle is practically completely melted in simulations involving Mars-sized (\(0.1M_\oplus\)) or larger bodies. The dominant non-solid phases change from melt (the region in phase space between the solidus and liquidus) to fully liquid to a supercritical fluid as impact energy increases. Highly oblique collisions do leave significant solid fragments, but these are a small minority of the total population of accretionary outcomes.}
    \label{fig:mant_melt}
\end{figure*}

\added{Our results are broadly consistent with \citet{Nakajima2021} after correcting the $T(P)$ melt curve equation in their melt model script. \citet{Nakajima2021} inadvertently coded a 0.011 prefactor to the $P^3$ term in the high-pressure branch of Equation~\ref{eqn:rubie} instead of the correct value of 0.00113, which increased the melting temperature and reduced the calculated amount of melt for the larger mass planets. There are several differences between our studies: the initial thermal profile, the equation of state, the method of calculating melt fraction, initial separation of bodies, and duration of simulations. Overall both studies demonstrate that above a certain specific impact energy, the accreting bodies will be completely melted and that the onset of complete melting is sensitive to the mass of the body. Other factors are discussed in section \ref{subsec:strength}.}

\subsubsection{Metal-Silicate Miscibility}\label{subsec:misc}
Potential miscibility between core and mantle materials has not yet been studied in detail, largely due to the uncertainties in temperatures calculated in SPH impact simulations using the previous generation of EOS models. With the revised iron alloy and pyrolite ANEOS equations of state, we can now more accurately evaluate post-impact temperatures and provide new insights, and cautionary statements, about the interpretation of temperatures in planetary collisions (see \S\ref{sec:discussion}). Here, we refer to materials becoming ``miscible'' as materials that have been heated above the solvus closure temperature for a given pressure, above which two separate chemical components become a single homogeneous solution for all mixing ratios.

\begin{figure}
    \includegraphics[width=0.45\textwidth]{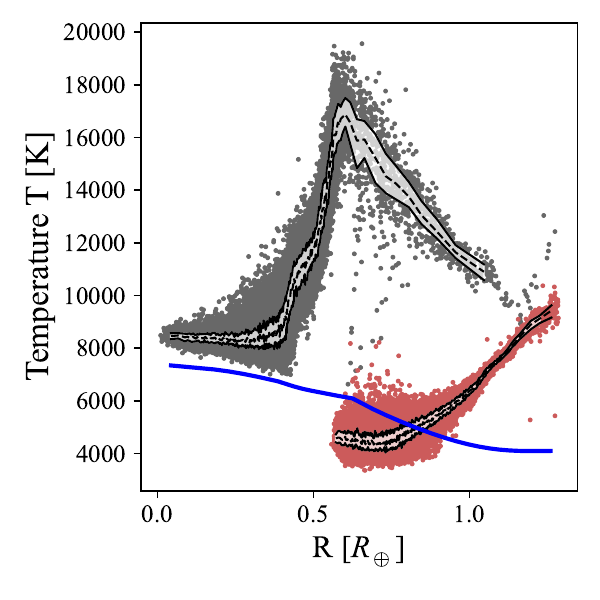}
    \caption{Temperature-equatorial radius plot for an example post-impact planet compared to the Fe-MgO solvus closure temperature from \citet{Wahl2015} (solid blue line). Silicate SPH particles are shown in red while iron alloy particles are shown in grey. The mantle contains ``superheated'' metal particles that are at significantly higher temperatures than the surrounding mantle material. 
    The median temperature values for a 100-particle moving bin are represented by the dotted black line, with the 25$^\textrm{th}$ and 75$^\textrm{th}$ percentiles represented by solid black lines. The grey region indicates that the majority of SPH particles are confined to a much smaller temperature range than the total spread. The ``equatorial'' region of the planet is defined as a region within $\pm0.1\,R_\text{planet}$ of the equator.
    The example here is a \(0.71M_\oplus\) body resulting from a \(0.5:0.25M_\oplus\) impact with \(\theta_i=30\degree\) and \(V_i/v_{esc}=1.15\). Pressure profiles for this planet are also displayed in Figure \ref{fig:rubie_herc_sph}.}
    \label{fig:tempprofile}
\end{figure} 

\begin{figure*}
    \centering
    \includegraphics[width=.9\textwidth]{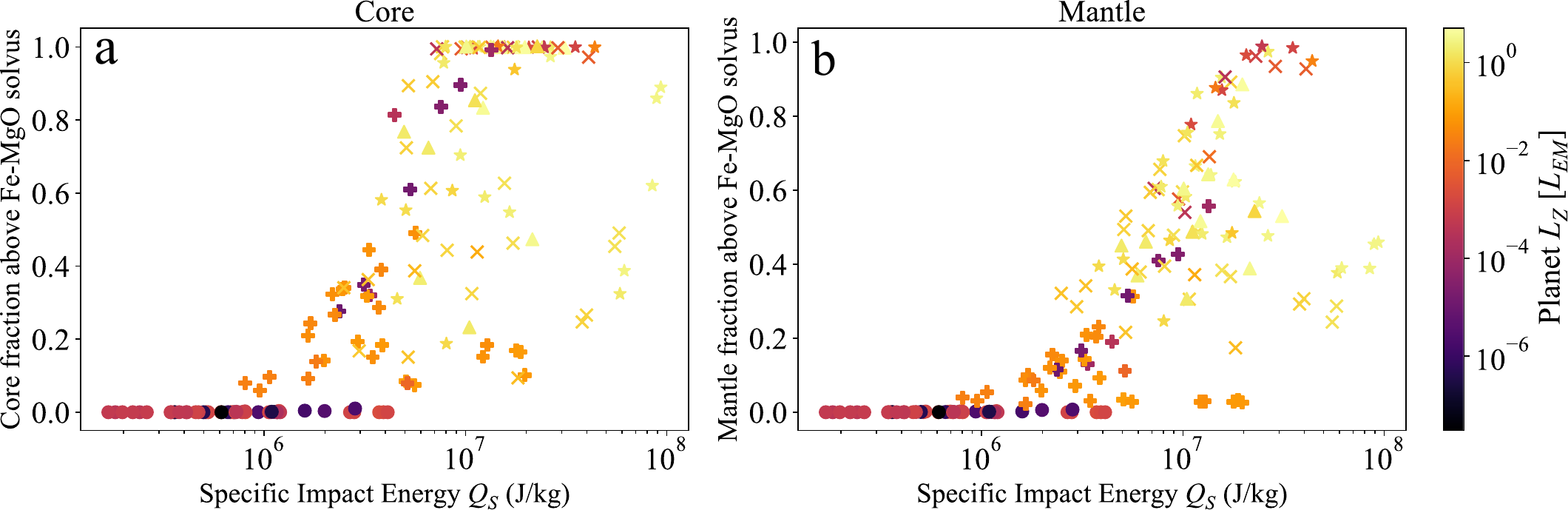}
    \caption{The fraction of core (\textbf{a}) and mantle (\textbf{b}) material in accretionary collisions that becomes miscible post impact, defined as attaining temperatures above the Fe-MgO solvus closure \citep{Wahl2015} at the respective pressure of the material. Data points are plotted as a function of specific impact energy \(Q_S\) and are color-coded by their angular momentum as fractions of the angular momentum of the Earth-Moon system [$L_{\rm EM}$]. Symbols indicate target planet mass: $\bullet$ -- moon (\(0.01 M_\oplus\)); \ding{58} -- Mars (\(0.1M_\oplus\)), $\times$ -- half-Earths (\(0.5 M_\oplus\)); $\star$ -- proto-Earths (\(0.91M_\oplus\)); and $\blacktriangle$ -- super-Earths (\(1.3 M_\oplus\)).}
    \label{fig:misc_Qs}
\end{figure*}

If miscibility occurs, 
the conditions for partitioning elements between metal and silicate may be dominated by the conditions at which the solution cools below the solvus and separate chemical phases begin to exsolve. The existence of miscible layers and the separation of those layers during cooling can also significantly affect the transport of heat through the planet and the total planetary heat budget \citep{jacobson2017,bailey_thermodynamically_2021}. Thus, material becoming miscible during formation, such as in giant impacts, is a critical process that must be included in studies of core formation and planetary thermal evolution. Here, we use the Fe-MgO solvus calculated by \citet{Wahl2015} as a proxy for terrestrial metal-silicate miscibility as comparable information is not yet available for more geologically relevant compositions. 
Recent updates to Fe-MgO phase equilibrium calculations indicate that solvus closure temperatures at higher pressures ($>$60 GPa) may be greater by 15-20\% than originally thought \citep{insixiengmay_mgo_2025}, but miscibility is still expected to be widespread. As the solvus closure of \citet{Wahl2015} better agrees with experimental data at low pressures, we choose to use it as a baseline estimate.
In this work, we do not include the effects of sub-resolution chemical mixing or diffusion, as there are currently no established methods of treating such processes during SPH impact simulations. 

Previous work has found that giant impacts can heat as much as 77\% of the mantle above the Fe-MgO solvus closure temperature in canonical moon-forming impacts (a 0.13$M_\oplus$ impactor hitting a 0.9$M_\oplus$ target at $\sim45\degree$) onto initially solid planets \citep{caracas2023}. We find similar results for collisions onto proto-Earth sized planets, including cases where close to 100\% of both the core and mantle exceed the solvus closure temperature. An example of such a case is shown in Figure~\ref{fig:tempprofile}, which present the post-impact equatorial temperature profile of a 0.71$M_{\oplus}$ body. The core and upper mantle are hotter than the Fe-MgO solvus closure temperature (blue line).

We calculated the core and mantle mass fractions that fall above the Fe-MgO solvus closure temperature for each impact in our study. Our results are summarized in Figure \ref{fig:misc_Qs} and data for each simulation is available in the accompanying GitHub repository. 
Collisions with targets that approach or exceed Mars-mass can result in a significant fraction ($>10$\%) of mantle and core material being above the solvus closure temperature. A supermajority of collisions involving Mars-sized targets produced significant amounts of core material above the solvus closure temperature, while over half produced significant mantle fractions above the solvus closure temperature. When target planet masses approach \(0.5M_\oplus\), over 50\% of both the core and mantle exceed the solvus closure. For larger mass planets, super-solvus core fractions quickly approach 100\% with more energetic collisions. For the mantle, we find similar results to \citet{caracas2023} for \(0.91M_\oplus\) mass targets, with higher specific impact energy cases regularly heating more than 70\% of the mantle above the solvus closure, with 3 of 37 half-Earth-sized and 6 of 37 proto-Earth-sized accretionary collisions in our survey producing mantle miscible fractions above 90\% (all of which also produced close to 100\% miscible core fractions). These estimates for miscible fractions are likely an upper limit for the core, and a lower limit for the mantle, as SPH codes do not provide a full treatment of heat transfer or chemistry between materials (\S \ref{subsec:temp}). 

The locations of these miscible fractions are in the upper portions of both the core and mantle, but these areas are likely an artifact of the SPH method for two reasons. First: the outer region of the core is unrealistically hot, while the lower mantle is unrealistically cold; moderation between the upper core and lower mantle would result in a larger mantle region heated above solvus closure temperatures. Second, in a fully non-solid mantle the regions that form metal-silicate solutions would be denser than nearby molten silicates and would sink to the base of the mantle. Thus, we expect the formation of a metal-silicate solution layer at the CMB regardless of where the constituent material originated.

\section{Separating thermal and rotational effects of giant impacts} \label{sec:rotational effects}
Core formation studies must make assumptions about the interior structure of the growing planet and the stage of cooling that dominates metal-silicate equilibration to determine the chemical evolution of the body. To construct a history of equilibration, a planet's growth rate may be parameterized analytically or utilize the collision histories from $N$-body planet formation simulations \citep[e.g.,][]{Rubie15,de_vries_impact-induced_2016}. Typically, the internal pressure-temperature profiles at the time of equilibration are assumed to correspond to static, non-rotating solutions with mantles at the freezing point (\S\ref{subsec:Splanet}). In this section, we focus on determining more realistic internal pressure-temperature profiles for planets during growth that incorporate the dramatic changes to their thermal and rotational states due to giant impacts. In particular, we will focus on separating the transient thermal effects from longer-lasting rotational effects. 

\begin{figure}
    \includegraphics[width=0.48\textwidth]{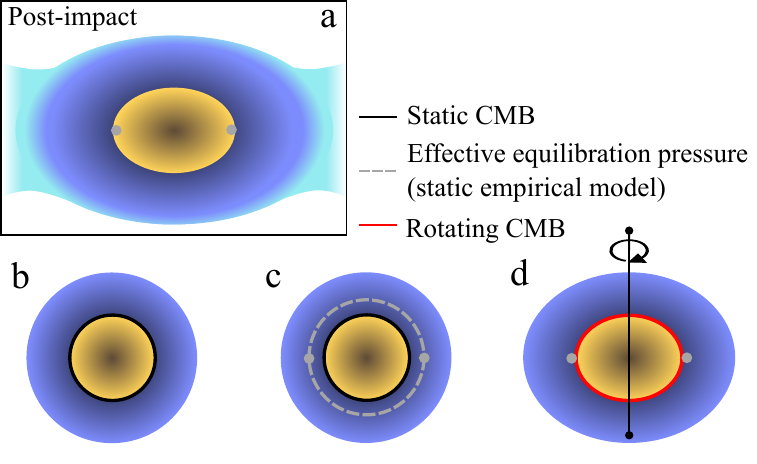}
    \caption{A schematic of the different planetary models used to calculate pressures at the CMB, as described in \S \ref{sec:rotational effects}. Panel \textbf{a} depicts a post-impact body with a fuzzy boundary between a mantle layer (darker blue) and a dense silicate atmosphere (lighter blue), extended into a synestia (not fully pictured). Panel \textbf{b} represents a static model planet. Panel \textbf{c} highlights a spherically-symmetric depth (dashed line) that corresponds to the ``static empirical model'' equilibration pressure inferred by modeling the observed composition of planetary mantles. Panel \textbf{d} depicts a rotating model planet. The red line highlights the CMB pressure of the rotating model, with grey points indicating a pressure similar to the grey points in Panel \textbf{c}. In Figures \ref{fig:Pvalues}, \ref{fig:Pvalue_mass}, \& \ref{fig:hercules_v2}, the black $\times$ corresponds to the black line in Panel \textbf{b}, the grey \ding{58} corresponds to the dashed grey line in Panel \textbf{c}, and red $\bullet$ corresponds to the solid red line in Panel \textbf{d}.}
    \label{fig:cartoon}
\end{figure}

\begin{figure}
  \includegraphics[width=0.45\textwidth]{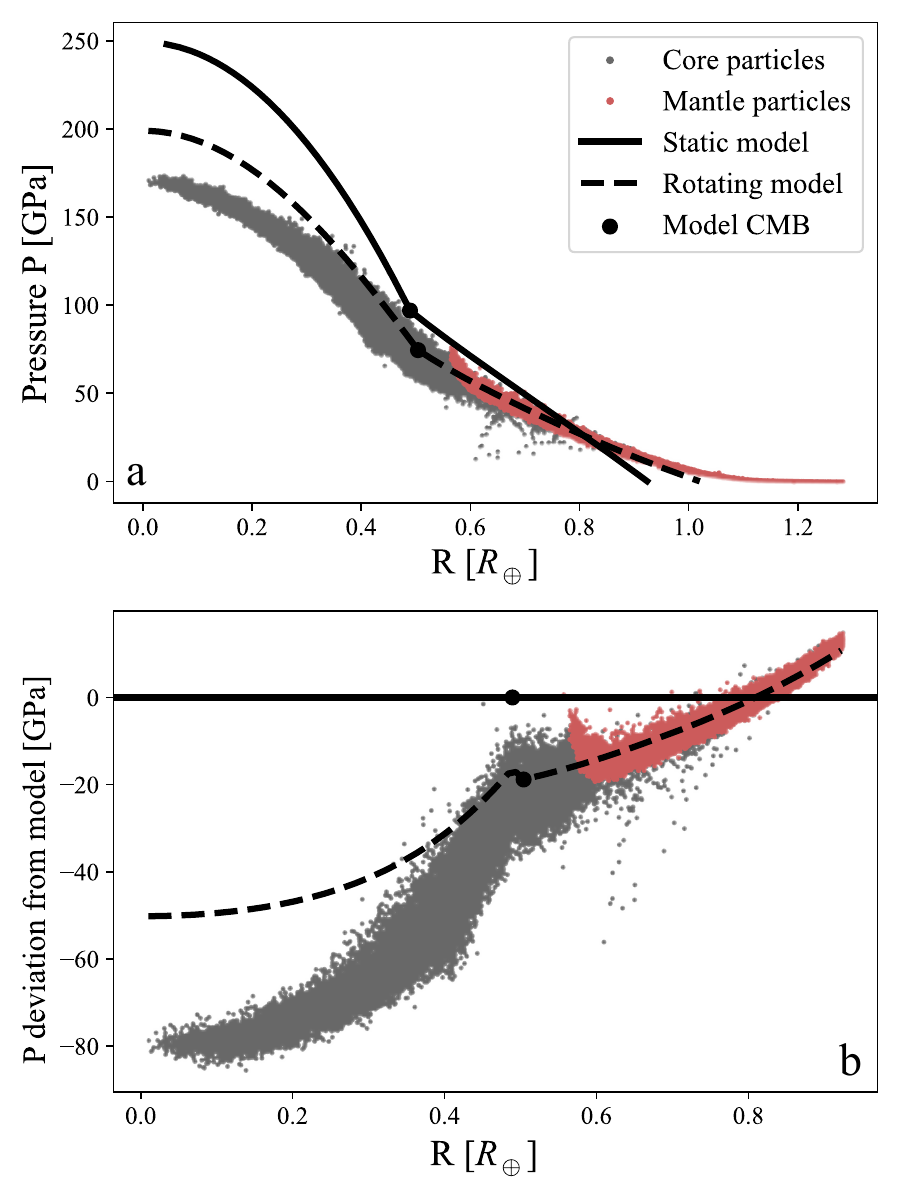}
    \caption{Example separation of the thermal and rotational effects from a giant impact on a body's pressure profile. An example post-impact body (co-rotating mantle SPH particles in red and core particles in grey, also Figure~\ref{fig:cartoon}\textbf{a}) is shown compared to a corresponding static model profile with the same mass and core mass fraction (solid black line; Figure~\ref{fig:cartoon}\textbf{b}), and a rotating model planet (dashed line; Figure~\ref{fig:cartoon}\textbf{d}) with the same mass, core mass fraction, and angular momentum. The post-impact temperature profile for this example is shown in Figure \ref{fig:tempprofile}. Panel \textbf{a} shows the absolute pressure while Panel \textbf{b} shows the pressure difference from the static model.}
    \label{fig:rubie_herc_sph}
\end{figure}

In this section, we describe a new approach to separate the thermal and rotational effects of giant impacts on the internal pressure of planetary bodies. 
The results from our SPH simulations describe the state of the body in the days after an impact. Post-impact bodies will cool to a magma ocean temperature profile similar to the static model in hundreds to thousands of years \citep{Lock2020}. The lifetime of the magma oceans can span millions of years \citep[e.g.,][]{Reese2006}. At present, robust techniques have not been developed that can directly model the coupled dynamical-thermal evolution from a hot, partially vaporized body to the subsequent solid body, but we can investigate and compare the states at different points in time with a small number of assumptions \citep{Lock2018,LockStewart2019,Lock2020}. 
Rather than beginning with a vaporous post-impact body and cooling to a condensed planet, it is more computationally straightforward to start with a reference internal model that represents the body's thermal state at the time of equilibration (e.g., solidification of lower mantle) and then add rotation. These reference models are shown schematically in Figure~\ref{fig:cartoon} where grey dots and dotted lines indicate where equilibration is occurring in each scenario.

We used the static model planets described in \S\ref{subsec:Splanet} to represent likely states during metal-silicate equilibration. We constructed a representative static model planet for each post-impact body in our dataset (Figure~\ref{fig:cartoon}a), taking into account the total mass and mantle-core ratios. The $P(R)$ structure of the static planets are given in Equations \ref{eqn:T&S12} and \ref{eqn:T&S1}, and the $T(P)$ profile uses Equation \ref{eqn:rubie}. We stress that, while we have chosen a specific reference thermal profile for comparison to previous works, our method for separating rotational effects is applicable for any choice of planetary profiles. Also, the results presented here will be broadly applicable for all choices of solid planet profiles, as the thermal expansivity of solid silicates is small compared to the effects of rotation.

While the heat from the giant impact radiates away, the angular momentum imparted to the growing planet will be conserved. We isolate the rotational consequences of giant impacts from the thermal effects by spinning up each static model planet (Figure~\ref{fig:cartoon}a) to match the angular momentum of its corresponding post-impact SPH body (Figure~\ref{fig:cartoon}d). We used the static model profiles (\S \ref{subsec:Splanet}) as initial conditions for HERCULES calculations (\S \ref{subsec:meth-HERC}) to solve for the internal structure of a freezing magma ocean planet with the same angular momentum as the corotating region of each post-impact SPH body. For generality, we neglect any dynamical interactions with satellites or other bodies during magma ocean freezing in this work.
As described earlier, these rotating planets and their CMB conditions we will be referred to as a ``rotating model'' or ``rotating CMB'' (Figure~\ref{fig:cartoon}c). 

We compare our calculated CMB conditions to the inferred effective equilibration pressures of metal-silicate equilibration from \citet{Rubie15}. These equilibration conditions, estimated by modeling the observed chemistry of planetary mantles, will be referred to as the ``static empirical model'' as it represents an empirical offset from the CMB conditions in the static model (discussed further in \S \ref{subsec:deficits}). The static empirical model pressures have generally been interpreted to represent the depth of post-impact melting \citep{Rubie15,deVriesLPSC2014,de_vries_impact-induced_2016}, which we illustrate in Figure~\ref{fig:cartoon}b. However, the inferred equilibration pressures could also be an amalgamation of several different effects.

Before we continue it is important to note that a number of assumptions were needed to construct our reference state models. The mass and core mass fraction of the static and rotating model planets correspond to the corotating region of the post-impact body. The coupled evolution of the disk and co-rotating region is not well understood at this time, and this assumption provides a standard point of comparison across our impact outcomes. Incorporation of the total bound mass is not physically possible as discussed below. Similarly, the rotating model planets do not include all of the angular momentum of the post-impact system, but only the angular momentum contained in the inner co-rotating region. The angular momentum in the disk was not included. 
Using the parameters of the co-rotating region, the HERCULES internal structure code was able to produce a rotating model planet for almost every impact in our dataset. However, we encountered lack of convergence when the post-impact state had a close-in moonlet.
Most of the time, this was resolved by letting the SPH simulation run for an additional length of time, but in some clumpy synestia-like cases the problem remained after $>$200 hours of simulation time and these 3 impact scenarios were excluded from further analysis. Attempts to include the mass and angular momentum in the disk in the rotating planet calculation often failed to converge to a solution due to the system exceeding the corotation limit \citep{Lock17}. Thus, moons must form in many of these systems before a stable corotating planet can be produced.

In most cases, the moment of inertia for the rotating model planet was similar to the moment of inertia of the corotating region of the post-impact bodies (Figure \ref{fig:dMOI} in Appendix~\ref{sec:subfigs}). However, the extended nature of the post-impact disks means that the moment of inertia of the total bound mass of the post-impact system is orders of magnitude larger, despite the disks containing only a small percentage of the total mass (Figure \ref{fig:dMOI_disc} in Appendix~\ref{sec:subfigs}).
As the disk's angular momentum is accreted to the planet, the rotation rate of the body will increase, which would further decrease the internal pressure. Thus our estimates of the CMB pressures in the rotating model planets can be considered an upper limit during the period between the giant impact and magma ocean freezing, when any moons are likely to remain close to the planet \citep{Zahnle_2015}. Ultimately, the magnitude of the pressure decrease will depend on the mass and angular momentum carried by the surviving moons, which we leave for future work.

Figure~\ref{fig:rubie_herc_sph} presents three pressure profiles for the example 0.71$M_{\oplus}$ body shown in Figure~\ref{fig:tempprofile}: 1) the post-impact equatorial profile of the corotating region taken directly from SPH (colored $\bullet$; corresponding to Figure~\ref{fig:cartoon}d); 2) the static model planet with the same mass (solid line; Figure~\ref{fig:cartoon}a); 3) the rotating model planet with the same mass and angular momentum (dashed line; Figure~\ref{fig:cartoon}c). 
The CMB pressure in the rotating model lies between the static model and the post-impact body. Compared to a static model, the rotating model profile contains only the contributions of rotation towards the internal pressure deficit, while the post-impact SPH profile contains the combined thermal and rotational effects.

\begin{figure*}
    \centering
    \includegraphics[width=1.0\textwidth]{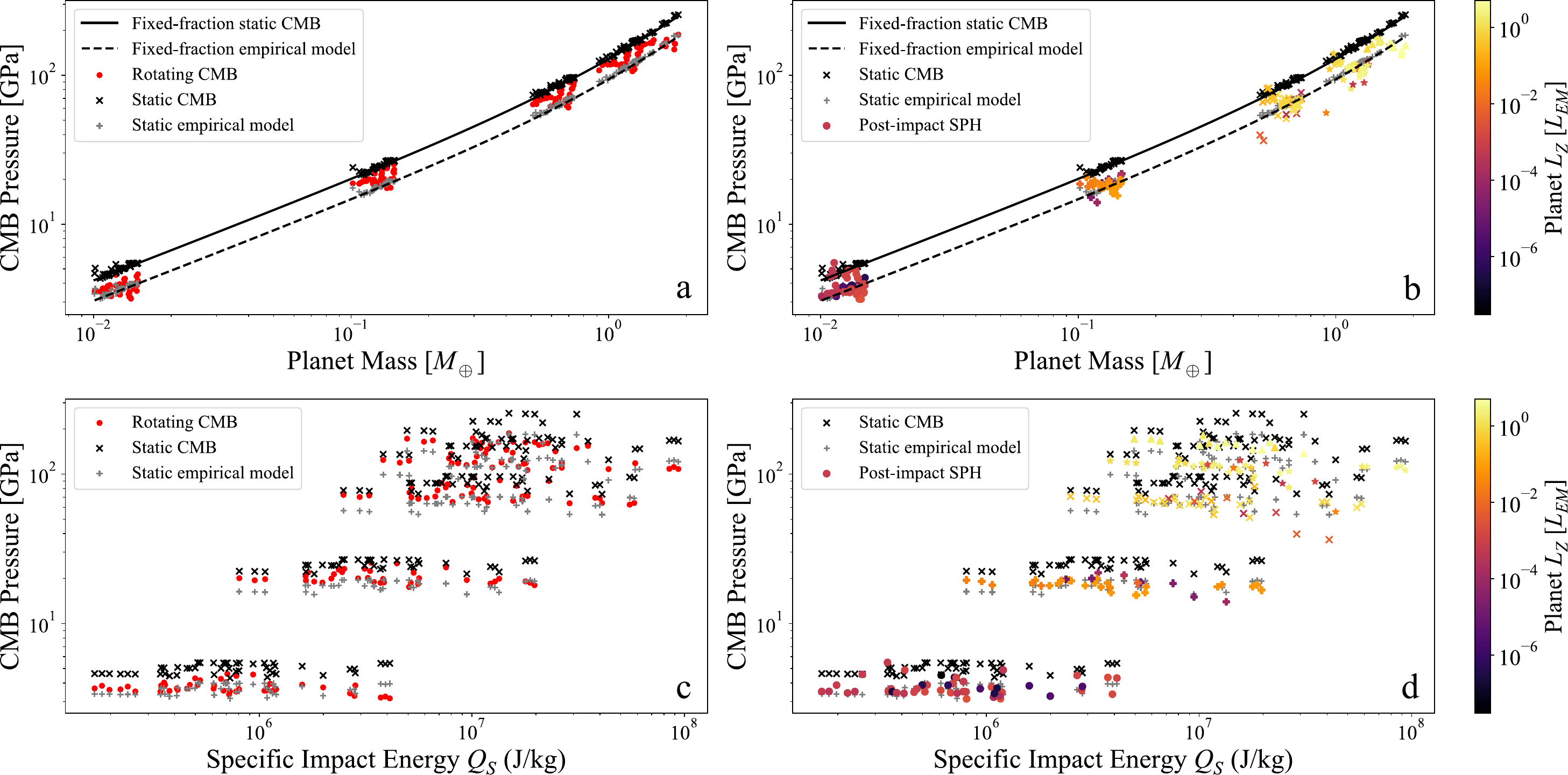}
    \caption{Comparing static, rotating, and empirical CMB pressures for conditions of metal-silicate equilibration. Panel \textbf{a}: The CMB pressures of the static and rotating models as a function of final planet mass. Each data point represents the model profile calculated for each impact in our dataset: black $\times$'s representing a static model, grey \ding{58}'s for the effective equilibration pressure fractional correction to the static model – referred to as the static empirical model, and red $\bullet$'s for the CMB pressures of the rotating models. Solid lines represent the analytic form of CMB pressure as a function of planet mass for the static model assuming a fixed 30\% core fraction; the CMB pressure for each impact case  varies slightly with the actual core mass fraction, causing a deviation from the analytic trendline. Panel \textbf{b}: Similar to the previous panel, but with the actual post-impact SPH CMB pressures plotted for comparison instead of the rotating models, portrayed and color coded in the same manner as in Figure \ref{fig:misc_Qs}. 
    Panel \textbf{c}: CMB pressures of the static, empirical, and rotating models as a function of modified specific impact energy \(Q_S\). Note that the data is largely grouped by initial target mass, as shown by the post-impact data markers in Panel \textbf{d}. Across both panels on the left, the rotating model CMB pressures generally fall between the static model and the empirical model for each simulation. Across both panels on the right, the post-impact data generally falls near the empirical model, but has a larger spread to lower and higher pressures (see Panels \textbf{b} \& \textbf{c} in Figure \ref{fig:hercules_v2} for more info).}
    \label{fig:Pvalues}
\end{figure*}

\begin{figure*}
    \centering
    \includegraphics[width=1.0\textwidth]{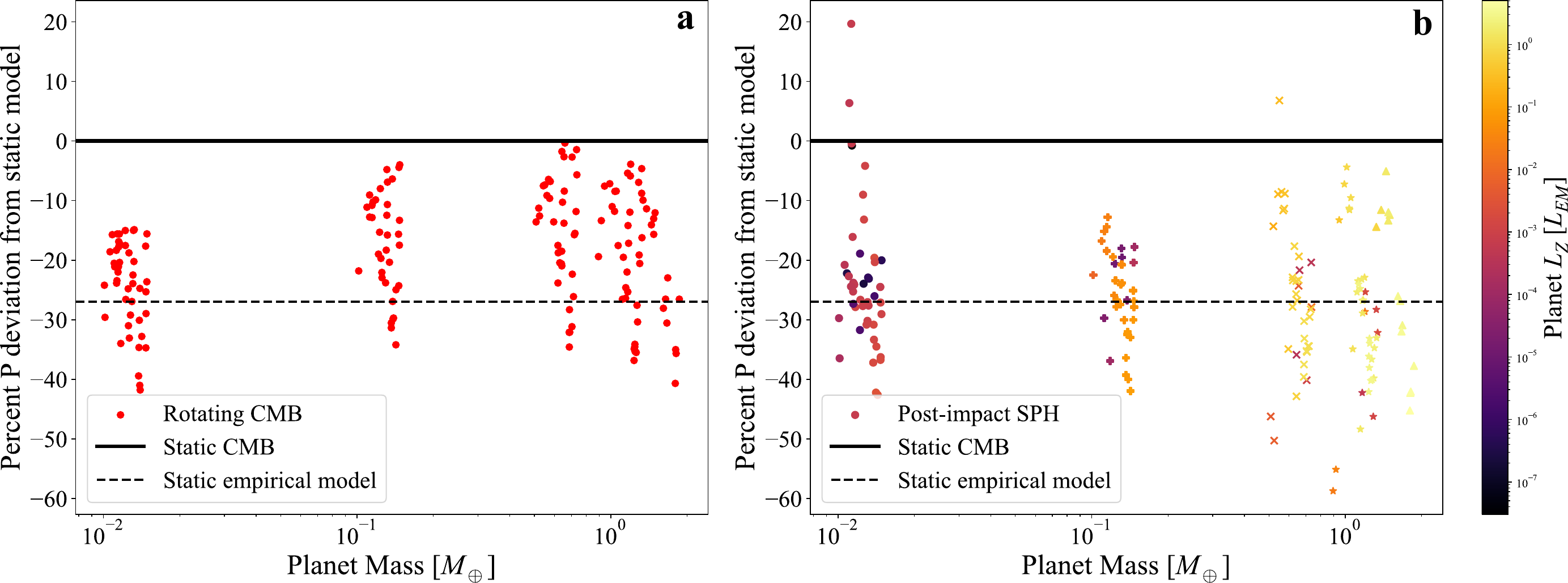}
    \caption{The same data as Panels \textbf{a} \& \textbf{b} in Figure \ref{fig:Pvalues}, except compared relative to the CMB pressures of the static model. Panel \textbf{a}: the CMB pressures of the rotating model (red points) relative to the static model (solid black line) and the effective equilibration pressure of the static empirical model (dotted black line). Panel \textbf{b}: the CMB pressures of the post-impact bodies (points colored by angular momentum of the final body, see Figure \ref{fig:misc_Qs} for a description of markers), also relative to the static model CMB pressures. Note that the CMB temperatures are generally much greater in the post-impact bodies than in the rotating planet model. Considering the mass in the corotating region only, the expected internal pressure increase due to the increase in density are approximately counteracted by the pressure reducing effects from the increase in rotation. Including some fraction of the bound mass and angular momentum of the post-impact disk would lead to a greater contribution from rotation.}
    \label{fig:Pvalue_mass}
\end{figure*}

\begin{figure*}[b]
    \includegraphics[width=1\textwidth]{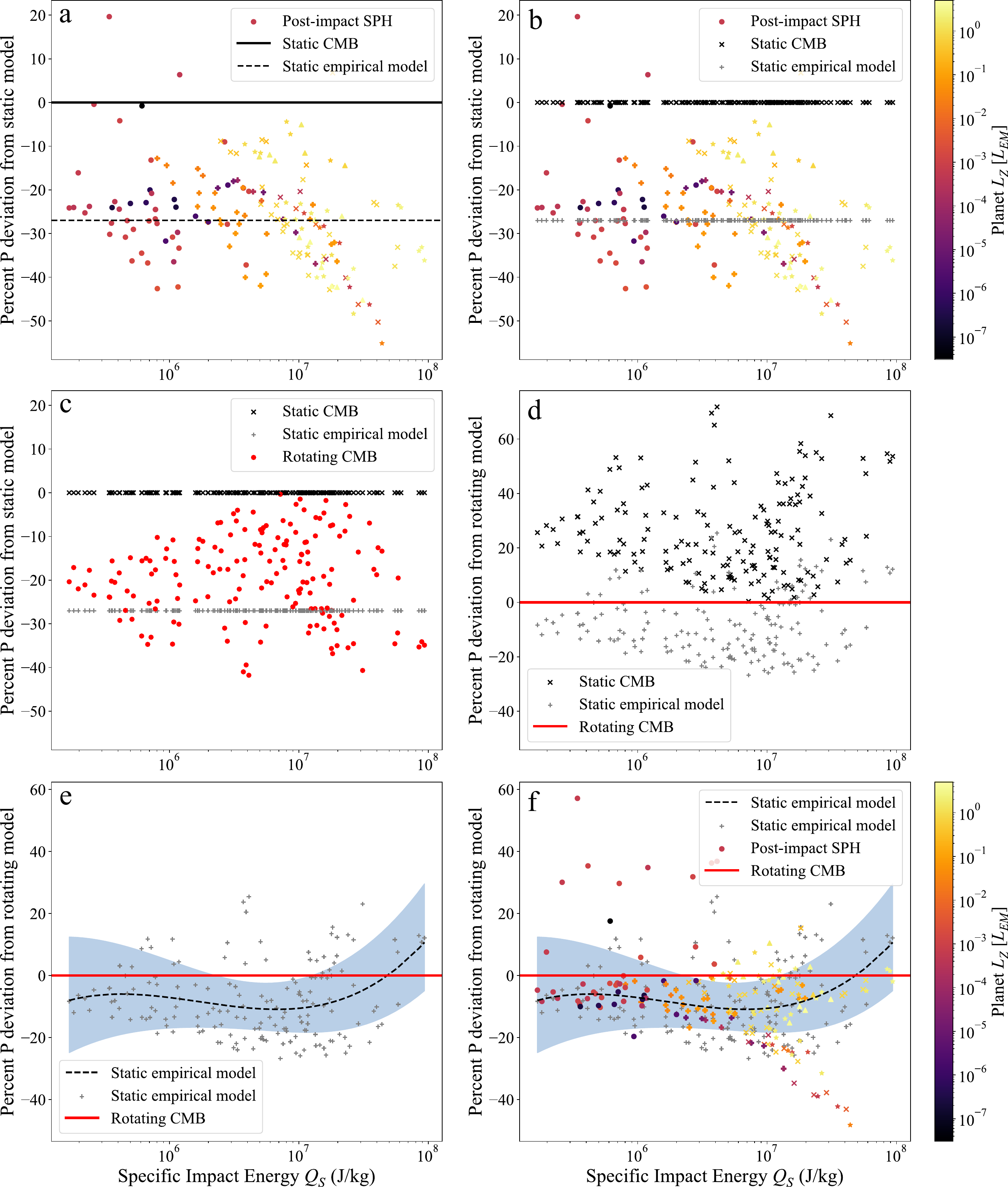}
    \caption{\textit{Caption on next page.}}
    \label{fig:hercules_v2}
\end{figure*}
\addtocounter{figure}{-1}
\begin{figure*}[t]
    \caption{\textit{Continued from previous page.} Comparing the CMB pressures of post-impact SPH data for accretionary collisions in our survey dataset with model profiles representing reference conditions for metal-silicate equilibration, both static and rotating. Panel \textbf{a}: The post-impact SPH data from this study is plotted as a function of specific impact energy and color-coded by final angular momentum in terms of $L_{\rm EM}$. Marker icons represent target planet size as described in Figure \ref{fig:misc_Qs}. The vertical axis indicates the CMB pressure difference from the static model profiles for each planet, referred to as the ``static CMB'' and represented by the solid black line. The dotted line represents the static empirical model. 
    Panel \textbf{b}: similar to Panel \textbf{a} except the pressure differences for the static CMB and the static empirical model for each SPH simulation are represented by black $\times$'s and grey \ding{58}'s, respectively, instead of lines. Panel \textbf{c}: similar to Panel \textbf{b}, except without the SPH data points shown and instead displaying the CMB pressure data points for the rotating thermal model profiles as calculated for each SPH simulation (in red) and referred to as ``rotating CMB'' data, as in Figure \ref{fig:Pvalues}. Panel \textbf{d}: the same underlying data from Panel \textbf{c}, except the red rotating CMB points are now the reference line and the vertical axis now represents the pressure difference from the rotating CMB (the solid red line). 
    Panel \textbf{e}: similar to Panel \textbf{d}, except only the static empirical model data is displayed, accompanied by a smoothed fit (dotted line) and corresponding 90\% confidence interval. Panel \textbf{f}: similar to Panel \textbf{e}, now including post-impact SPH CMB pressures shown in Panels \textbf{a} \& \textbf{b}, except this time the post-impact data is relative to the rotating CMB for each data point.}
\end{figure*}

The CMB pressure for each reference state for all impacts is presented in Figure \ref{fig:Pvalues}.  For each impact outcome, we extracted the equatorial CMB pressure for (i) the post-impact body (colored by corotating angular momenta in Panels \textbf{c} and \textbf{d}), (ii) the static model (black $\times$ in all panels), (iii) the static empirical model (grey \ding{58} in all panels), and (iv) the rotating model (red $\bullet$ in Panels \textbf{a} and \textbf{b}). Figure \ref{fig:Pvalue_mass} shows the same data as a relative difference in CMB pressure compared to the static model. In most cases, the rotating model CMB pressure (red $\bullet$ in a) lie between the static model (solid line) and the static empirical model (dashed line) pressures. The post-impact CMB pressures span a wide range around the static empirical model (colored symbols in b). 


Since most collisions with planets Mars-sized or larger have substantial regions with temperatures above the metal-silicate solvus, the post-impact state has to cool below the solvus closure temperature for there to be metal-silicate equilibration in order to produce the equilibration signatures observed today. The dominant signatures of elemental partitioning then begin when the metal and silicate solution begins to separate into distinct compositional phases, as the Fe-MgO solvus near the closure point is quite flat \citep{Wahl2015}. Note that existing measurements of partitioning coefficients have been made at temperatures far below the solvus temperatures, and future work at higher temperatures are needed to understand the evolution of chemical equilibration during cooling.

We therefore assume that the observed metal-silicate equilibration signature is dominated by the end stages of cooling of the magma ocean. The goal is to reframe the empirical observations of the pressure of metal-silicate equilibration in comparison to the rotating planet model instead of the static planet model as the former is more realistic depiction of the expected magma ocean planets. 
Each step is depicted in Figure~\ref{fig:hercules_v2}: 
\begin{enumerate}[label={\Alph*)}]
    \item We begin with a standard comparison between the CMB pressures of the post-impact state and the static model represented by the horizontal solid line in Panel \textbf{a}. The empirical model value is a mean pressure deficit of 27\% as inferred by \citet{Rubie15}. The data are shown against specific impact energy, covering the typical range of giant impact scenarios during terrestrial planet formation.
    \item For each post-impact body in our dataset, we calculate the CMB pressures for each corresponding static model (black $\times$'s) and static empirical model (grey \ding{58}'s), shown in Panel \textbf{b}. These calculations have core mass fractions corresponding to each impact outcome (e.g., see pressure variations in Figure~\ref{fig:Pvalue_mass}\textbf{a} and \textbf{b}).
    \item As the bodies cool and contract after a giant impact, conservation of angular momentum increases the rotational velocity. Panel \textbf{c} replaces the post-impact CMB pressure for each body with the corresponding rotating model planet value as described above.
    \item We shift the reference standard for all the points from the static model (black $\times$) to the rotating model (red $\bullet$) in Panel \textbf{c} to derive the red line in Panel \textbf{d}. In other words, the values for the CMB pressure offsets for each model have been recalculated relative to the red line in Panel \textbf{d}.
    \item We approximately fit the new static empirical model CMB pressure offset as a polynomial function of specific impact energy (see Table \ref{tbl:hercules_fit}), shown by the dashed line and a 2.5 standard deviation shaded envelope in Panel \textbf{e}. We find that the empirically determined CMB pressures of equilibration overlap with the CMB conditions in the rotating model planets. The form of this polynomial is: $100\% \cdot (P_\text{CMB\,e.m.} - P_\text{CMB\,rot.})/ P_\text{CMB\,rot.} = A(\text{log}_{10}[Q_S])^3+B(\text{log}_{10}[Q_S])^2+C(\text{log}_{10}[Q_S])+D$.
    \item Panel \textbf{f} adds the post-impact CMB pressures (symbols colored by angular momentum) referenced to the rotating planet model. 
\end{enumerate} 


In the previous interpretation of the empirical equilibration pressures, the depth of melting would correspond to a substantial fraction of the depth to the CMB, shown by the dashed circle in Figure \ref{fig:cartoon}b. However, in a rotating planet, the pressures for equilibration are much closer to the CMB. This result is more consistent with the post-impact temperatures and phases calculated in this work (\S \ref{sec:melting}), which demonstrate that very little mantle remains solid in the giant impact stage of terrestrial planet growth, which would create no mechanism for metal to pond and equilibrate in the mid-mantle.

In this section, we have demonstrated that the pressure deficits from rotation are substantial. We find that the CMB pressures in the rotating planet models are a more physically robust estimate for the conditions of metal-silicate equilibration after a giant impact than static planet models. Because planets are expected to be rotating throughout planet formation, the pressure reductions due to rotation must also be included in smaller accretionary events during planet formation. Empirical scaling laws for the relationships between planet mass, angular momentum, and internal pressure will be included in a forthcoming companion paper.


\section{Discussion}\label{sec:discussion}

\subsection{Interpretations of temperature in SPH simulations: Implications for miscibility and heat transport}\label{subsec:temp}
Here we discuss the key features observed in the thermal profiles of our SPH post-impact bodies. In this work, we have improved on previous impact studies by running simulations using an EOS model that includes more accurate, experimentally constrained, temperatures. 
However, one of the shortcomings of all current SPH impact simulations is that materials are modeled as single-components, and particles that come into contact cannot chemically or thermally exchange material when they would realistically at least partially equilibrate.
As a result, some features in our simulations are a result of limitations in the SPH technique and would not occur in a real collision. 

For example, a substantial portion of the post-impact bodies in our dataset contain ``superheated'' iron particles entrained within the silicate mantle with temperatures over 5000~K higher than the surrounding mantle material (Figure~\ref{fig:tempprofile}). These particles are usually highly shocked impactor core material that reach temperatures where the density of the iron alloy overlaps with the silicate mantle, and can constitute close to 10\% of the planet's overall iron budget. In SPH, without any heat or chemical transfer, buoyancy dominates the radial positions of particles in the co-rotating region of the post-impact body and very low density iron remains in the mantle. Realistically, these iron particles would chemically react with the surrounding silicate material and come to a local thermal equilibrium. The lower silicate mantle of post-impact bodies would therefore be hotter and more iron rich than in the raw simulation output.
 

In many cases, this chemical and thermal equilibrium would lead to layers of the mantle where silicates and metal would form a single chemical solution. As described in \S \ref{subsec:misc}, we have found that the temperatures of a significant portion of post-impact bodies exceed the Fe-MgO solvus closure temperature after typical giant impacts (see  Figures~\ref{fig:tempprofile} and \ref{fig:misc_Qs}). Chemical equilibration in these regions would produce a solution, a thermodynamic mixture, of the two materials preventing separation of the iron to the core. Even in areas of the mantle with average temperatures beneath the solvus closure, silicates and metals could still form multi-component mixtures as the exact solvus temperature is dependent on the local average composition. Regions of the colliding bodies reach temperatures above the local solvus early in the impact and metal that in our simulations falls under buoyancy to the core could have realistically been retained by chemical solution in the mantle. 

It is important to note that the mixing we are discussing here is different to the process often referred to as ``mixing'' elsewhere in the literature. Recently, work has been done to improve how SPH describes interactions between particles with significantly different densities, to allow for finer scale physical ``mixing'' that better captures interfacial boundaries and instabilities \citep[e.g.,][]{sandnes_remix_2025,ruiz-bonilla_dealing_2022,Pearl_2022}. However, these works might better be described as improving \textit{stirring} of materials. Material boundaries can become more stretched out and disrupted and different materials are therefore more intermingled than in standard SPH. However, if shear is removed and the system is allowed to evolve just under buoyancy then the materials would separate out into layers depending on their density \citep[as seen in simulations of disruption of Jupiter's proto-core,][]{sandnes_no_2024}. Here, we are discussing thermodynamic mixing of materials, where initially separate materials form a single solution that cannot be separated by buoyancy forces.   

The existence of thermodynamically mixed layers in the mantle would delay some degree of core formation. Any free metal falling from the mantle into the core after the first few hours of the impact would have to pass through a thermodynamically mixed layer with the possibility that it would be dissolved into such a layer. That metal would remain in the mixed layer until the lower mantle cooled sufficiently to allow for phase separation of the metal-silicate solution and removal of the denser metal-rich phase to the core. The partitioning of elements between the metal and silicate would be set by the chemistry of the system at the solvus, allowing for a large amount of light elements to be added to the core which could provide a source of energy for Earth's magnetic field \citep{badro_early_2016, ORourke2016}.

A metal-silicate solution layer at the CMB could substantially influence the thermochemical evolution of planetary interiors and the development of dynamos \citep{jacobson2017,bailey_thermodynamically_2021,helled_fuzzy_2024,garnero_fuzzy_2000}. A metal-silicate solution layer at Earth's CMB early in its history could also play a role in the origin of the present-day Large Low Shear Velocity Provinces and Ultra Low Velocity Zones \citep[LLSVPs and ULVZs, volumes of material at the base of the mantle that have anomalous seismic velocities:][]{khan_thermally_2015,mcnamara_llsvp_2019} and the chemical anomalies that may be retained in the lower mantle \citep[e.g., noble gases,][]{williams_sujoy_2019,mukhopadhyay_early_2012}. Even after initial cooling and delayed core formation from the super-solvus layers, these layers would still retain a composition enriched in iron compared to the average mantle. If this iron enrichment produced a large enough density excess, then such layers could be stable at the bottom of the mantle for all of Earth's history and would be a long-lived seismic and chemical anomaly \citep{tackley_strong_2002}. Further, the timing for formation and exsolution of a mixed layer could have significant consequences for radiogenic isotopic systems such as Hf-W \citep[e.g.,][]{deguen_turbulent_2014,nimmo_tungsten_2010,fischer_sensitivities_2017} \added{because it would change the conditions for metal silicate equilibration compared to the standard assumptions made in the literature.} 

The strongly stratified cores and mantles produced in our simulations, where each layer has a large density and specific entropy gradient, would also significantly affect the post-impact evolution of the body. Given the computational limitations described above, the strength of the thermal gradients should be considered upper limits as some thermal equilibration by mechanical mixing of hotter (higher entropy) and cooler (lower entropy) regions would realistically have occurred during the impact event. The extent to which our simulations correctly capture chemical stratification is harder to determine, but may be a lower estimate. Our simulations suggest that the temperature on the core side of the CMB will be substantially (thousands of K) greater than the mantle side after the impact, providing a strong driver for heat transport out of the core. However, the post-impact temperature inversion of the core could counteract this driver and suppress dynamo activity for tens of millions of years \citep{arkanihamed_giant_2010}. 

Chemical and thermal equilibration during giant impacts have significant consequences for the chemical and thermal history of Earth and the interpretation of geochemical interpretation of accretion, but determining the extend of mixing will require significant computational and EOS developments.

\subsection{Pressure deficits in post-impact bodies and conditions for rapid core formation} \label{subsec:deficits}
We have found that the pressures in the interior of post-impact bodies and condensed rotating planets, specifically at the CMB, are lower than would be expected in static planets of the same mass. 
We now compare our results with standard approaches in order to discern the thermal and rotational effects of giant impacts on core formation in growing planets.

In studies of core formation, elemental partitioning between the mantle and core was previously assumed to occur at pressure-temperature conditions that lie between the mantle liquidus and solidus, conceptually representing the idea that closure of equilibration is defined by the $P-T$ conditions when the magma ocean freezes \citep{RUBIE201131,Rubie15,Wade2005}. These studies found that it is difficult to reproduce the moderately siderophile elemental abundances in the Earth's mantle when assuming that equilibration occurred at the CMB, e.g., in cases of a full mantle magma ocean.
This pressure deficit from the CMB led to the idea that equilibration occurred at an ``effective equilibration pressure'' (previously mentioned in \S \ref{sec:rotational effects}) that is a fraction of CMB pressures (often assumed to be a fixed fraction, $f$) in growing planets such that \(P_{\text{eq}}=fP_{\text{CMB}}\), where (for this example) \(P_{\text{CMB}}\) is described by Equation \ref{eqn:T&S12} when \(r=r_\text{CMB}\) \citep{Rubie15,fischer_sensitivities_2017}. \citet{Rubie15} found that CMB pressure fractions between 0.70 and 0.73 for Mars-sized or larger protoplanets can reproduce the present-day terrestrial mantle chemistry. Meanwhile, \citet{de_vries_impact-induced_2016} found that these pressures of equilibration can depend on the presence of an atmosphere affecting the crystallization time of a magma ocean, broadly agreeing with a value of \(f\sim0.7\) for slow-cooling planets with atmospheres, though they found that the required equilibration pressures increased with thinner atmospheres due to faster cooling. These results have been interpreted as representing the pressures at the bottom of an average melt pond in the mantle during collisional growth with embryo-embryo impacts \citep{Rubie15,de_vries_impact-induced_2016,Wade2005}. 
As our study did not include an additional atmospheric parameter space in the initial conditions of our SPH impacts, we chose to use these standard values rather than the more detailed treatment of fractional equilibration pressures described in \citet{de_vries_impact-induced_2016}, given they found that \(f\sim0.7\) was first-order accurate in embryo-embryo collisions. 

When examining our data, we find a significant pressure difference between our simulated post-impact SPH bodies and reference static models, as shown in Figures \ref{fig:Pvalues} \& \ref{fig:Pvalue_mass}.
This is due to both mechanisms described in \S \ref{subsec:meth-HERC}: thermal differences (bodies immediately post-impact are necessarily hotter and less dense than freezing magma oceans), and rotational contributions (as the reference temperature profiles are in static, non-rotating bodies). As post-impact bodies can cool rapidly \citep[over hundreds to thousands of years,][]{Lock2020}, the conditions immediately post-impact do not necessarily reflect the conditions at which chemical equilibration occurs if core formation continues throughout the cooling of the magma ocean. Only core formation processes that occur rapidly after the giant impact will experience these exact pressures. As described previously, to determine the pressures of later metal-silicate equilibration it is necessary to separate the thermal and rotational effects on internal pressure. 


\subsection{Implications for core formation}
Including the rotational effects of giant impacts as described in \S \ref{sec:rotational effects} has two major implications on core formation processes, including those that come after the giant impact. First, metal-silicate equilibration may have taken place closer to CMB conditions, or physically closer to the CMB, than previously thought. This is not because the pressure and temperature conditions of equilibration are necessarily different, but instead because pressures at the CMB were lower than previously estimated due to rotational effects. In terms of the models included in this work, the pressures of the static empirical model (the inferred conditions of equilibration) are very close to the rotating model CMB pressures. This indicates that as the planet becomes extended, the physical location of the CMB will be closer to the region in the post-impact body where the pressures and temperatures of equilibration occur (see Figure \ref{fig:cartoon}).

Second, the previously discussed static empirical model – the fractional correction for effective equilibration depth – is unlikely to represent a depth of melting. Any core formation analysis must necessarily include a rotational contribution, since all realistic impacts are oblique and impart significant angular momentum. For many impacts in our dataset, the pressure deficit due to rotation alone can match the equilibrium pressures inferred from moderately siderophile element data in Earth's mantle. Further, almost all accretionary impacts with targets larger than Moon-sized bodies result in complete mantle melting. 

It has been suggested that much greater fractional corrections of \(f\sim0.5\) could be more appropriate than those used in this discussion for collisions between Moon-sized bodies \citep{Rubie15}. For such impact scenarios studied in this work, these collisions do not fully melt the mantle, and there is often a small amount of solid mantle material left, with a supermajority in a melted state slightly below the liquidus. Additionally, almost no mantle material in bodies of this size is heated past the Fe-MgO solvus closure.
It is reasonable that in these cases there could be a contribution relating to the shallow depth of fully-liquid mantle material or lack of miscibility in the upper mantle, along with the effect of rotation in such a manner that sums to the much lower fractional correction of 50\% of static model CMB pressures.
Further investigations of metal-silicate equilibration will likely benefit from determining the fractional equilibration pressures necessary to reproduce elemental abundance constraints on a more size-specific or even planet-specific basis, as both melt depth and equilibrium pressure structures vary with impact energy, impact angle, and planet size \citep{Lock17,Nakajima2015}, while equilibration conditions could vary significantly with the presence of an atmosphere \citep{de_vries_impact-induced_2016}.

\added{\subsection{Effects of material strength and other assumptions}}\label{subsec:strength}
In this study we have considered collisions between bodies as small as 0.01~$M_{\rm Earth}$ using purely hydrodynamic simulations. However, the effects of material strength on giant impact outcomes have been found to be potentially non-negligible in impact cases involving bodies smaller than $\sim 0.1$~$M_\oplus$ \citep{emsenhuber_sph_2018,emsenhuber_sph_2024}. Impact simulations including and without material strength showed differences in the distribution of ejecta and the distribution of post-impact temperatures, and so it is important to consider the potential effect of material strength on our results. 

As discussed by \citet{emsenhuber_sph_2024} the transition into a strength-negligible regime is not only controlled by the planetary mass, but also by the temperature-dependent thermal weakening of the material yield strength. During solar system formation, bodies were considerably hotter than at present and stayed at elevated temperatures near the solidus for long periods of time \citep[e.g.,][]{Solomatov1993,Andrault2016}. The effect of material strength during impacts involving these warm, early bodies would therefore be limited, with much of the solid mantle melting soon after first impact. Strength-dependent processes are therefore likely secondary compared to the shock and hydrodynamic processes captured in our simulations, and the results of this study are likely widely applicable during much of planet formation, even for the smallest-mass bodies we considered. However, more work is needed to understand the effects of material strength in giant impacts involving cooler and smaller bodies.

\added{In this work, we held the core mass fraction constant. As observed in \citet{Carter15}, the smallest bodies may have a greater diversity of core mass fraction compared to the largest bodies. The change in core mass fraction will need to be considered in future work.

In late giant impacts, mantles may have cooled so that strength and other factors become more important. As discussed in \citet{Solomatov1993} and \citet{Andrault2016}, the Earth's mantle cools slowly after solidification. Small planetary bodies may cool more quickly and have a larger variety of outcomes than calculated here. However, current observations of Mars infer a melt layer at the core-mantle boundary, suggesting that cooling may be slow even in smaller planetary embryos \citep{samuel2023geophysical,khan2023evidence}. Future studies should focus specifically on the sensitivities to the initial thermal profile.

Our resolutions tests determined that the particle number in this study was sufficient for calculating the largest remnant and energy deposition; however future work interested in details of the miscible region at the core mantle boundary should conduct a dedicated resolution test.
}


\section{Conclusions}
Accurately determining how giant impacts perturb the thermal state of planetary bodies is necessary to trace the physical and chemical state of Earth during its formation. Here we have presented the results of a wide-ranging suite of giant impact simulations using state-of-the-art EOS models, providing greatly improved estimates for heating efficiency, heat distribution, and core-mantle-boundary conditions. We have determined how the initial efficiency of converting impact energy into internal energy, the distribution of that internal energy within the resulting planet, and the temperature and pressure changes at the core-mantle boundary after a giant impact varies between impacts and respond to initial conditions.

Our results show that giant impacts typically produce radically different conditions of metal-silicate equilibration than previously assumed in core formation models. Firstly, the extreme heating during giant impacts produces large volumes of material on either side of the CMB in post-impact bodies where metal and silicates are miscible and high mixing ratio iron-silicate solutions would be present and stable. Core formation from these regions would be delayed by at least hundreds of years as the lower mantle must first cool below the Fe-MgO solvus closure, and hence exsolve into separate phases, before metal can be extracted to the core. Metal-silicate equilibration at near-solvus conditions would radically change the partitioning of  elements into the core and vestiges of super-solvus regions could still be recorded in the present-day mantle as LLSVPs, ULVSs, or regions with isotopic anomalies.

Secondly, we find that the rotation of planetary bodies immediately after giant impacts and during the subsequent magma ocean phase significantly decreases the CMB pressure to such an extent that metal-silicate equilibration near the CMB in rotating planets is likely consistent with the abundance of moderately siderophile elements in the present-day mantle. We conclude that previous work has probably misattributed the effect of rotation in the aftermath of giant impacts on equilibration conditions to a relatively modest extent of impact melting, which is incompatible with the high degrees of melting found in our giant impact simulations. 

Moving forward, core formation models that accurately incorporate these two effects are needed to be able to interpret the geochemical tracers of core formation. As a step towards achieving this goal, we have provided scaling laws that predict the pressure and temperature of the CMB for a given set of impact parameters so that future core formation models can incorporate realistic post-impact conditions. Eventually, accretion models that trace the rotation rate of planets through accretion and include the formation and exsolution of super-solvus layers will be able to more completely elucidate the history of Earth and other planets.

\begin{acknowledgements}
This work was supported by NASA Emerging Worlds grant 80NSSC21K0386 (ANP, STS) and UK Natural Environment Research Council grant NE/V014129/1 (SJL). This work was performed on the NASA Pleiades and UC Davis HPC Peloton \& Impact computing clusters. This work was partially supported by the Center for Matter at Atomic Pressures (CMAP), a National Science Foundation (NSF) Physics Frontier Center, under Award PHY-2020249. Any opinions, findings, conclusions or recommendations expressed in this material are those of the author(s) and do not necessarily reflect those of the National Science Foundation.
\end{acknowledgements}
\section*{Data Availability}

Companion Jupyter notebooks for planet initialization, SPH simulation setup, data analysis, and figure plotting are included in a GitHub repository at https://github.com/PlanetSim/Postema-giant-impacts

Full data files and initial condition files for the SPH simulations performed in this study are included in a Dryad repository (DOI: 10.5061/dryad.vdncjsz44).

Updates to the HERCULES code used in this project will be made available on the HERCULES GitHub repository and archived on Zenodo.

The preparation of this study utilized the following software packages for calculations and data presentation:
\begin{enumerate}
    \item Python version 3.13.9
    \item NumPy version 2.3.5
    \item SciPy version 1.16.3
    \item Numba version 0.62.1
    \item SwiftSimIO version 10.7.2
    \item WoMa version 1.4.4
    \item SeaGen version 1.5
    \item Astropy version 7.2.0
    \item H5Py version 3.15.1
    \item Matplotlib version 3.10.8
    \item Unyt version 3.0.4
    \item SWIFT version 0.9.0
    \item HERCULES version 1.1
    \item ANEOS Pyrolite version SLVTv0.2G1
    \item ANEOS Fe-8.15wt\%Si version SLVTv0.2G1
\end{enumerate}




\clearpage
\appendix

\section{Impact Parameter Space}

The impact parameters for the giant impact simulations are summarized in Table \ref{tbl:ICs}.

\section{Smoothed rotating model CMB points}
The polynomial coefficients representing the smoothed data points for the rotating model CMB conditions included in Figure \ref{fig:hercules_v2}e,f are summarized in Table \ref{tbl:hercules_fit}.

\begin{table*}[]
\begin{tabular}{llllll}
\toprule
$M_T/M_i$ & $M_T$ & $\theta_i$ [deg] & $V_i/V_{\rm esc}$ ($Q_S$ [MJ/kg]) \\ \midrule
 2:1 &
  \small Moon (0.01 $M_\oplus$)&
  \scriptsize \begin{tabular}[c]{@{}l@{}}0\\ 10\\ 20\\ 30\\ 45\\ 60\\ 75\\ 89.7\end{tabular} &
  \scriptsize \begin{tabular}[c]{@{}l@{}}1 (0.71), 1.15 (0.93), 1.5 (1.59), 2.0 (2.83)\\ 1.15 (0.78)\\ 1 (0.53), 1.15 (0.70), 1.5 (1.18), \textit{2.0 (2.11)}\\ 1 (0.51), 1.15 (0.68), \textbf{1.45 (1.08)}, \textit{1.5 (1.16), 2.0 (2.05)}\\ 1 (0.62), \textbf{1.15 (0.81)}, \textit{1.35 (1.11), 1.5 (1.37), 2.0 (2.44)}\\ \textbf{1 (1.06), 1.05 (1.16)}, \textit{1.15 (1.40), 1.5 (2.38), 2.0 (4.22)}\\ \textbf{1 (3.72), 1.025 (3.91), 1.05 (4.11)}, \textit{1.1 (4.57)}\\ \textbf{1, 1.025, 1.05}\end{tabular} \\ \midrule
 &
  \small Mars (0.1 $M_\oplus$)&
  \scriptsize\begin{tabular}[c]{@{}l@{}}0\\ 20\\ 30\\ 45\\ 60\\ 75\\ 89.7\end{tabular} &
  \scriptsize\begin{tabular}[c]{@{}l@{}}1 (3.36), 1.15 (4.44), 1.5 (7.55), 2.0 (13.4)\\ 1 (2.51)*, 1.15 (3.32)*, 1.5 (5.64)*, \textit{2.0 (10.0)*}\\ 1 (2.54)*, 1.15 (3.24)*, \textbf{1.45 (5.15)}, \textit{1.5 (5.51)*, 2.0 (9.80)}\\ 1 (2.92), \textbf{1.15 (3.86)\textsuperscript{\textdagger}}, \textit{1.35 (5.31), 1.5 (6.56), 2.0 (11.7)}\\ \textbf{1 (5.06), 1.05 (5.58)}, \textit{1.15 (6.69), 1.5 (11.4), 2.0 (20.2)}\\ \textbf{1 (17.9)\textsuperscript{\textdagger}, 1.025 (18.8)\textsuperscript{\textdagger}, 1.05 (19.7)}\textsuperscript{\textdagger}, \textit{1.1 (21.6)}\\ \textbf{1, 1.025, 1.05}\end{tabular} \\ \midrule
 &
  \small Half-earth (0.5 $M_\oplus$) &
  \scriptsize\begin{tabular}[c]{@{}l@{}}0\\ 20\\ 30\\ 45\\ 60\\ 75\\ 89.7\end{tabular} &
  \scriptsize\begin{tabular}[c]{@{}l@{}}1 (10.2), 1.15 (13.5), 1.5 (23.0), 2.0 (40.9)\\ 1 (7.68)*, 1.15 (10.2)*, 1.5 (17.3)*, \textit{2.0 (30.7)*}\\ 1 (7.53)*, 1.15 (9.95)*, \textbf{1.45 (15.8)*}, \textit{1.5 (16.9)*, 2.0 (30.1)*}\\ 1 (8.98)\textsuperscript{\textdagger}, \textbf{1.15 (11.9)\textsuperscript{\textdagger}}, \textit{1.35 (16.4), 1.5 (20.2), 2.0 (35.9)}\\ \textbf{1 (15.6)\textsuperscript{\textdagger}, 1.05 (17.2)\textsuperscript{\textdagger}}, \textit{1.15 (20.7), 1.5 (35.1), 2.0 (62.5)}\\ \textbf{1 (55.2)\textsuperscript{\textdagger}, 1.025 (58.0)\textsuperscript{\textdagger}, 1.05 (60.9)}, \textit{1.1 (66.8)}\\ \textbf{1\textsuperscript{\textdagger}, 1.025\textsuperscript{\textdagger}, 1.05\textsuperscript{\textdagger}}\end{tabular} \\ \midrule
 &
  \small Proto-earth (0.91 $M_\oplus$)&
  \scriptsize\begin{tabular}[c]{@{}l@{}}0\\ 20\\ 30\\ 45\\ 60\\ 75\\ 89.7\end{tabular} &
  \scriptsize\begin{tabular}[c]{@{}l@{}}1 (15.6), 1.15 (20.7), 1.5 (35.1), 2.0 (62.5)\\ 1 (11.8)*, 1.15 (15.8)*, 1.5 (26.5)*, \textit{2.0 (47.0)*}\\ 1 (11.5)*, 1.15 (15.2)*, \textbf{1.45 (24.2)*}, \textit{1.5 (25.9)*, 2.0 (46.1)*}\\ 1 (13.8)*, \textbf{1.15 (18.2)\textsuperscript{\textdagger}}, \textit{1.35 (25.1), 1.5 (31.0), 2.0 (55.1)}\\ \textbf{1 (24.0)*, 1.05 (26.4)\textsuperscript{\textdagger}}, \textit{1.15 (31.7), 1.5 (54.0), 2.0 (96.0)}\\ \textbf{1 (84.8)\textsuperscript{\textdagger}, 1.025 (89.0)*, 1.05 (93.5)\textsuperscript{\textdagger}}, \textit{1.1 (102.6)}\\ \textbf{1\textsuperscript{\textdagger}, 1.025*, 1.05\textsuperscript{\textdagger}}\end{tabular} \\ \midrule
 &
  \small Super-earth (1.3 $M_\oplus$)&
  \scriptsize\begin{tabular}[c]{@{}l@{}}30\\ 45\\ 60\end{tabular} &
  \scriptsize\begin{tabular}[c]{@{}l@{}}1 (14.9)*, 1.15 (19.7)*, \textit{1.5 (33.6)*, 2.0 (59.7)*}\\ 1 (17.9), \textbf{1.15 (23.6)*}, \textit{1.5 (40.1), 2.0 (71.4)}\\ \textbf{1 (31.1)}, \textit{1.15 (41.1), 1.5 (70.0), 2.0 (124.4)}\end{tabular} \\ \midrule
\end{tabular}
\caption{Initial conditions of impact simulations included in this study. We tested five target planet sizes, each impacted by three different sizes of impactors at mass ratios of 1:2, 1:3, and 1:6. At each set of target-impactor sizes we tested at least three more-direct impact angles (30$\degree$, 45$\degree$, and 60$\degree$). For those impact angles, each target-impactor pair was tested over at least four scaled impact velocities ($V_i/V_{esc}$): 1, 1.15, 1.5, and 2.0. We also tested highly glancing impacts at 75$\degree$ and close to 90$\degree$ (actually 89.7$\degree$, the highest angle recorded in the N-body database of \citet{Carter_dataverse_2020}), each at $V_i/V_{esc}=$ 1, 1.025, and 1.05 (the highest scaled impact velocity predicted to be accretionary by \citet{genda_merging_2012}). The modified specific impact energy ($Q_S$) values for each collision are given in parentheses (though $Q_S$ values for the 89.7$\degree$ cases are nonphysical and not included, see \S \ref{subsec:QS}). Standard font face indicates simulations that produced merging outcomes, \textbf{bolded font} indicates graze-and-merge outcomes, while \textit{italicized font} indicates hit-and-runs. Asterisks(*) indicate impact events that produce synestias rather than corotating planets with disks, while daggers(\textsuperscript{\textdagger}) indicate events that produced post-impact bodies very close to the corotation limit that show both solid-planet- and synestia-like traits \citep[referred to as ``co-CoRoL'' bodies by ][]{Lock17}.}
\end{table*}
\addtocounter{table}{-1}

\begin{table*}[]
\begin{tabular}{llllll}
\toprule
$M_T/M_i$ & $M_i$ & $\theta_i$ [deg] & $V_i/V_{\rm esc}$ ($Q_S$ [MJ/kg]) \\ \midrule
3:1 &
  \small Moon &
  \scriptsize\begin{tabular}[c]{@{}l@{}}0\\ 20\\ 30\\ 45\\ 60\\ 75\\ 89.7\end{tabular} &
  \scriptsize\begin{tabular}[c]{@{}l@{}}1 (0.50), 1.15 (0.66), 1.5 (1.12), 2.0 (2.00)\\ 1 (0.36), 1.15 (0.47), 1.5 (0.80), \textit{2.0 (1.43)}\\ 1 (0.35), 1.15 (0.46), \textbf{1.5 (0.78)}, \textit{2.0 (1.38)}\\ 1 (0.41), \textbf{1.15 (0.55)}, \textit{1.5 (0.93), 2.0 (1.65)}\\ \textbf{1 (0.72)}, \textit{1.15 (0.95), 1.5 (1.62), 2.0 (2.88)}\\ \textbf{1 (2.54), 1.025 (2.67), 1.05 (2.80)}, \textit{1.1 (3.07)}\\ \textbf{1, 1.025, 1.05}\end{tabular} \\ \midrule
 &
  \small Mars &
  \scriptsize\begin{tabular}[c]{@{}l@{}}0\\ 20\\ 30\\ 45\\ 60\\ 75\\ 89.7\end{tabular} &
  \scriptsize\begin{tabular}[c]{@{}l@{}}1 (2.36), 1.15 (3.13), 1.5 (5.32), 2.0 (9.46)\\ 1 (1.70), 1.15 (2.24), 1.5 (3.82)*, \textit{2.0 (6.79)}\\ 1 (1.65), 1.15 (2.18), \textbf{1.5 (3.71),} \textit{2.0 (6.60)}\\ 1 (1.98), \textbf{1.15 (2.62)}, \textit{1.5 (4.45), 2.0 (7.92)}\\ \textbf{1 (3.46)}, \textit{1.15 (4.58), 1.5 (7.79), 2.0 (13.8)}\\ \textbf{1 (12.2), 1.025 (12.8), 1.05 (13.5)}, \textit{1.1 (14.8)}\\ \textbf{1, 1.025, 1.05}\end{tabular} \\ \midrule
 &
  \small Half-earth &
  \scriptsize\begin{tabular}[c]{@{}l@{}}0\\ 20\\ 30\\ 45\\ 60\\ 75\\ 89.7\end{tabular} &
  \scriptsize\begin{tabular}[c]{@{}l@{}}1 (7.20), 1.15 (9.52), 1.5 (16.2), 2.0 (28.8)\\ 1 (5.21)*, 1.15 (6.88)*, 1.5 (11.7)*, \textit{2.0 (20.8)*}\\ 1 (5.08)*, 1.15 (6.72)*, \textbf{1.5 (11.4)*}, \textit{2.0 (20.3)*}\\ 1 (6.13)\textsuperscript{\textdagger}, \textbf{1.15 (8.11)\textsuperscript{\textdagger}}, \textit{1.5 (13.8), 2.0 (24.5)}\\ \textbf{1 (10.8)*}, \textit{1.15 (14.2), 1.5 (24.2), 2.0 (43.0)}\\ \textbf{1 (38.0)*, 1.025 (40.0)*, 1.05 (41.9)*}, \textit{1.1 (46.0)}\\ \textbf{1, 1.025*, 1.05}\textsuperscript{\textdagger} \end{tabular} \\ \midrule
 &
  \small Proto-earth &
  \scriptsize\begin{tabular}[c]{@{}l@{}}0\\ 20\\ 30\\ 45\\ 60\\ 75\\ 89.7\end{tabular} &
  \scriptsize\begin{tabular}[c]{@{}l@{}}1 (11.0), 1.15 (14.5), 1.5 (24.7), 2.0 (43.9)\\ 1 (7.97)*, 1.15 (10.5)*, 1.5 (17.9)*, \textit{2.0 (31.9)*}\\ 1 (7.80)*, 1.15 (10.3)*, \textbf{1.5 (17.5)*}, \textit{2.0 (31.2)*}\\ 1 (9.43)*, \textbf{1.15 (12.5)\textsuperscript{\textdagger}}, \textit{1.5 (21.2), 2.0 (37.7)}\\ \textbf{1 (16.6)*}, \textit{1.15 (21.9), 1.5 (37.3), 2.0 (66.2)}\\ \textbf{1 (58.6)*, 1.025 (61.5)\textsuperscript{\textdagger}, 1.05 (64.6)\textsuperscript{\textdagger}}, \textit{1.1 (70.9)}\\ \textbf{1*, 1.025*, 1.05\textsuperscript{\textdagger}} \end{tabular} \\ \midrule
 &
  \small Super-earth &
  \scriptsize\begin{tabular}[c]{@{}l@{}}30\\ 45\\ 60\end{tabular} &
  \scriptsize\begin{tabular}[c]{@{}l@{}}1 (10.1)*, 1.15 (13.6)*, \textbf{1.5 (22.7)*}, \textit{2.0 (40.4)*}\\ 1 (12.2)*, \textbf{1.15 (16.2)}, \textit{1.5 (27.5), 2.0 (48.9)}\\ \textbf{1 (21.5)\textsuperscript{\textdagger}}, \textit{1.15 (28.4), 1.5 (48.4), 2.0 (86.0)}\end{tabular} \\ \bottomrule
\end{tabular}
\caption{{\it Continued.}}
\end{table*}

\addtocounter{table}{-1}

\begin{table*}[]
\begin{tabular}{llllll}
\toprule
$M_T/M_i$ & $M_T$ & $\theta_i$ [deg] & $V_i/V_{\rm esc}$ ($Q_S$ [MJ/kg]) \\ \midrule
6:1 &
  \small Moon &
  \scriptsize\begin{tabular}[c]{@{}l@{}}0\\ 20\\ 30\\ 45\\ 60\\ 75\\ 89.7\end{tabular} &
  \scriptsize\begin{tabular}[c]{@{}l@{}}1 (0.27), 1.15 (0.36), 1.5 (0.61), 2.0 (1.09)\\ 1 (0.18), 1.15 (0.24), 1.5 (0.41), 2.0 (0.73)\\ 1 (0.17), 1.15 (0.22), 1.5 (0.38), \textit{2.0 (0.67)}\\ 1 (0.20), \textbf{1.15 (0.26)}, \textit{1.5 (0.44), 2.0 (0.79)}\\ \textbf{1 (0.34)}, \textit{1.15 (0.45), 1.5 (0.77), 2.0 (1.37)}\\ \textbf{1 (1.20), 1.025 (1.26), 1.05 (1.32)}, \textit{1.1 (1.45)}\\ \textbf{1, 1.025, 1.05 }\end{tabular} \\ \midrule
 &
  \small Mars &
  \scriptsize\begin{tabular}[c]{@{}l@{}}30\\ 45\\ 60\\ 75\\ 89.7\end{tabular} &
  \scriptsize\begin{tabular}[c]{@{}l@{}}1 (0.80), 1.15 (1.06), 1.5 (1.81), \textit{2.0 (3.22)}\\ 1 (0.95), \textbf{1.15 (1.26)}, \textit{1.5 (2.14), 2.0 (3.80)}\\ \textbf{1 (1.66)}, \textit{1.15 (2.19), 1.5 (3.73), 2.0 (6.62)}\\ \textbf{1 (5.81)}, \textit{1.1 (7.03)}\\ \textbf{1, 1.025, 1.05} \end{tabular} \\ \midrule
 &
  \small Half-Earth &
  \scriptsize\begin{tabular}[c]{@{}l@{}}30\\ 45\\ 60\\ 75\\ 89.7\end{tabular} &
  \scriptsize\begin{tabular}[c]{@{}l@{}}1 (2.49)*, 1.15 (3.29)*, 1.5 (5.59)*, \textit{2.0 (9.95)*}\\ 1 (2.96)\textsuperscript{\textdagger}, \textbf{1.15 (3.92)}, \textit{1.5 (6.67), 2.0 (11.9)}\\ \textbf{1 (5.19)}, \textit{1.15 (6.86), 1.5 (11.7), 2.0 (20.8)}\\ \textbf{1 (18.2), 1.025 (19.2), 1.05 (20.1)}, \textit{1.1 (22.1)}\\ \textbf{1, 1.025, 1.05} \end{tabular} \\ \midrule
 &
  \small Proto-Earth &
  \scriptsize\begin{tabular}[c]{@{}l@{}}30\\ 45\\ 60\\ 75\\ 89.7\end{tabular} &
  \scriptsize\begin{tabular}[c]{@{}l@{}}1 (3.82)*, 1.15 (5.06)*, 1.5 (8.60)*, \textit{2.0 (15.3)*}\\ 1 (4.58)*, \textbf{1.15 (6.05)}, \textit{1.5 (10.3), 2.0 (18.3)}\\ \textbf{1 (8.04)\textsuperscript{\textdagger}}, \textit{1.15 (10.6), 1.5 (18.1), 2.0 (32.1)}\\ \textbf{1 (28.3), 1.025 (29.7), 1.05 (31.2)}, \textit{1.1 (34.2)}\\ \textbf{1, 1.025, 1.05} \end{tabular} \\ \midrule
 &
  \small Super-Earth &
  \scriptsize\begin{tabular}[c]{@{}l@{}}30\\ 45\\ 60\end{tabular} &
  \scriptsize\begin{tabular}[c]{@{}l@{}}1 (4.95)*, 1.15 (6.55)*, 1.5 (11.1)*, \textit{2.0 (19.8)*}\\ 1 (5.95)*, \textbf{1.15 (7.87)*}, \textit{1.5 (13.4), 2.0 (23.8)}\\ \textbf{1 (10.5)\textsuperscript{\textdagger}}, \textit{1.15 (13.8), 1.5 (23.5), 2.0 (41.8)}\end{tabular} \\ \bottomrule
\end{tabular}
\caption{{\it Continued.}}
\label{tbl:ICs}
\end{table*}

\section{Heating in head-on accretionary impacts} \label{subsec:head-on}
Head-on impacts (\(\theta_i=0\)) are often used as a fiducial case in planetary science, despite them being vanishingly unlikely in an ideal swarm of planetesimals \citep{holsapple_scaling_1993}, or extraordinarily rarely observed in N-body simulations \citep{Carter_dataverse_2020}. For this reason alone it might be deemed necessary to omit them as unrealistic edge cases for giant impact simulations, but most impact studies, including this current study, include them as a standard endmember test case.

We found that head-on collisions are notable outliers when it comes to core-mantle heat partitioning and impact heating efficiency. Unlike either oblique accretionary collisions or hit-and-run impacts, accretionary head-on collisions partitioned 80\% of their internal energy into the mantle for all impact cases and showed no decrease in this ratio towards mantle-core mass ratios at higher impact energies
(Table \ref{tbl:HE_coeff} and Figure \ref{fig:Qs_IE_hnr}) and the slope of the mantle and core scaling law fits in log-log space is largely parallel. At higher impact energies, the head-on cases thus begin to diverge more from oblique accretionary impacts. Including head-on impacts in global scaling laws would significantly decrease the predictive ability of those scaling laws. Because both oblique accretionary impacts and accretionary head-on impacts are internally consistent, we treat them separately when it comes to heat partitioning and producing scaling laws for heating efficiency.

Fitting comparable scaling laws to those described in Table \ref{tbl:HE_coeff} for accretionary head-on impacts produced \(R^2\) values for 0.943 and 0.960 for the mantle and core, respectively. These corresponded to fit errors of \(\pm\)0.036 and \(\pm\)0.030 in log-log space and percent errors in non-log space of \(\pm\)8.0\% and \(\pm\)6.7\% for the mantle and core, respectively. Fitting parameters are given in Table \ref{tbl:HE_coeff_0}.

\begin{figure}
    \centering
    \includegraphics[width=.7\textwidth]{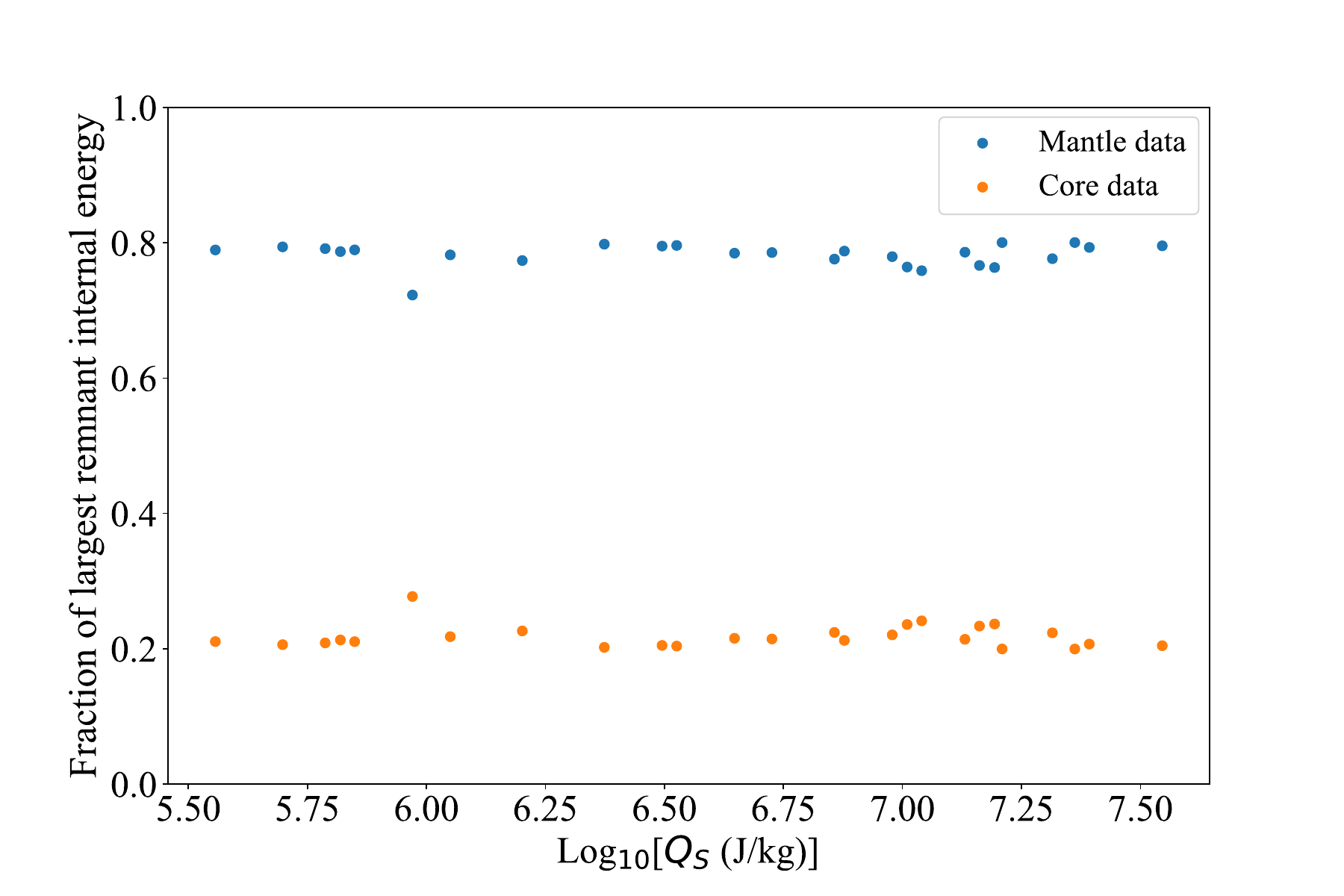}
    \caption{Mantle-core internal energy partitioning for accretionary head-on collisions \(\theta_i=0\). For all collisions, post-impact mantles receive 80\% of the internal energy budget of the planet, unlike for oblique accretionary and hit-and-run collisions (contrast with Figure \ref{fig:IE-partition}}
    \label{fig:IE-partition-0}
\end{figure}

\begin{figure}
    \centering
    \includegraphics[width=.7\textwidth]{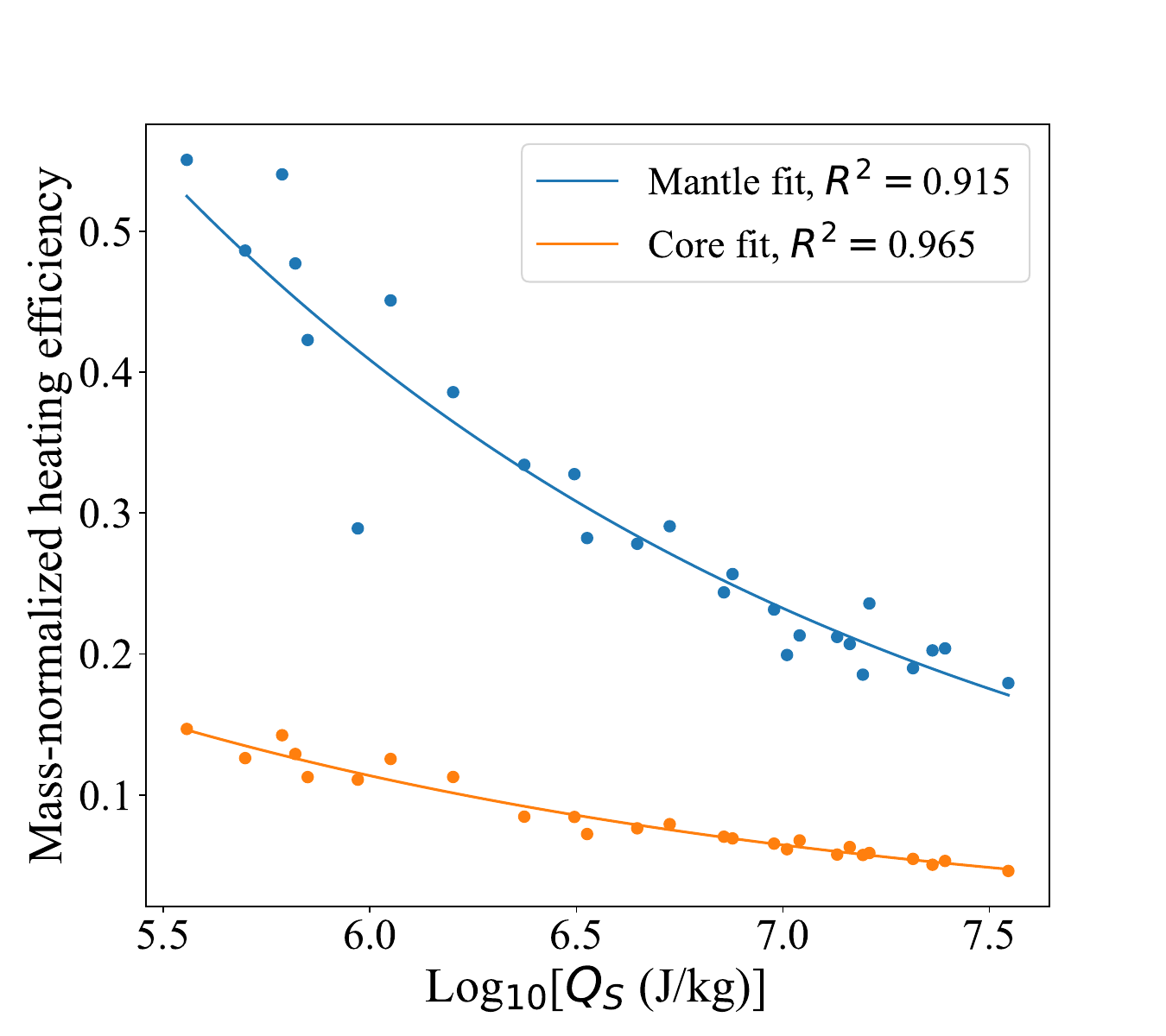}
    \caption{Mass-normalized heating efficiency for accretionary head-on impacts, similar to Panel \textbf{b} in Figure \ref{fig:Qs_IE_hnr}.}
    \label{fig:Qs_IE_0}
\end{figure}

\begin{table}[]
\centering
\begin{tabular}{lll}
\multicolumn{3}{l}{\(\text{IE}_{\text{norm}}/E_{\text{tot}}=(M_{\text{bnd}}/M_T)\cdot10\string^[AQ_S+B]\)} \\ \hline
Accretion, only head-on & $A$ & $B$ \\  \hline
Mantle & -0.24872469 & 1.11282925 \\
Core & -0.24985274 & 0.56389473 \\ \hline 
\end{tabular}%

\caption{Scaling law coefficients for the accretionary head-on scaling law fits depicted in Figure \ref{fig:Qs_IE_0}. The mantle data fit produces an \(R^2\) value of 0.964 while the core fit produces an \(R^2\) of 0.925. Note that unlike the fits described in Table \ref{tbl:HE_coeff}, that the value of the \(A\) coefficients are very similar for the mantle and core. This is due to the constant level of internal energy partitioning between the mantle and core over the range of impact energies included in this study.}
\label{tbl:HE_coeff_0}
\end{table}

\section{Modified Specific impact energy $Q_S$ and pre-impact deformation in extremely glancing collisions} \label{subsec:QS}
Specific impact energy $Q_S$ \citep{Lock17} was developed as a correction to the interacting specific impact energy term $Q'_R$, developed in \citet{Leinhardt12} and meant to describe the amount of impact energy concentrated in the interacting mass of an oblique collision. The original impact simulations that were the basis for $Q_S$ were performed on spherical bodies that began immediately tangent to each other, which represents the time immediately before impact for two solid spheres. A central part of the derivation of $Q'_R$ is the introduction of an ``interacting mass'' fraction term $\alpha$, which is ``the mass fraction of the projectile estimated to be involved in the collision''\citep{Leinhardt12} and is described by Equation 11 in \citet{Leinhardt12}:
\begin{equation}
    \frac{M_{\text{interact}}}{M_i}=\frac{3R_il^2-l^3}{4R_i^3}\equiv\alpha\quad,
\end{equation}
where $M_i$ and $R_i$ are the mass and radius of the impactor, and $l$ is the projected length of the projectile overlapping the target in the $y$-axis of the collision plane (where the projectile velocity vector is parallel to the $x$-axis at the moment of impact). In the case where the target and projectile are solid spherical bodies, $l$ will approach zero as the impact angle approaches 90$\degree$, as the surface of the target and impactor will be perfectly tangent at \(\theta_i=90\degree\). Thus, $\alpha$ also approaches zero as impact angles approach 90$\degree$. The modified reduced mass of the target and the interacting impactor is then described by Equation 12 in \citet{Leinhardt12}:
\begin{equation}
    \mu_\alpha=\frac{\alpha M_iM_T}{\alpha M_i+M_T}\quad,
\end{equation}
where $M_T$ is the mass of the target planet. It follows that as \(\theta_i\to90\degree\) and \(l,\alpha \to 0\), $\mu_\alpha$ also approaches zero. A problem then appears with the definition of $Q'_R$ as a modification of impact energy $Q_R$ (which is always a nonzero, finite, positive quantity), given in Equation 13 of \citet{Leinhardt12}:
\begin{equation}
    Q'_R=\frac{\mu}{\mu_\alpha}Q_R\quad,
\end{equation}
where $\mu$ is the standard reduced mass of the target-impactor system. Because \(\mu_\alpha \to 0\) as \(\theta \to 90\degree\), it is evident that $Q'_R$ then tends to infinity as its denominator approaches zero. In the regime of solid spherical planets, this makes sense, as the kinetic energy of the impacting planet and gravitational potential well of the close pass between the target and impactor still remain, but the amount of mass that actually interacts in the initial collision becomes smaller and smaller, thus making the impact energy per interacting mass greater and greater. This issue breaks the usage of this term as a catch-all energy value to describe the general shock effects of the collision, as a totally glancing planet will have demonstrably lower physical and thermal processing than a more direct impact. This is primarily due to the fact that during extremely glancing impacts, gravity-dominated bodies experience tidal deformation, leading to a significant mass fraction of the bodies interacting even in cases where solid spheres would barely touch each other.

\citet{Lock17} introduced $Q_S$, a corrected version of $Q'_R$ that more accurately describes how efficiently impact energy is coupled to the impact-induced shock pressure field, aiming for the parameter to be ``proportional to the entropy increase in the impacting bodies'':
\begin{equation}
    Q_S=Q'_R\left(1+\frac{M_i}{M_T}\right)(1-b)
\end{equation}
where $b$ is the sine of impact angle $\theta_i$. The first correction factor accounts for the fact that a larger volume of material is shocked in more equal-sized impacts. The second correction factor is more relevant here, and attempts to account for the fact that grazing impacts couple impact energy to shock heating with decreasing efficiency. By itself, this second correction term appears that it might address the issue of \(\lim\limits_{\theta_i \to 90\degree}Q'_R=\infty\), as \(b=1\) for \(\theta_i=90\degree\). However, either a l'Hôpital's rule test or a numeric investigation will still show that $Q_S$ tends towards infinity. For this reason, although our collision dataset includes impact angles very close to 90$\degree$, we omit those collisions from our analysis and do not present their $Q_S$ values in Table \ref{tbl:ICs} as their unrealistically large $Q_S$ defeats the purpose of using such a term as a proxy for thermal impact processing. We caution that care should be taken when using specific impact energy terms such as $Q_S$ or $Q'_R$ for impact angles significantly above 75$\degree$, as extrapolation from scaling laws built around lower impact angles will be inaccurate.

The main cause of this failure of $Q_S$ or $Q'_R$ to accurately capture large impact angles stems from the assumption of solid spherical bodies. Especially when dealing with standard SPH (which includes no treatment for material strength), giant impact simulations for gravity-dominated bodies will involve planets that largely behave as fluids and will show a significant amount of tidal distortion as the target and impact make their initial approach. A value for $l$ based off of solid spheres will never be accurate for highly oblique angles, as even in totally glancing cases there will be significant non-zero overlap due to tidal forces. Because tidal distortion means these objects are no longer spheres and cannot be accurately corrected by placing a floor value on $l$, a proper treatment for $\alpha$ will likely require a future numerical investigation of SPH simulations to characterize the amount of distorted interacting mass at various planet sizes, masses, impact velocities, and target-impactor mass ratios. Standard SPH should be seen as one endmember for realistic impact scenarios, as realistic treatments of material strength will result in behavior somewhere between fluid-only deformation for larger or weaker planets and solid spheres without any deformation for small icy bodies.

\begin{table*}[h]
    \centering
        \begin{tabular}{lllll}
        \multicolumn{5}{r}{\(100\% \cdot (P_\text{CMB\,e.m.} - P_\text{CMB\,rot.})/ P_\text{CMB\,rot.} = A(\text{log}_{10}[Q_S])^3+B(\text{log}_{10}[Q_S])^2+C(\text{log}_{10}[Q_S])+D\)} \\
        \hline
        & \(A\)& $B$ & $C$ & $D$ \\
        \hline
        Value & 5.64543622 & -105.51448005 & 651.23892659 & -1335.41975242 \\
        \end{tabular}%
    \caption{Fit coefficients for the polynomial describing the static empirical model CMB pressure deficits compared to the rotating model as shown in Figure \ref{fig:hercules_v2}e as a function of specific impact energy \(Q_S\) in units of J/kg.} 
    \label{tbl:hercules_fit}
\end{table*}

\section{Polynomial pressure extrapolations across the CMB}\label{subsec:polynomial}
\begin{figure}
    \centering
    \includegraphics[width=0.8\linewidth]{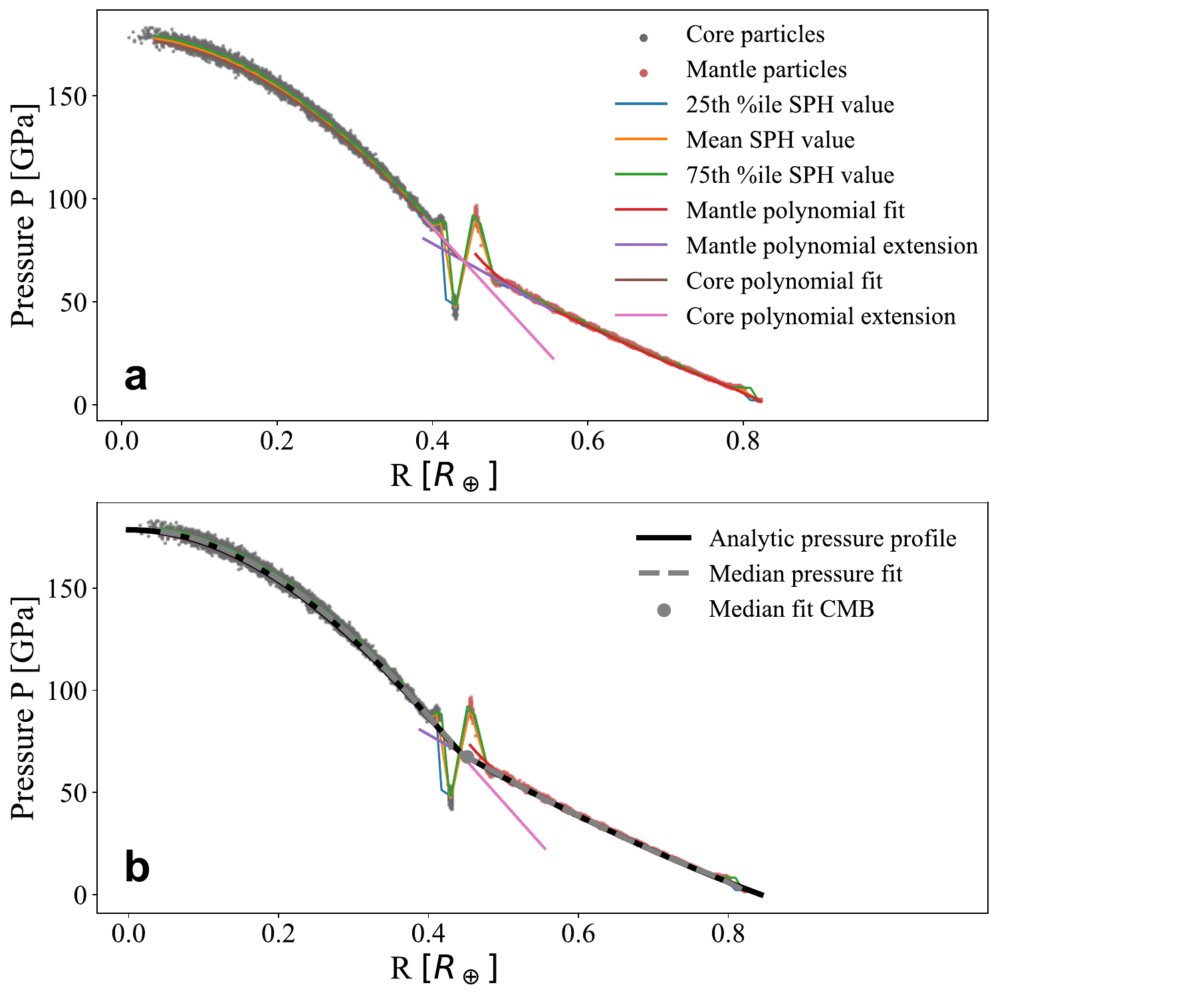}
    \caption{A comparison between the polynomial extrapolation method for finding pressure-radius relationships across CMB discontinuities in an initialized SPH body, the SPH data that the polynomial extrapolation is fitting, and the ``true'' analytic pressure profile that the polynomial method is expected to reproduce. Panel \textbf{a}: a comparison between SPH data (raw data points and moving averages of the range of SPH values) and the polynomial fit and extension regions. Panel \textbf{b}: the data from Panel \textbf{a}, now overlaid with the combined polynomial and extrapolation regimes chosen to represent the SPH pressure-radius profile (dashed grey line), the CMB of that pressure-radius relation (grey dot), and the analytic profile used to create the pre-initialization SPH planet (black line).}
    \label{fig:polyfit}
\end{figure}

As described in \S\ref{sec:methods:SPH_init}, planets generated via the SPH method tend to exhibit perturbations in pressure space across density discontinuities (Figure~\ref{fig:polyfit}). It is therefore necessary to adopt some method of extrapolating pressure between regions of continuous pressure-radius relationships either side of density boundaries in order to approximate a realistic pressure (and other properties) near density discontinuities (e.g., the CMB). In this study, as in others \citep[e.g.,][]{Lock2018,Stewartinprep}, an approximate, continuous pressure relationship that crossed the CMB was performed by fitting two quartic polynomials in radius-pressure space for the mantle and core regions that did not display any pressure excursions. The discontinuous regions are defined as locations where additional pressure values significantly diverge from the moving average of the rest of the pressure-radius data. At the edge of the two continuous fit regions, the pressure profile was linearly extrapolated based on the average slope of the polynomial fit points of the 1000 equatorial SPH particles closest to the CMB (but still within the continuous regions). The point where the two linear extrapolations met across the CMB pressure discontinuity was taken as the transition between the mantle and core pressure profiles. The final pressure profiles was then formed from the combination of the two polynomial and two linear profiles. 

To gauge the accuracy of this method, we compared the pressure profiles and the pressure value of the CMB produced from fitting a relaxed SPH body to the smooth analytic pressure profile that was used to produce the initial, pre-relaxation SPH body. In this scenario, the analytic profile is the ``true'' pressure-radius relationship that the SPH body is supposed to represent and which the fitting method is attempting to reproduce (see Figure \ref{fig:polyfit}). It is important to note that a pre-impact planet is a ``worst case'' scenario, as simulated post-impact bodies typically have much smoother pressure profiles across the CMB due to the presence of super-heated iron at the top of the core which reduces the density contrast at the CMB (e.g., Figure~\ref{fig:tempprofile}a). We found that the polynomial fits and extension closely approximated the analytic profile, indicating that the methodology is valid to apply to post-impact bodies. For each post-impact body studied in the final survey, the polynomial extrapolation result was visually inspected to assure that it functioned correctly, did not diverge to unrealistic values not represented in the SPH data, and monotonically increased from surface to center. Out of all of the impact simulations included in this study, only one Moon-mass impact had such extreme pressure discontinuities that a coherent radius-pressure profile could not be algorithmically resolved, and this impact was removed from further analysis. 

\section{Supplementary Figures} \label{sec:subfigs}
\begin{figure}[h]
    \centering
    \includegraphics[width=1.05\textwidth]{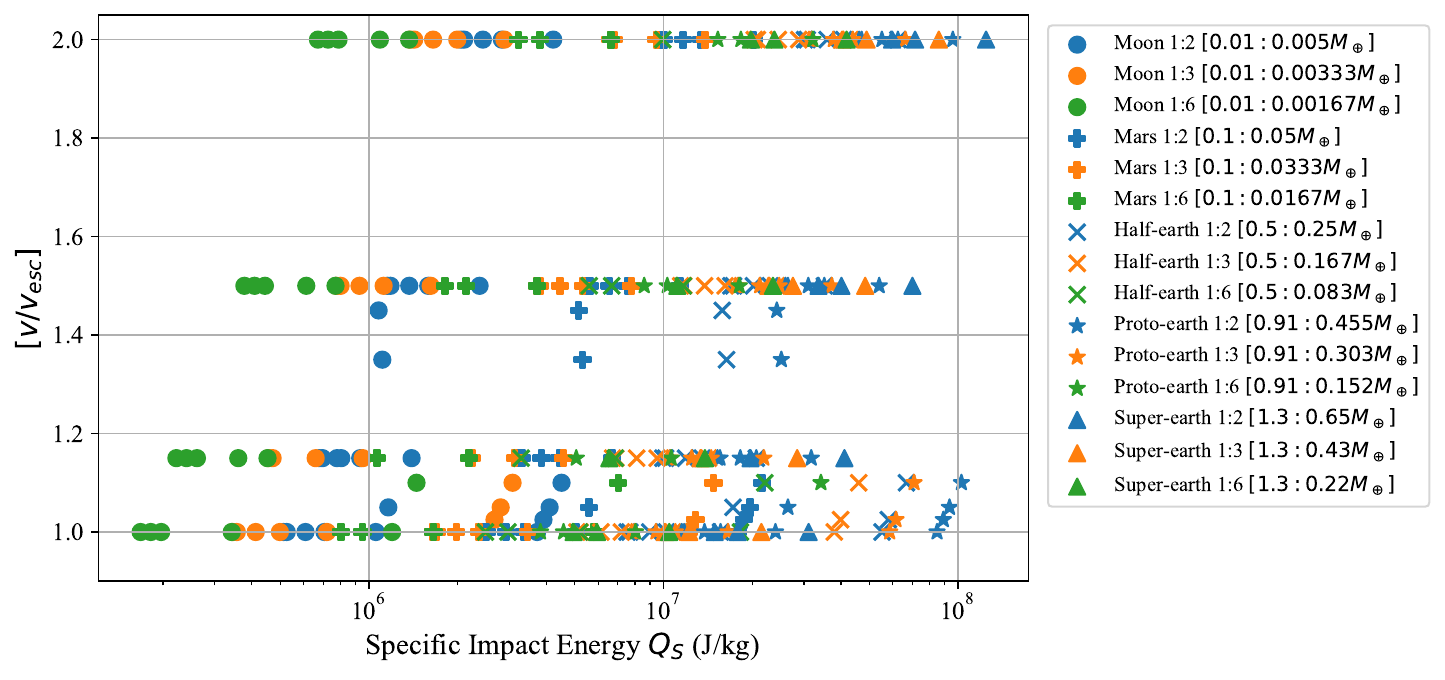}
    \caption{Initial impact conditions in our study, comparing specific impact energy \(Q_S\) to scaled impact velocity \(v/v_{esc}\). The legend describes the masses and mass ratios of the collision dataset, with the actual mass values of each target-impactor pair given in brackets.}
    \label{fig:Qs_v_ICs}
\end{figure}

\begin{figure}[h]
    \centering
    \includegraphics[width=1.05\textwidth]{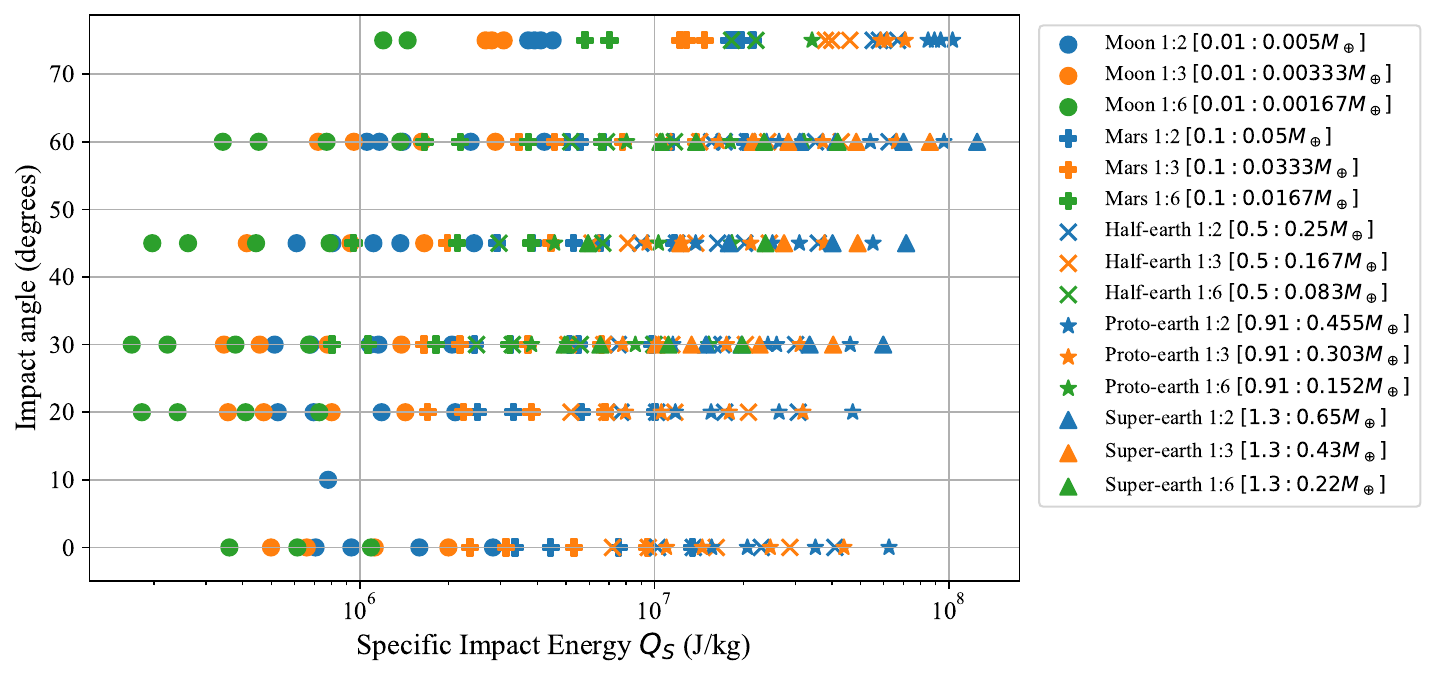}
    \caption{A corollary of \ref{fig:Qs_v_ICs} for impact angle and specific impact energy \(Q_S\).}
    \label{fig:Qs_theta_ICs}
\end{figure}

\begin{figure}
    \centering
    \includegraphics[width=0.75\textwidth]{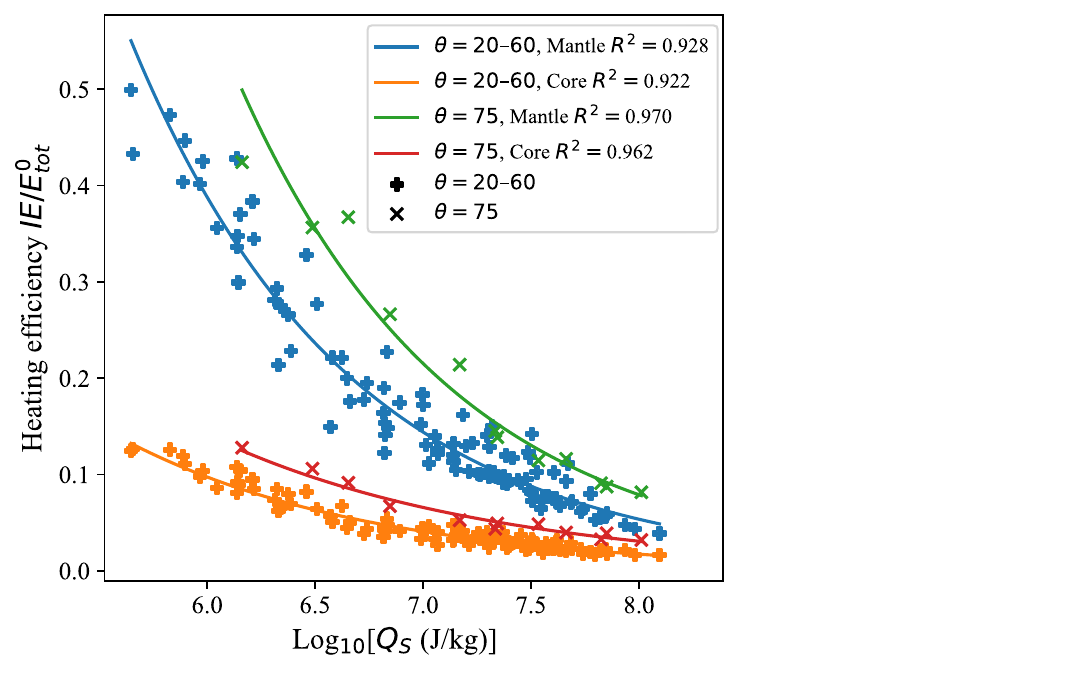}
    \caption{
    Hit-and-run impact heating efficiency relationships as a function of specific impact energy $Q_S$: the final mantle or core internal energy of the largest remnant as a fraction of total initial system energy (\(\text{IE}_{\text{mantle}}/E_{\text{tot}}\) or \(\text{IE}_{\text{core}}/E_{\text{tot}}\)). Similar to the accretionary groupings, this data falls into two groups: the bulk of impact angles tested ($\theta=20\degree$ to $60\degree$), and significantly oblique impacts ($\theta\sim75\degree$). For impact angles much higher than \(75\degree\), $Q_S$ cannot accurately capture the amount of interacting mass in the collision due to the importance of pre-impact tidal deformation.}
    \label{fig:Qs_IE_hnr_supp}
\end{figure}

\begin{figure}
    \centering
    \includegraphics[width=.8\textwidth]{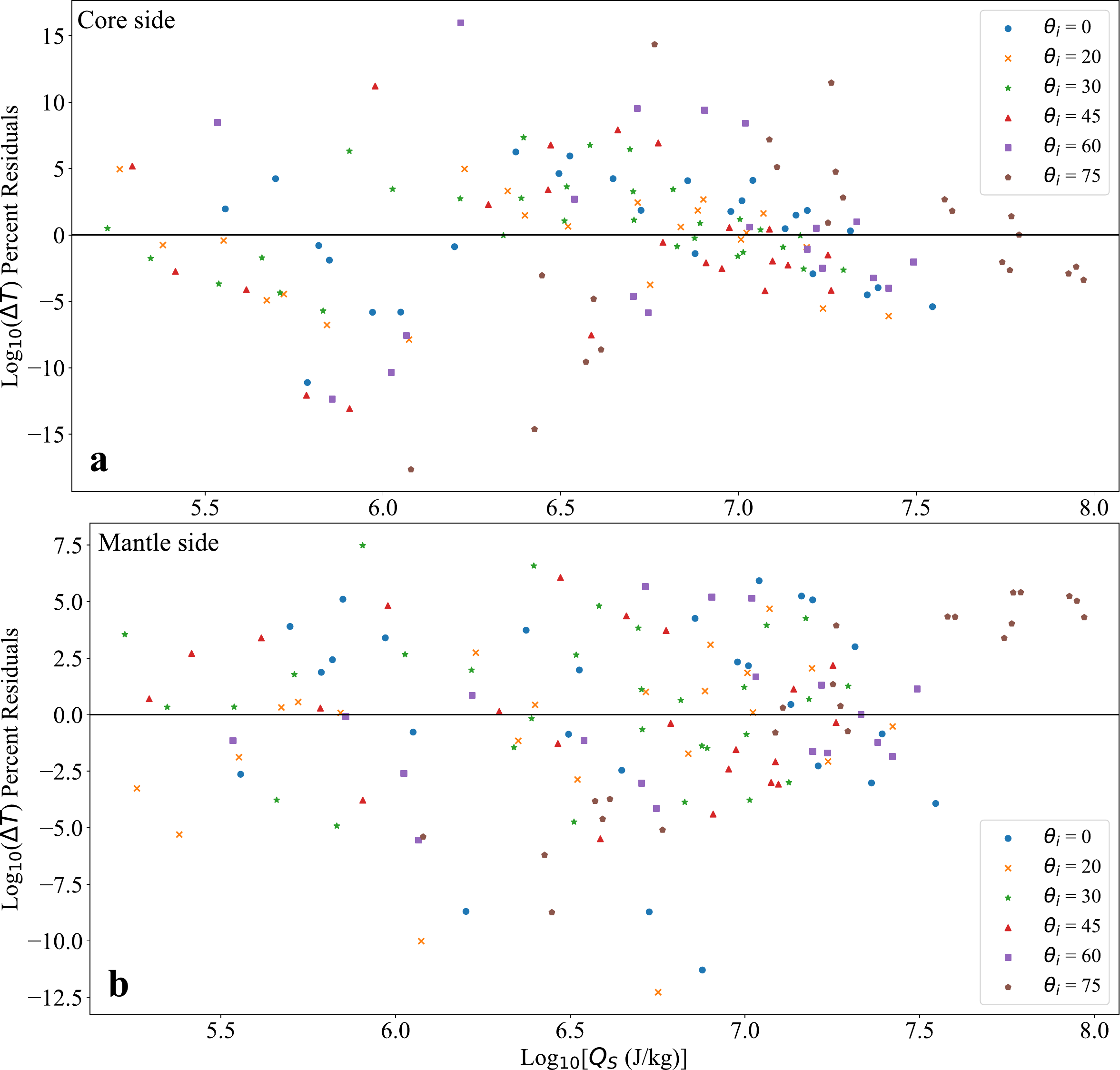}
    \caption{Percent residuals for the core-mantle boundary temperature change fits depicted in Figure \ref{fig:3D_dT_fit}, \added{for core (Panel \textbf{a}) and mantle (Panel \textbf{b}) sides of the CMB}.}
    \label{fig:CMB_T_resids}
\end{figure}

\begin{figure}
    \includegraphics[width=1\textwidth]{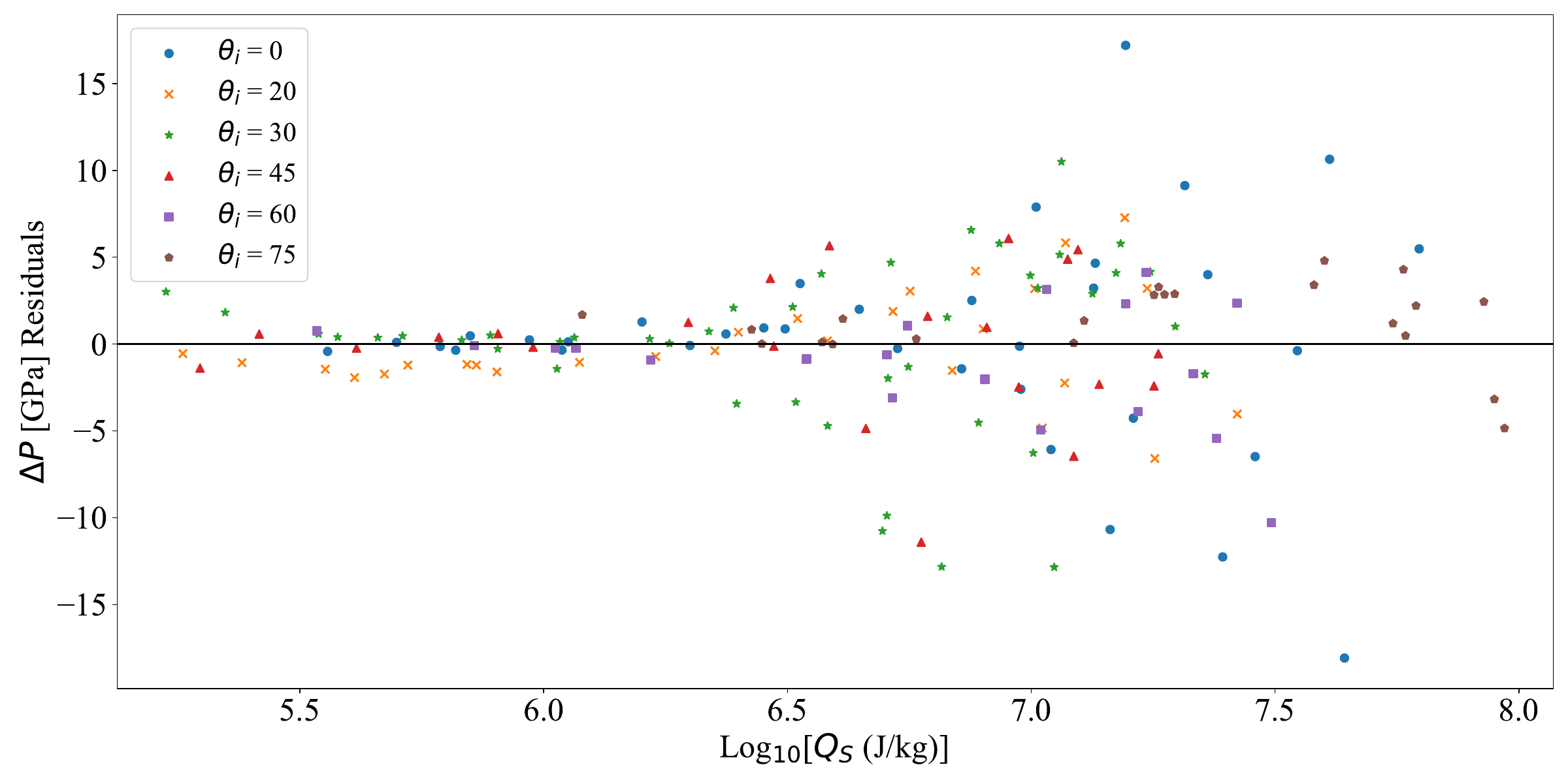}
    \caption{Absolute residuals for the core-mantle boundary pressure change fits depicted in Figure \ref{fig:3D_dP_fit}. Absolute values instead of percentages are depicted due to data at smaller specific impact energies being both positive and negative but small in overall magnitude, which creates extreme percent errors for small absolute residuals. The importance of predicting small near-zero values is to indicate that CMB pressures were relatively constant.}
    \label{fig:CMB_P_resids}
\end{figure}

\begin{figure}
    \centering
    \includegraphics[width=0.8\linewidth]{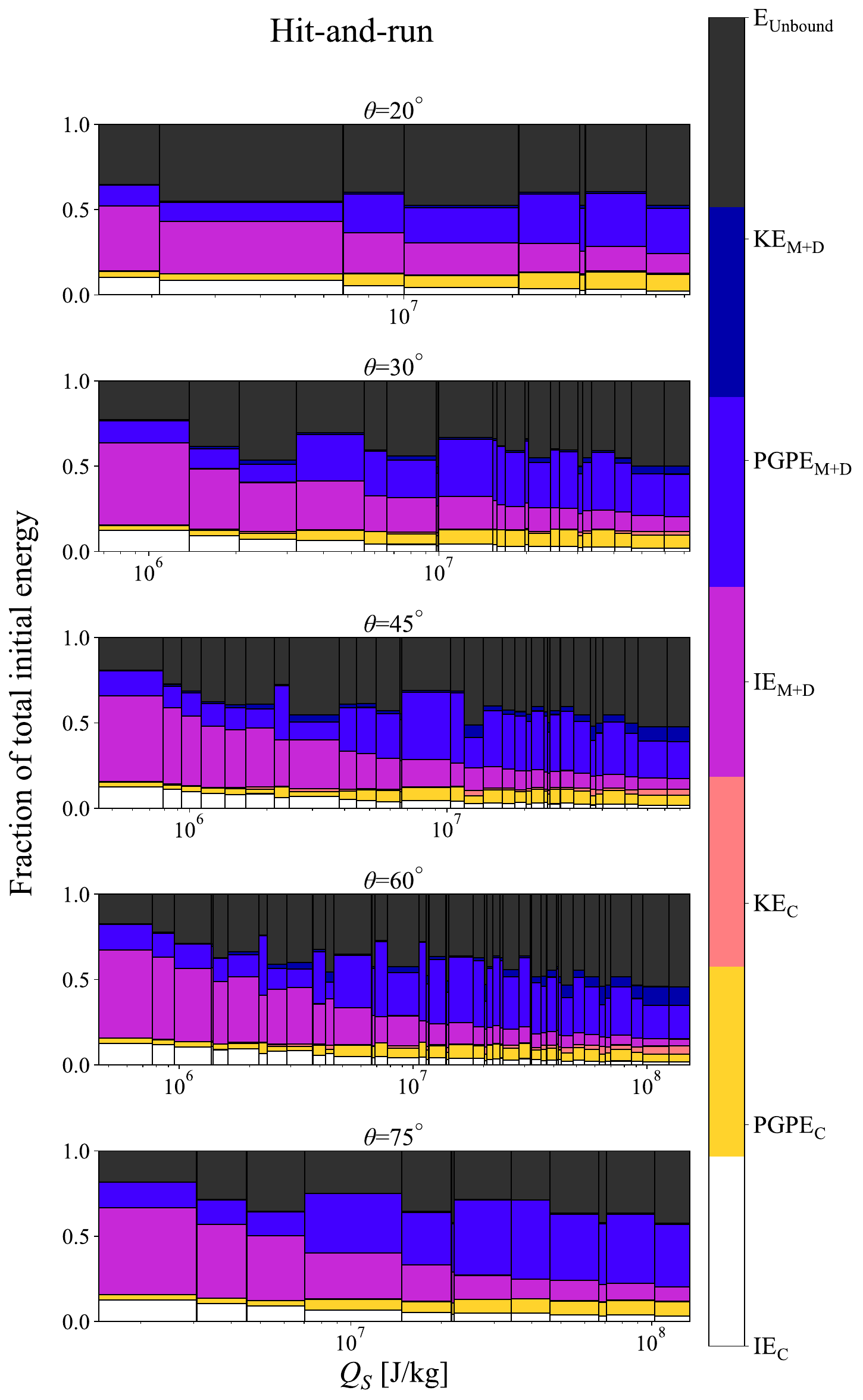}
    \caption{Energy budgets for hit-and-run impacts included in this study, similar to Figure \ref{fig:IE_plus_PGPE}.}
    \label{fig:energy_hnr}
\end{figure}

\begin{figure}
    \centering
    \includegraphics[width=0.8\linewidth]{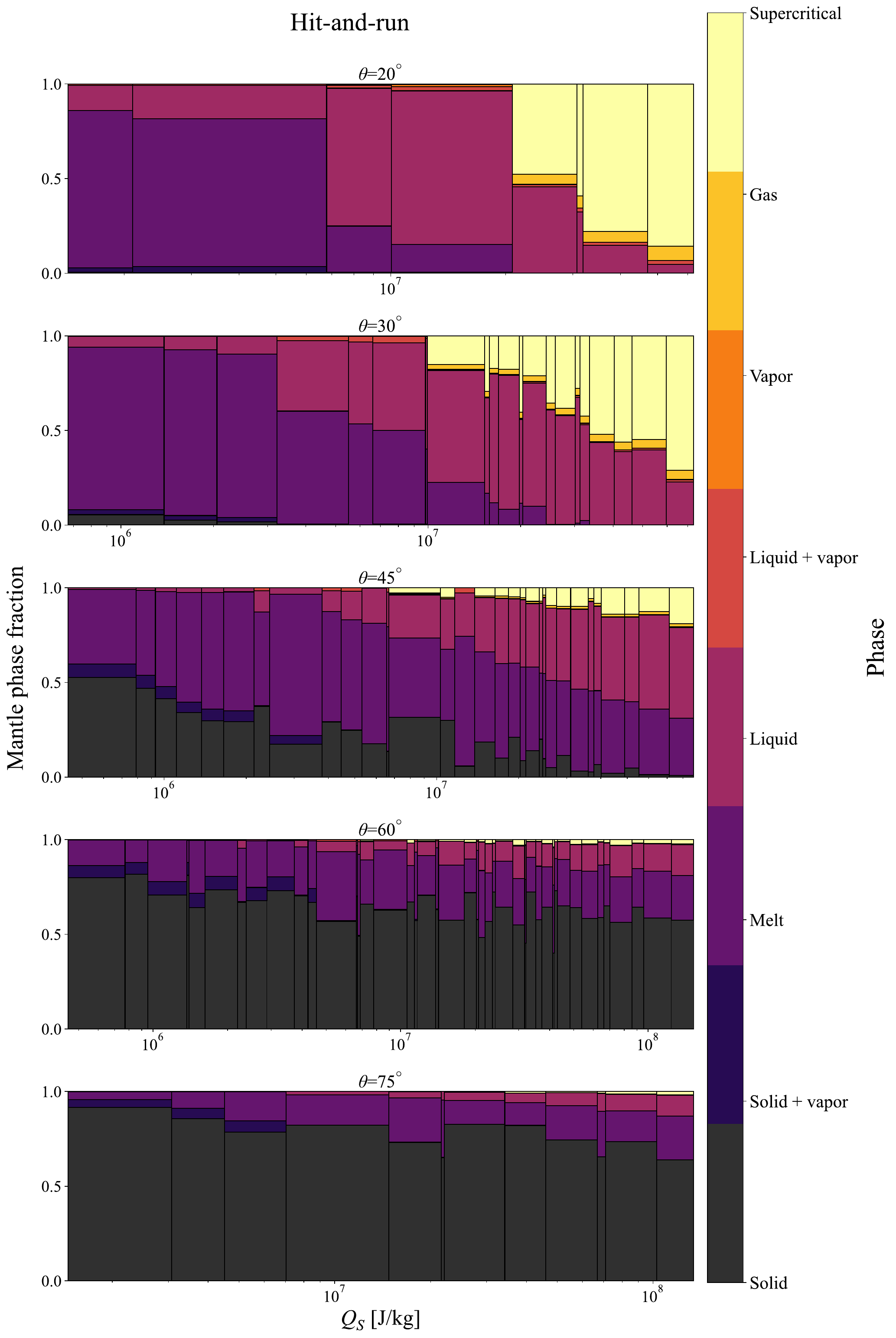}
    \caption{Mantle phase fractions for hit-and-run impacts included in this study, similar to Figure \ref{fig:mant_melt}. Low-angle hit-and-run impact still produce full-mantle melting similar to accretionary impacts, largely due to the higher velocities that are necessary to achieve a hit-and-run outcome. High impact angles leave significantly more solid material. No head-on impacts were able to produce hit-and-run outcomes regardless of velocity.}
    \label{fig:melt_hnr}
\end{figure}

\begin{figure}
    \centering
    \includegraphics[width=0.8\linewidth]{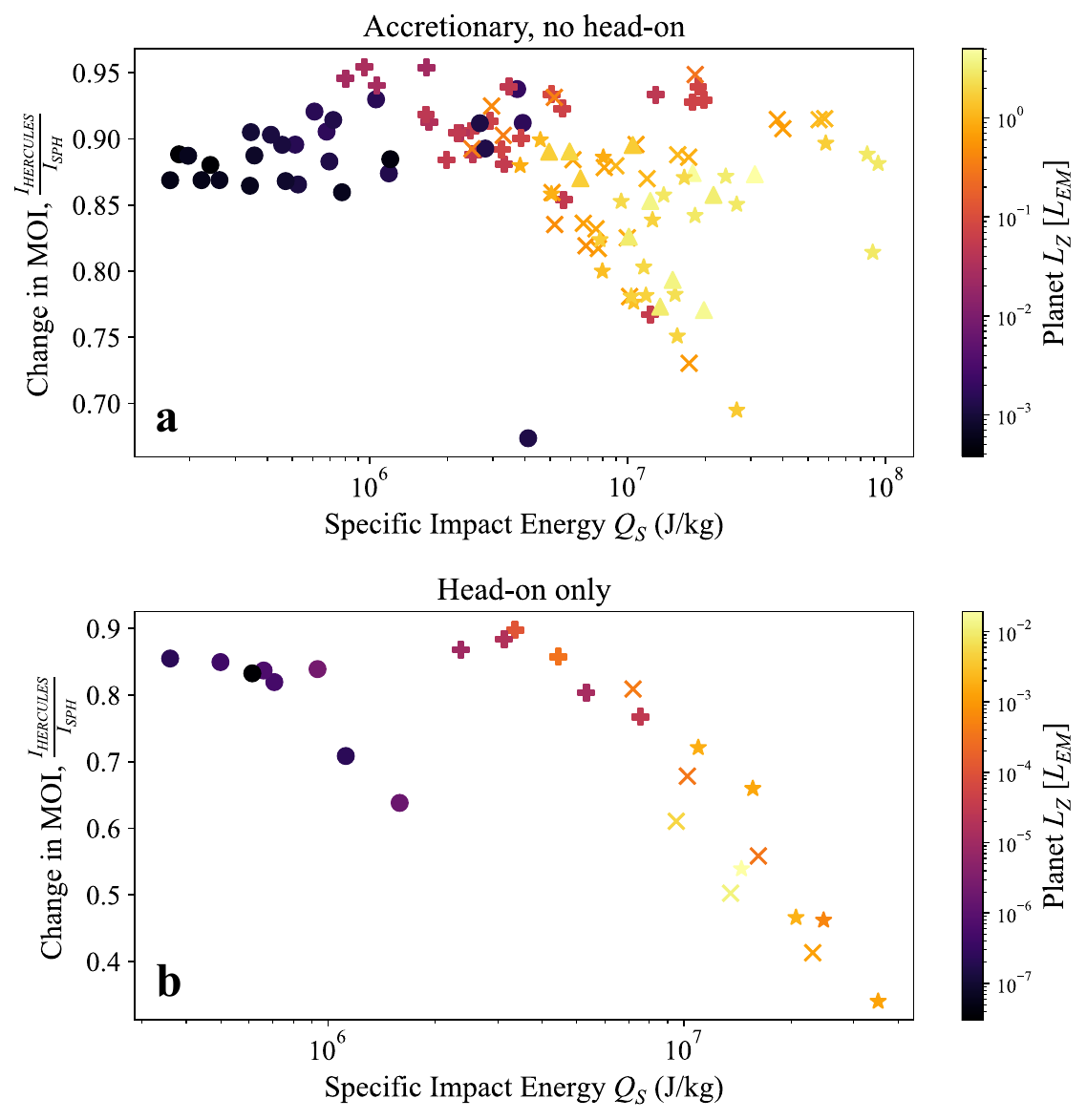}
    \caption{Fractional change in moment of inertia between the co-rotating portion of the final post-impact SPH bodies and their corresponding rotating planet profiles for each accretionary event in this study. Panel \textbf{a} shows oblique impacts while Panel \textbf{b} shows head-on events. Markers indicate the size of the target body as described in Figure \ref{fig:misc_Qs}.}
    \label{fig:dMOI}
\end{figure}

\begin{figure}
    \centering
    \includegraphics[width=0.8\linewidth]{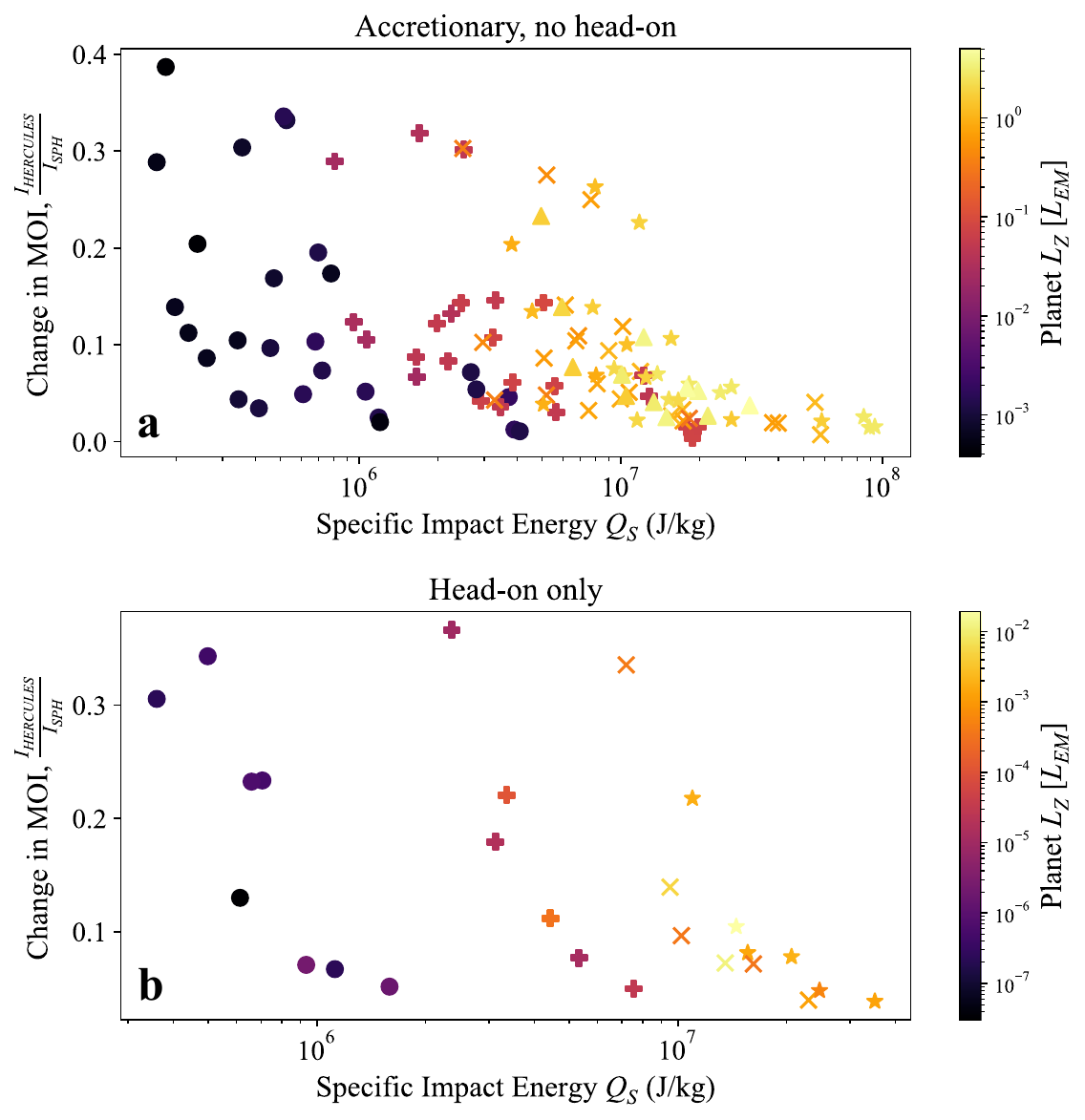}
    \caption{Similar to Figure \ref{fig:dMOI}, this figure shows fractional change in moment of inertia between the final post-impact SPH bodies including disc mass and their corresponding spinning rotating planet profiles for each accretionary event in this study. Panel \textbf{a} shows oblique impacts while Panel \textbf{b} shows head-on events. 
    Due to the highly extended nature of post-impact bodies (many of which form synestias), the moment of inertia of the discs are extremely large despite containing less than 5\% of the overall bound mass.
    Similar to Figure \ref{fig:dMOI}, markers indicate the size of the target body as described in Figure \ref{fig:misc_Qs}.}
    \label{fig:dMOI_disc}
\end{figure}

This section contains supplementary material referenced in the text.
Figures \ref{fig:Qs_v_ICs}\&\ref{fig:Qs_theta_ICs} contain information about the impact parameter space included in this study. Figure \ref{fig:Qs_IE_hnr_supp} depicts hit-and-run impact heating efficiency comparable to Figure \ref{fig:Qs_IE_hnr} for accretionary collisions. Figures \ref{fig:CMB_T_resids}\&\ref{fig:CMB_P_resids} displays the residuals for the fits shown in Figures \ref{fig:3D_dT_fit}\&\ref{fig:3D_dP_fit}. Figure \ref{fig:energy_hnr} shows the energy budgets for the largest remnant of hit-and-run collisions, comparable to Figure \ref{fig:IE_plus_PGPE} for accretionary collisions. Likewise, Figure \ref{fig:melt_hnr} shows phase budgets for hit-and-run collisions comparable to Figure \ref{fig:mant_melt}. Finally, Figures \ref{fig:dMOI}\&\ref{fig:dMOI_disc} show changes in the moment of inertia between accretionary post-impact SPH bodies and their corresponding rotating model planets in HERCULES for the corotating region and full planet plus disk, respectively.

\clearpage
\bibliographystyle{aasjournal.bst} 
\bibliography{refs-rev2}





\end{document}